\documentclass{revtex4}
\usepackage{graphicx}% Include figure files
\usepackage{amsmath}
\begin{document}

\title{Obtaining a scalar fifth force\\
 via a broken-symmetry couple between the scalar field and matter}

\author{Hai-Chao Zhang}
\email{zhanghc@siom.ac.cn}

\affiliation{Key Laboratory for Quantum Optics, \\
Shanghai Institute of Optics and Fine Mechanics, Chinese Academy of Sciences, Shanghai 201800}

\date{\today}

\begin{abstract}
A matter-coupled scalar field model is presented in obtaining a scalar fifth force when the constraint of the current cosmological constant is satisfied. The interaction potential energy density between the scalar field and matter has a symmetry-breaking form with two potential wells. The cosmological constant is proven to be a value of the scalar-field's self-interaction potential energy density at the minimum of an effective matter-density-dependent potential energy density. The effective potential is a sum of the interaction potential and the self-interaction potential of the scalar field. The scalar field can stably sit at the minimum and then the time-dependent cosmological `constant' behaves like a constant. The scheme does not conflict with chameleon no-go theorems. However, since the quintessence is trapped by one of the interaction potential wells the observed cosmic acceleration can be accounted for the scalar field. The scalar field is also extrapolated to account for inflation at the inflationary era of the Universe. In this era matter fluid is relativistic and then the interaction potential wells vanish. The unconfined quintessence therefore dominates the evolution of the Universe. It is concluded that `Planck 2018 results' favour the closed space of the Universe. The reasons are not only the measured value of the current Hubble constant, but also the observed feature of a concave potential in the framework of single-field inflationary models. By invoking a pseudo-potential in the inflationary era, the concave feature can be attributed to the pseudo-potential although the self-interaction potential is a convex function. The pseudo-potential is defined by a sum of the self-interaction potential and the energy density scale of the curvature of the Universe. It is that the positive curvature leads to the concave feature of the pseudo-potential. Within the constraints of the cosmological constants including the maximum cosmological constant in the inflationary era, a large strength of the fifth force compared with gravity is obtained. Due to the short range of the interaction, the local test of gravity is satisfied. The coupling coefficient denoting the force strength is inversely proportional to ambient density, while the interaction range is inversely proportional to square root of the density. For the current matter density $\sim {10^{ - 27}}\,\rm{kg / {m^{3}}}$ of the Universe the corresponding interaction range is $\sim 5 \,\rm{\mu m}$ and the coupling coefficient is $\sim{10^{ 31}}$. Since the fifth force is localized in an extreme thin shell, testing experiments might be designed so as to the test objects can pass through the thin shell.
\end{abstract}

\maketitle
\section{Introduction}\label{Introduction}

The acceleration of the cosmic expansion has now been firmly established \cite{z27,z28}, and the cosmological parameters are constrained at a sub-percent level \cite{z1,z1i}. A possible origin of this repulsive gravitational effect is new scalar fields coupling to matter \cite{z2,z3,z4,z5,z6,z7}. According to quantum field theory, the coupled scalar fields could produce new fifth forces \cite{z33,z36,z37,z38,z39,z40}. However, this setting still lacks specification of how to depict the fifth forces in a precise mathematical mode with the constraint of the cosmological observations and laboratory experiments, such as the cosmological constant, the ratio of matter density to the total energy density in the Universe, the precision measurements of hydrogenic energy levels \cite{z15}, etc. Since the fifth forces have not yet been observed in laboratory \cite{z8} and in solar-system experiment, modified gravity models, such as scalar field theories including chameleon \cite{z3}, symmetron \cite{z5,z6} and dilaton \cite{z7}, introduce screening mechanisms to suppress the coupling strength and/or the interaction range through dense environments. The initial motivations of introducing scalar field to understand dark energy, especially to obtain naturally the cosmological constant, remain as open questions in physics.

The idea of quintessence for solving the problem of the cosmological constant is that the potential energy of a single scalar field dynamically relaxes with time \cite{z34,z9,z10,z35}. It argues that, since our universe is old enough, the cosmological constant becomes smaller from its `natural' value of Planck energy scale \cite{z30,z31,z32,z29}. Owing to its dynamic property \cite{z104,z106,z112}, this scenario is considered as one of the possible models to overcome Weinberg's no-go theorem \cite{z11,z12}. However, the problem is \cite{z54}: if the dark energy evolves slowly on the cosmological time scale, the requisite potential of the scalar field is often regarded as very shallow. The shallowness implies that the mass of the scalar field is smaller than $\hbar H_0\sim 10^{-33} \, \rm{eV}$ with $H_0$ denoting the current Hubble constant. And then the scalar field leads to a long range interaction \cite{z66} if it couples to ordinary matter. The absence of observable interaction and the constraint of the equivalence principle imply the existence of some suppressing mechanism. Sometimes, one even describes that the scalar field does not couple to baryons but only couple to dark matter.

The other `natural' value of the cosmological constant is zero \cite{z2}. The coupled scalar filed theory argues that the coupling of the scalar field to matter may lead its potential energy dynamically to evolve from zero to the observed cosmological constant \cite{z7,z52}. For chameleon-like scalar field (e.g., symmetron and varying-dilaton), chameleon no-go theorems \cite{z101,z13} have been proven based on the assumption that the strength of chameleon-like force is comparable to gravity. Unfortunately, from chameleon no-go theorems, an misleading corollary is extensively accepted that chameleon-like scalar field cannot account for the cosmic acceleration except as some form of dark energy \cite{z101,z13,z57}. This seems to imply that the chameleon-like model cannot simultaneously screen and drive dark energy.

The misleading corollary at least results from one of the additional requirements that the scalar field should mediate a long-range interaction in low-density regions \cite{z80} (e.g., the current density of the Universe). The restrictive requirement may come from the consideration that the field should be potent to enact cosmological effects (such as the acceleration of the cosmic expansion) through the long-range interaction. From the $\Lambda$CDM model \cite{z26}, we know that, it is the cosmological constant that drives the Universe acceleration rather than any long-range force \cite{z100}. When a scalar field is used to explain the cosmological constant, it is worth noting that, the potential density of the scalar field links to the cosmological constant rather than a long-range force. Although the existence of scalar field will generate scalar fifth force, the effect of the potential density is not equivalent to the effect of the scalar fifth force. Therefore, the requirement that the scalar field should be light to mediate a long-range interaction is not necessary.

The other recognition also takes effect in deriving the misleading corollary. One of the chameleon no-go theorems precludes the possibility of self-acceleration over the last Hubble time. However this preclusion cannot used to rule out the other chameleon-like model to mimic the cosmological constant if quintessence or vacuum energy can naturally emerge in the model.

We need only to pay much more attention to the three things as follows: the extremely small fifth force in the current precision tests of gravity \cite{z80}, the cosmological constant and inflation of the Universe. If the fifth force does really exist, the less effect of observable interaction \cite{z80} means an extreme short interaction-range or/and an extreme weak strength.

There is lots of literature in working to evade the chameleon no-go theorems. It has been suggested that using a symmetry breaking self-interaction potential as a phase transition switch and another scalar field to drive dark energy \cite{z53}. By introducing scalar field dependent masses of neutrinos \cite{z52}, it has been proven that the potential of the scalar field becomes positive from its initial zero value to drive the Universe acceleration. But, why does not the mechanism of mass varying neutrinos apply to baryons? The reason may be the same as mentioned above: to avoid gravitational problems such as a long range fifth force. This discrimination indicates that the equivalence principle no longer holds. By applying Gaussian potentials and its asymptotic behavior \cite{z52}, however, the adiabatic instability \cite{z61} can be avoided. It is explored that there exists an adiabatic regime in which the dark energy scalar field instantaneously tracks the minimum of its effective potential \cite{z70}. However, the adiabatic regime is always subject to an instability if the coupling strength is much larger than the gravitational although the instability can be evaded at weaker couplings \cite{z61}. The screening effect in the chameleon-like models suppresses efficiently the strength of the scalar force so as to be in agreement with precision tests of gravity \cite{z2}.

In this paper a broken-symmetry interaction is introduced to keep the minimum of the effective potential nearly invariant and then to alleviate the adiabatic instability problem in the most cosmic epochs except for the inflationary era. The adiabatic instability is one of the most important features in inflation.  In order to derive a mathematical expression for the scalar fifth force under the constraint of the cosmological constant, it is necessary to use a symmetry-breaking coupling function rather than using symmetry breaking in the self-interaction potential as in \cite{z53,z62}. The symmetry-breaking interaction between matter and the scalar field can localize a vacuum expectation value (VEV) of the scalar field in the effective potential minimum. The effective potential is a sum of the interaction potential and the self-interaction potential \cite{z5,z6}. The parameters in the model are determined by using `Planck 2018 results' \cite{z1,z14} and the theory naturalness.

It should be emphasized here that the theory of the chameleon-like scalar field has introduced a very important and crucial conception \cite{z5,z38}: a scalar-field-independent energy density of matter. Furthermore, potentially, it has introduced a corresponding field-independent pressure and has proven a conservation law of the energy density. The conservation law constructs one of the investigation foundations of this paper. Since matter couples to the scalar field, the energy density of matter in the Universe no longer conserves itself. Therefore, this scalar-field-independent matter density should be introduced to reflect a conserved quantity, such as, a non-relativistic particle number in the Universe. Only the new conserved quantity is included in the model and distinguished with the real physics energy density of matter so can we use the results of astronomical observation to fit the parameters of the model. The real physics energy density of matter includes both the scalar-field-independent density and the coupling energy density with the scalar field.

To be clear, a scalar-field-independent but temperature-dependent equation of state for matter is also introduced and discussed. We regard the equation of state as a hypothesis which needs to be further confirmed by cosmological observations. The setting with the symmetry-breaking coupling function can not only drive dark energy without adding a cosmological constant to the self-interaction potential, but also suppress the interaction range of the fifth force to satisfy the local tests of gravity. The force strength and the interaction range are dependent on the ambient matter density. The force strength is considerably larger compared with gravity under an ultrahigh-vacuum environment, which makes it is possible to detect the fifth force in laboratory. For the further constraint of the scalar fifth force model, we extrapolate the scalar field to drive inflation at the inflationary era of the Universe.

For the closed space in our scenario, the Universe will contract in the future and then the Universe will become hotter and hotter. The ultrahigh frequency oscillation of the scalar field behaves like a pressure-less fluid and then enhances the contracting rapidly. When the the kinetic energy density far larger than the potential energy density, the scalar field can generate a great deceleration effect, which may be called deflation. With the temperature increasing, the interaction between matter and the scalar field will approach to a decoupling phase. The vigorous scalar field will climb up along its self-interaction potential to its maximum value when the kinetic energy is exhausted, and then roll down from the maximum to cause the Universe's rapid growth. As a result, a systematic description of the cosmic acceleration expansion at the present epoch and the very rapid expansion at the inflationary epoch is possible. However, in current literature \cite{z1i} the most probable candidate of the self-interaction potential might be a concave shape, while the self-interaction potential used in our model is a convex one. This seeming paradox results from the assumption in literature that $\dot \phi $ does not pass through zero (not change sign) during inflation in deriving the parametrization of the self-interaction potential $ V( \phi)$ \cite{z22,z23,z48,z55,z59,z60,z84}, where $\phi$ denotes the scalar field and overdot indicates derivative with respect to cosmic time. The result of the parametrization of $ V( \phi)$ depends on the initial value of $\dot \phi $. One often chooses either $\dot \phi>0$ or $\dot \phi<0$ throughout, which is obviously not valid for the case with a turning point from the climbing-up phase to the rolling-down one. The concave feature means that the curvature of the Universe plays an important part in the inflationary era.

This paper is organized as follows. In section \ref{preliminaries}, the technical preliminaries are listed. The expression of the fifth force is reviewed. Both a scalar-field-independent matter density and a scalar-field-independent equation of state for matter are introduced. The temperature dependence of the equation of state is also discussed. In section \ref{thecosimiconstant}, the acceleration equation of the Universe is rewritten in the scalar field coupling case, and then the cosmological constant is described by a special value of the self-interaction potential density of the scalar field. For characterizing the fact that the cosmological constant is nearly fixed with the dynamical model, the symmetry-breaking interaction potential is introduced and discussed. In addition, a negative damping motion is presented, which collects energy in the scalar field during the contracting of the Universe. In this section, we also distinguish the adiabatic condition and the oscillation condition. In section \ref{expansion}, the important parameter of the setting and the current matter density of the Universe are determined by using the model with the current astronomical observation data. It is proven in this section that the cosmological constant is nearly fixed as long as matter density is large enough. When matter density becomes extremely small due to the expansion, the cosmological constant is proportional to square of the density. Comparing the total energy density of the Universe to the critical density calculated with the Hubble constant in `Planck 2018 results', the Universe might be a closed universe. The maximum radius of the Universe and some transition redshifts are also calculated. In section \ref{Nogo}, we show why our setting can avoid the physical corollary of chameleon no-go theorems and the over shooting problem. Interestingly, our model does not conflict with the no-go theorems, at least mathematically. But the model can break through their unreasonable corollary that chameleonlike scalar field cannot account for the cosmic acceleration. When the symmetry-breaking couple with matter is used, the appropriate value of the self-interaction potential can be easily acquired and then can drive the acceleration. In the end of the section \ref{Nogo}, the problem of the zero-point energy density is discussed briefly. In section \ref{contraction}, the contraction of the Universe is discussed. The minimum radius of the Universe is estimated to be falling in a large range. The Universe might undergo a climbing-up and rolling-down process near the minimum radius. In this section, by introducing a pseudo-potential, which is a sum of the self-interaction potential density of the scalar field and the energy density scale of the curvature of the Universe, the feature of the observed concave potential is explained as the attribute of the pseudo-potential. Then, the feature implies a closed space. In section \ref{force}, the screening effect and the strength of the scalar fifth force are discussed. It is shown that the matter-coupled scalar field model satisfies the constraint of the precision measurements of hydrogenic energy levels\cite{z15} due to the density-dependent screening effect. Approximate expressions of the scalar fifth force are derived in this section, which can help experimental physicists design experiments in testing the scalar fifth force. Conclusions are presented in section \ref{discussion}.

\section{Technical preliminaries}\label{preliminaries}

Consider the action governing the dynamics of a scalar field as follows\cite{z52,z61}:

\begin{equation}
S = \frac{1}{{{\hbar ^3}{c^4}}}\int {{d^4}x\sqrt { - g} } \left[ {\frac{{{\hbar ^3}{c^7}}}{{16\pi G}}R - \frac{1}{2}{\hbar ^2}{c^2}{\partial _\mu }\phi {\partial ^\mu }\phi  - V\left( \phi  \right)} \right] + \sum\limits_i {{S_i}\left( {{g_{\mu \nu }}{A^2}\left( \phi  \right),{\psi _i}} \right)} ,\label{equ1}
\end{equation}
where $\phi $ is the scalar field with self-interaction potential $V\left( \phi  \right)$ , and ${\psi _i}$  denotes matter fields, such as spinor field. The coupling between the scalar field and ${\psi _i}$  is given by the conformal coupling ${A^2}\left( \phi  \right){g_{\mu \nu }}$  where the coupling function $A\left( \phi  \right) > 0$. Since forces relate to the spatial gradient of some potential, one may work in Newtonian gauge with the perturbed line element about Minkowskian space-time as
\begin{equation}
d{s^2} = {g_{\mu \nu }}d{x^\mu }d{x^\nu } =  - \left( {1 + 2\Phi } \right){c^2}d{t^2} + \left( {1 - 2\Psi } \right)d{\vec {x}}^2 , \label{equ2minus}
\end{equation}
where the metric potentials $\Phi $ and $\Psi $ are space-dependent but time-independent. For the source of a static, pressureless, non-relativistic matter distribution, apart from the Newtonian force, a test particle is subject to a new fifth force \cite{z2,z39,z40,z13}:

\begin{equation}
\vec {a} =  - {c^2} \nabla \ln A\left( \phi  \right)=- {c^2}\frac{{{A_{,\phi }}\left( \phi  \right)}}{{A\left( \phi  \right)}}\nabla\phi.\label{equ2}
\end{equation}
Be careful not to confuse the unfortunate notation $\vec {a}$ for acceleration of a test particle and $a\left( t \right)$ for scale factor of the Universe. The scalar fifth force is strongly dependent on the form of the coupling function $A\left( \phi  \right)$ besides the gradient of the scalar field. The mathematical expression of $A\left( \phi  \right)$ will be speculated and the physical parameters of $A\left( \phi  \right)$  will be given based on the constraint of the cosmological constant in section \ref{symetrybreaking}. The validness of the scheme is tested in the rest part of this paper.

Astronomical observations have not found that dark energy evolves with time \cite{z1}. Consequently, if dark energy does really originated from a dynamic scalar field, the most probable candidate of $A\left( \phi  \right)$ might be a symmetry-breaking form. The broken-symmetry shape for $A\left( \phi  \right)$ can localize the VEV of the scalar field, which will be discussed in details in section \ref{thecosimiconstant}. In order to infer the form of the coupling function from the constraints of cosmological observation data, one can begin from the consideration of a homogeneous, isotropic universe with a scale factor $a\left( t \right)$ described by the line element:

\begin{equation}
d{s^2} =  - {c^2}d{t^2} + {a^2}\left( t \right)\left[ {\frac{{d{r^2}}}{{1 - K{r^2}}} + {r^2}\left( {d{\theta ^2} + {{\sin }^2}\theta d{\phi ^2}} \right)} \right],\label{equ3}
\end{equation}
where the values $ K=1,\,0,\,{\rm{or}\,  - 1}$ corresponding to closed, flat or open spaces, respectively.
Variation of the action (\ref{equ1}) with respect to the metric yields the Friedmann equation \cite{z52,z61}:
\begin{equation}
{H^2} = \frac{{8\pi G}}{3}\left[ {\sum\limits_i {{\rho _i}{A^{1 - 3{w_i}}}\left( \phi  \right)}  + \frac{1}{{{\hbar ^3}{c^5}}}\left( {V\left( \phi  \right) + \frac{{{\hbar ^2}}}{2}{{\dot \phi }^2}} \right)} \right] - \frac{{K{c^2}}}{{{a^2}}},\label{equ4}
\end{equation}
where $H \equiv {{\dot a}/a }$ is the Hubble parameter which defines the cosmic expansion rate; $G$ is the gravitational constant; overdots indicate derivatives with respect to cosmic time $t$; $i$  denotes several species of non-interacting perfect fluids of matter sources; ${\rho _i}$ is scalar-field-independent matter density \cite{z2,z3,z4,z5,z6,z7,z12,z13} and its equation of state is
\begin{equation}
{w_i} \equiv \frac{{{p_i}}}{{{\rho _i}{c^2}}},\label{equ5}
\end{equation}
with ${p_i}$ being the pressure of the fluid component. According to statistical mechanics, both the energy density ${\rho _i}{c^2}$ and the pressure are functions of the system temperature $T$. Then the equation of state ${w_i}$ is $T$-dependent, \emph{i.e}., ${w_i}\left( T \right)$ . It is worth noting that, regardless whether the temperature value is large or not, the equation of state must be calculated by relativity statistical mechanics so that both thermal energy and rest energy are included \cite{z52,z25,z26}. Then we can conclude that, for dust including cold dark matter (CDM) \cite{z26}, ${w_i} = 0$; for radiations and relativistic particles, ${w_i}=1/3$; in general, $0 \le {w_i}\left( T \right) \le 1/3$. As temperature increases continuously, we will see that ${w_i}\left( T \right) $ gradually from $0$ approaches to $1/3$  and then the final value of ${w_i}\left( T \right) = 1/3$ will result in the decoupling of the scalar field to matter. It should be emphasized that the definition-like choice of ${\rho _i}$ and ${p_i}$ is independent of the scalar field and satisfies the conservation law:
\begin{equation}
{\dot \rho _i} =  - 3H{\rho _i}\left( {1 + {w_i}} \right).\label{equ6}
\end{equation}
The choice of Eq. (\ref{equ6}) describes that both the number density and the corresponding entropy are conserved. The number of the particles (or the distribution numbers for energy levels) is not altered but the masses of the particles (or the energy eigenvalues) are shifted due to the coupling of matter to the scalar field. Consequently, ${\rho _i}$ (${p_i}$ ) denotes the mass densities (pressures) in the decoupled cases, such as ${w_i} = 1/3$, or $A(\phi ) = 1$. Thus, Eq. (\ref{equ6}) reflects that the corresponding entropy is conserved. Actually, Eq. (\ref{equ6}) is supposed validly not only for non-relativistic particles but also for relativistic particles (see also Appendix \ref{aveappb}). To distinguish between the expansion and the contraction by Hubble parameter, we rewrite Eq. (\ref{equ4}) as
\begin{equation}
\frac{{\dot a}}{a} \equiv {H^ \pm } =  \pm {\left( {\frac{{8\pi G}}{3}\left[ {\sum\limits_i {{\rho _i}{A^{1 - 3{w_i}}}\left( \phi  \right)}  + \frac{1}{{{\hbar ^3}{c^5}}}\left( {V\left( \phi  \right) + \frac{{{\hbar ^2}}}{2}{{\dot \phi }^2}} \right)} \right] - \frac{{K{c^2}}}{{{a^2}}}} \right)^{{1 \mathord{\left/
 {\vphantom {1 2}} \right.
 \kern-\nulldelimiterspace} 2}}}.\label{equ7}
\end{equation}
Then ${H^ + }$ and ${H^ - }$ denote the expansion and the contraction of the Universe, respectively.
Variation of the action (\ref{equ1}) with respect to $\phi $ gives \cite{z52,z61}:
\begin{equation}
{\hbar ^2}\ddot \phi  + 3{H^ \pm }{\hbar ^2}\dot \phi  + {V_{{\rm{eff}},\phi }}\left( \phi  \right) = 0,\label{equ8}
\end{equation}
where the subscript `$,\phi $' denotes a partial derivative with respect to $\phi $; the effective potential density is
\begin{eqnarray}
V_{{\rm{eff}}}\left( \phi  \right) = V\left( \phi  \right) + V_{{\mathop{\rm int}} }, \label{equ8plus}
\end{eqnarray}
with
\begin{eqnarray}
V_{{\mathop{\rm int}} } \equiv \sum\limits_i {{\rho _i}{\hbar ^3}{c^5}\left[ {{A^{1 - 3{w_i}}}\left( \phi  \right) - 1} \right]}\label{equ8plusa}
\end{eqnarray}
indicating the interaction with matter \cite{z5}. From Eq. (\ref{equ8}), one can deduce that when the Universe contracts the scalar field may grow rapidly. The detail will be discussed in section \ref{contraction}. According to statistical mechanics, as the temperature approaches infinity, the equation of state approaches 1/3 and then the interaction potential vanishes. As a contrast, for cold but extremely dense matter perfect fluid, the value $A({\phi _{\min }})$ of a symmetry-breaking coupling function shown in next section later at the minimum of the effective potential trends to 1, the interaction potential also vanishes [see Eqs. (\ref{equ16b}) and (\ref{equ18-a})].

In summary, the scalar field must be required to account for the observed cosmic acceleration so as to obtain logically the fifth force from the coupled scalar field. To achieve this goal, a scalar-field-independent matter density and the corresponding conservation law are introduced in the definition-like manner. The conservation law describes that both the number of the particles and the corresponding entropy are conserved regardless whether matter couples to the scalar field or not. Since the masses of the particles are shifted due to the coupling, the real physics matter density depends on the scalar field, and the corresponding entropy is no longer conserved due to the $T$-dependence of ${w_i}$ (see also Appendix \ref{ceappb}). When a new degree of freedom is introduced, it is necessary to add accordingly a new energy form. ${\rho _i}$  and the corresponding conservation law are introduced to reflect the aspect of the scalar-field-independence of matter.

\section{A quartic self-interaction potential energy density with a symmetry-breaking interaction to make the Universe accelerate}\label{thecosimiconstant}

It has always been deemed that only when the scalar field leads to a long range fifth force \cite{z7,z66} can it represent dark energy evolving on cosmological time scales. However, the requirement of the long range of the interaction is not necessary, which will be discussed in section \ref{Nogo}. In this section we introduce a broken-symmetry interaction between the scalar field and matter to localize the minimum of the effective potential and then the fifth force is very short ranged. Our setting differs from \cite{z54,z53,z61,z62,z49} in which the broken-symmetry is only related to the self-interaction potential of the scalar field and then adiabatic instability occurs or reacts very sensitively to the changes of the background density \cite{z39,z57,z61,z41}. It will be proven in this section that a value of the self-interaction potential around the minimum of the effective potential acts a constant-like dark energy (or equivalently, the cosmological constant) due to the broken-symmetry interaction. The broken-symmetry couple also results in a density-dependent and short ranged fifth force, which will be discussed in section \ref{force}.

\subsection{Driving cosmic acceleration via the coupled scalar field}\label{cosmic acceleration}

From Eqs. (\ref{equ4}) and (\ref{equ8}), the acceleration equation of the Universe is obtained as:
\begin{equation}
\frac{{\ddot a}}{a} = \frac{{4\pi G}}{{3{\hbar ^3}{c^5}}}\left[ {2V\left( \phi  \right) - 2{\hbar ^2}{{\dot \phi }^2} - \sum\limits_i {{\rho _i}\left( {1 + 3{w_i}} \right){\hbar ^3}{c^5}{A^{1 - 3{w_i}}}\left( \phi  \right)} } \right].\label{equ9}
\end{equation}
Since Eq. (\ref{equ9}) constructs one of our investigation foundations, it is derived in detail in Appendix \ref{appA}. It is noteworthy that ${w_i}$, the equation of state for matter, is temperature-dependent. With temperature growing, the coupling effect of matter to the scalar field decreases. For pressureless matter sources ${w_i} = 0$, the acceleration of the Universe becomes
\begin{equation}
\frac{{\ddot a}}{a} = \frac{{4\pi G}}{{3{\hbar ^3}{c^5}}}\left[ {2V\left( \phi  \right) - 2{\hbar ^2}{{\dot \phi }^2} - \rho {\hbar ^3}{c^5}A\left( \phi  \right)} \right],\label{equ10}
\end{equation}
where ${\rho} = \sum\nolimits_i {{\rho _{i}}} $. We emphasize again that, ${\rho}$ is a decoupled total matter density which is independent on the scalar field, and the total physics matter density in the pressureless case should be $\rho A( \phi )$ which includes the interaction energy of matter with the scalar field. The energy exchange between the scalar field and matter is discussed in detail in Appendix \ref{appB}.

Eq. (\ref{equ9}) shows that the self-interaction potential energy density $V\left( \phi  \right)$ of the scalar field drives the Universe accelerating expansion, while both the kinetic energy density of the scalar field and the any form energy density of matter lead to a decelerating expansion. From Eq. (\ref{equ8}) one sees that the evolution of the scalar field is a damping oscillation in the expansion period of the Universe ($H \equiv {H^ + } > 0$). Therefore, if the scalar field evolves to the minimum of the effective potential and can stably sit at the minimum, one might obtain a cosmological constant. Substitute the field value ${\phi _{\min }}$ at the minimum into Eqs. (\ref{equ4}) and (\ref{equ10}), and neglect the kinetic energy term of the scalar field, we get simple expressions for Friedmann equation (\ref{equ4}) and the acceleration equation (\ref{equ10}) as follows:
\begin{subequations}\label{equ11and12revise}
\begin{eqnarray}
{H^2} \equiv \frac{{{{\dot a}^2}}}{{{a^2}}} = \frac{{8\pi G}}{3}\left[ {\frac{{V\left( {{\phi _{\min }}} \right)}}{{{\hbar ^3}{c^5}}} + \rho A\left( {{\phi _{\min }}} \right)} \right] - \frac{{K{c^2}}}{{{a^2}}},\label{equ11}\\
\frac{{\ddot a}}{a} = \frac{{4\pi G}}{3}\left[ {\frac{{2V\left( {{\phi _{\min }}} \right)}}{{{\hbar ^3}{c^5}}} - \rho A\left( {{\phi _{\min }}} \right)} \right].\label{equ12}
\end{eqnarray}
\end{subequations}
Comparing Eq. (\ref{equ12}) with the acceleration of the Universe in the $\Lambda$CDM model \cite{z26,z100} of
\begin{equation}
\frac{{\ddot a}}{a} = \frac{{\Lambda {c^2}}}{3} - \frac{{4\pi G}}{3}\rho_{\rm{m}} ,\label{equ13}
\end{equation}
where $\Lambda$ and $\rho_{\rm{m}}$ are the cosmological constant and the real physics matter density, respectively. One can see that the value of the self-interaction potential at ${\phi _{\min }}$ acts as the cosmological constant,
\begin{equation}
\Lambda  = \frac{{8\pi GV\left( {{\phi _{\min }}} \right)}}{{{\hbar ^3}{c^7}}},\label{equ14}
\end{equation}
rather than the minimum of the effective potential as one sometimes used \cite{z7,z8}. Also, the mass density $\rho_{\rm{m}}$ of matter is equal to $\rho A( \phi_{\min} )$ and becomes $\phi$-dependent, \emph{i.e}.,
\begin{equation}
\rho_{\rm{m}} = \rho A( \phi_{\min} ).\label{equ14rep}
\end{equation}
That the value of the self-interaction potential at the minimum plays the role of dark energy has also been demonstrated through post-Newtonian approximation \cite{z24}.

Physicists often use ${\Lambda _{\rm{E}}}$, the energy scale of the dark energy density $\Lambda _{\rm{E}}^4$, to describe the cosmological constant, which is defined by:
\begin{equation}
\Lambda _{\rm{E}}^4 \equiv \frac{{\Lambda {\hbar ^3}{c^7}}}{{8\pi G}} \equiv V\left( {{\phi _{\min }}} \right).\label{equ15}
\end{equation}

Since the scalar field always trends towards the minimum of the effective potential due to the positive damping coefficient of $ 3H^{+}>0$ in the expansion period of the Universe, to obtain the cosmological constant is strongly dependent on the adiabatic condition that guarantees the stability of the scalar field sitting at the minimum of the effective potential. Consequently, large effective masses of the scalar field and nearly invariant minimums of the effective potential are necessary. This can be achieved by invoking a broken-symmetry couple function, which will be shown in next subsection \ref{symetrybreaking}.

\subsection{Symmetry-breaking coupling function}\label{symetrybreaking}

\begin{figure}
\centering
\includegraphics[width=250pt]{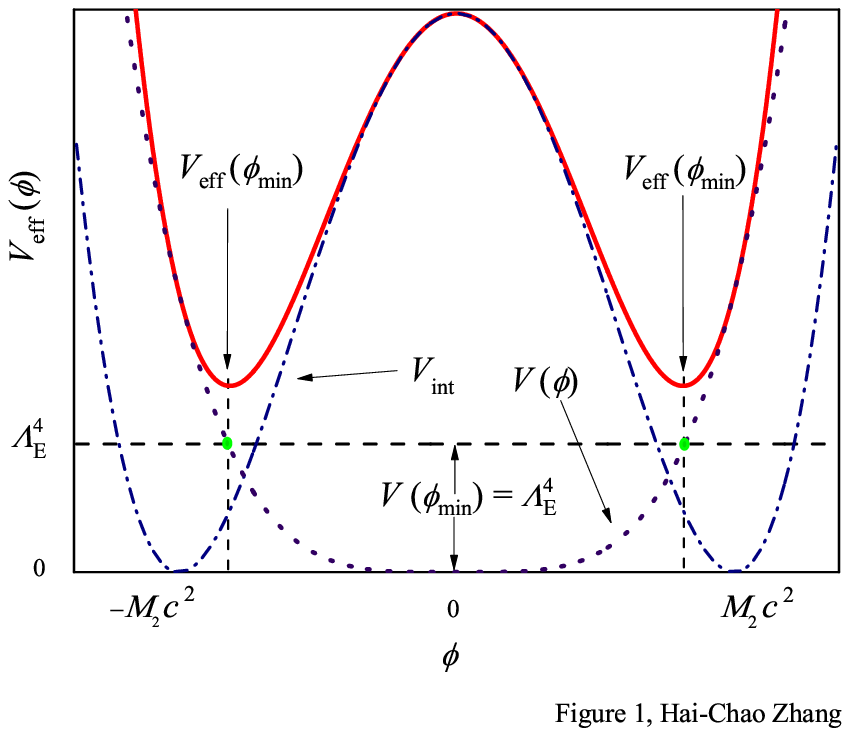}
\caption{The effective potential ${V_{{\rm{eff}}}}\left( \phi  \right)$ (solid curve) is the sum of a scalar potential $V\left( \phi  \right)$ (dotted curve) and a matter-density-dependent interaction term ${V_{{\mathop{\rm int}} }}$ (dot-dashed curve). The value $V\left( {{\phi _{\min }}} \right)$ (dashed line) of the self-interaction potential at one of the symmetry-breaking vacuums ${\phi _{\min }}$ acts as the cosmological constant to drive late-time cosmic acceleration.}\label{figure1}
\end{figure}

In order to obtain the fifth force under the cosmic constraints, we choose a quartic self-interaction potential and a symmetry-breaking coupling function as follows:
\begin{subequations}\label{equ16revise}
\begin{eqnarray}
V(\phi ) &=& \frac{\lambda }{{\rm{4}}}{\phi ^4};\label{equ16a}\\
A(\phi ) &=& 1 + \frac{{\rm{1}}}{{{\rm{4}}{M_1}^{\rm{4}}{c^{\rm{8}}}}}{\left( {{\phi ^{\rm{2}}} - {M_2}^{\rm{2}}{c^{\rm{4}}}} \right)^2},\label{equ16b}
\end{eqnarray}
\end{subequations}
where ${M_1}$, ${M_2}$ are parameters with mass dimension and $\lambda $ is a dimensionless parameter. Both the self-interaction potential and the coupling function have ${\mathbf{Z}_{\rm{2}}}$ ($\phi  \to  - \phi $) symmetry. All the parameters above can be determined by the constrains of the cosmological observations and the theoretical naturalness (the detail discussion is shown in Appendix \ref{appC}). Concise display of the parameters is shown in Eq. (\ref{equ17}):
\begin{equation}
\lambda  = \frac{1}{6},\, {M_1} = \frac{M_2}{2^{3}},\, {M_2} = 4.96168 \; {\rm{meV/}}{c^2}=8.845\times {{10}^{ - 39}}\,\rm{kg}.\label{equ17}
\end{equation}
According to Eqs. (\ref{equ8plus}) and (\ref{equ16revise}), the effective potential energy density has a very simple form in the case of ${w_i} = 0$ as follows:
\begin{equation}
V_{{\rm{eff}}}\left( \phi  \right) = \frac{\lambda }{{\rm{4}}}{\phi ^4} + \frac{{\rho {\hbar ^3}}}{{{\rm{4}}{M_1}^{\rm{4}}{c^{\rm{3}}}}}{\left( {{\phi ^{\rm{2}}} - {M_2}^{\rm{2}}{c^{\rm{4}}}} \right)^2}.\label{equ17rep}
\end{equation}
The effective potential density versus the scalar field is shown in Fig. \ref{figure1}.

\subsubsection{\label{masseff} The $\lambda $-dependent minima and the $\lambda $-independent effective mass}

The two degenerate minima of the effective potential of Eq. (\ref{equ17rep}) and the effective mass around the minima are obtained as follows (see Appendix \ref{appC}):
\begin{subequations}\label{equ18}
\begin{eqnarray}
{\phi _{\min }} &=&  \pm {\left( {\frac{{\rho {\hbar ^3}{M_2}^{\rm{2}}{c^4}}}{{\lambda {M_1}^{\rm{4}}{c^3} + \rho {\hbar ^3}}}} \right)^{1/2}}, \label{equ18-a}\\
m_{{\rm{eff}}}^{\rm{2}}& \equiv &\frac{{{V_{{\rm{eff}},\phi \phi }}\left( {{\phi _{\min }}} \right)}}{{{c^4}}} = \frac{{{\rm{2}}\rho {\hbar ^3}{M_2}^{\rm{2}}}}{{{M_1}^{\rm{4}}{c^3}}}.\label{equ18-b}
\end{eqnarray}
\end{subequations}
The scalar field has to choose only one of the minima and then the ${\mathbf{Z}_{\rm{2}}}$ symmetry is spontaneously broken.

From Eqs. (\ref{equ18}) above, one sees that the effective mass of the scalar field is not dependent on $\lambda $, but $\phi _{\min }$ is dependent on $\lambda $. The two properties are main purposes of the choice shown in Eq. (\ref{equ16revise}). These properties describe that the self-interaction potential of the scalar field has nothing to do with the effective mass but can move the position of the minimum of the effective potential. These important properties guarantee that the observed cosmic acceleration stems entirely from the scalar field rather than any static vacuum energy, which will be discussed in the end of section \ref{Nogo}. The $\lambda $-independent effective mass in the general case of ${w_i} \neq 0$ is obtained in Appendix \ref{prappc}.

\subsubsection{\label{adiacondition} The condition of adiabatic tracking}

One sees that through Eq. (\ref{equ18}) the scalar field can adiabatically track the minimum of the effective potential. The changing rate ${{\dot \phi }_{\min }}/{\phi _{\min }}$ of the minimum position due to the change of the matter density can be described by Eq. (\ref{s8plus2}) in Appendix \ref{appC}. The adiabatic condition guarantees that, if the field is initially at the minimum, it will follows the minimum adiabatically during the later evolution. Since the reciprocal of $\left| {3H/2} \right|$ is the characteristic time of the evolution of the Universe, the adiabatic condition can be expressed as follows:
\begin{equation}
\left| {\frac{{{{\dot \phi }_{{\text{min}}}}}}{{{\phi _{{\text{min}}}}}}} \right| \leq \left| {\frac{{3H}}{2}} \right|.\label{equ18replus}
\end{equation}
The smaller the changing rate of the minimum position, the stronger the stability of the scalar field sitting at the minimum. For pressure-less matter source the scalar field in our scheme can adiabatically follow the minimum, which is proven by Eq. (\ref{s8plus2}) in Appendix \ref{appC}.

If the field is initially not at the minimum, one should take into account the oscillation condition. The response time for the scalar field to adjust itself to the position of the minimum is characterized by $1/\omega _{\text{c}}$ with the Compton frequency ${\omega _{\rm{c}}} \equiv {m_{{\rm{eff}}}}{c^2}/\hbar$. The decay time that for the evolution of the scalar field is characterized by $2/(3H)$. In general cases, the Compton frequency is considerably larger than the Hubble expansion rate and then the oscillation condition
\begin{equation}
{\omega _c} \geq \left| {\frac{{3H}}{2}} \right|  \label{equ18revise2}
\end{equation}
is satisfied. For example, the energy scale of the Compton frequency in the present matter density of the Universe is estimated to be $\hbar \omega _{{\text{c0}}} \sim 60\,{\text{meV}}$, which is about six times of the cosmological constant. The energy scale of the Hubble expansion rate in the present time is $\hbar H_0\sim 10^{-33} \, \rm{eV}$. However, it is worth noting that the oscillation condition and the adiabatic condition cannot be satisfied in the inflationary era when the Hubble rate, the effective mass of the scalar field and the energy density of the Universe vary extremely violently. Particularly, Eq. (\ref{equ18}) is no longer valid in the inflationary era since the assumption of the equation of state $w_i=0$ is invalid.

\subsubsection{\label{comwavelength} The Compton wavelength of the scalar field}

By the way, the Compton wavelength of the scalar field is defined by ${\mathchar'26\mkern-10mu\lambda _{\rm{c}}} \equiv \hbar /\left( {{m_{{\rm{eff}}}}c} \right)$, which describes the interaction range between matter and the scalar field. Using Eqs. (\ref{equ17}) and (\ref{equ18}), the Compton wavelength is obtained as
\begin{equation}
{\mathchar'26\mkern-10mu\lambda _{\rm{c}}}\left[ {\rm{m}} \right] = \frac{{1.648 \times {{10}^{ - 19}}}}{{{{\left( {\rho \left[ {{\rm{kg/}}{{\rm{m}}^{\rm{3}}}} \right]} \right)}^{1/2}}}}.\label{equ19}
\end{equation}
The larger the ambient matter density, the shorter the Compton wavelength of the coupled scalar field. This interaction range is so short that it is difficult to detect even in a low-density condition of empty space. The further discussion will be shown in section \ref{force}.

\subsubsection{\label{negadamping} The negative-damping oscillation of the scalar field}

Considering the scalar field around the minimum to obtain an approximation for the equation of motion, the effective potential can be expanded into: ${V_{{\rm{eff}}}}\left( \phi  \right) = {V_{{\rm{eff}}}}\left( {{\phi _{\min }}} \right) + {V_{{\rm{eff,}}\phi }}\left( {{\phi _{\min }}} \right)\left( {\phi  - {\phi _{\min }}} \right) + m_{{\rm{eff}}}^{\rm{2}}{c^4}{\left( {\phi  - {\phi _{\min }}} \right)^2}/2 $,
and then the equation (\ref{equ8}) of motion becomes:
\begin{equation}
 \ddot \phi  + 3{H^ \pm }\dot \phi  + \omega _c^2\left( {\phi  - {\phi _{\min }}} \right) = 0,\label{equ20}
\end{equation}
which describes a damped (negative-damped) oscillation for ${H^+}> 0$ (${H^-}< 0$). The damped (negative-damped) oscillation can be classified into three cases: \paragraph{$\left| {3H/2} \right| > {\omega _{\text{c}}}$} over-damping (over-negative-damping); \paragraph{$\left| {3H/2} \right| = {\omega _{\text{c}}}$} critically-damping (critically-negative-damping); \paragraph{$\left| {3H/2} \right| < {\omega _{\text{c}}}$} under-damping (under-negative-damping).

\begin{figure}
\centering
\includegraphics[width=200pt]{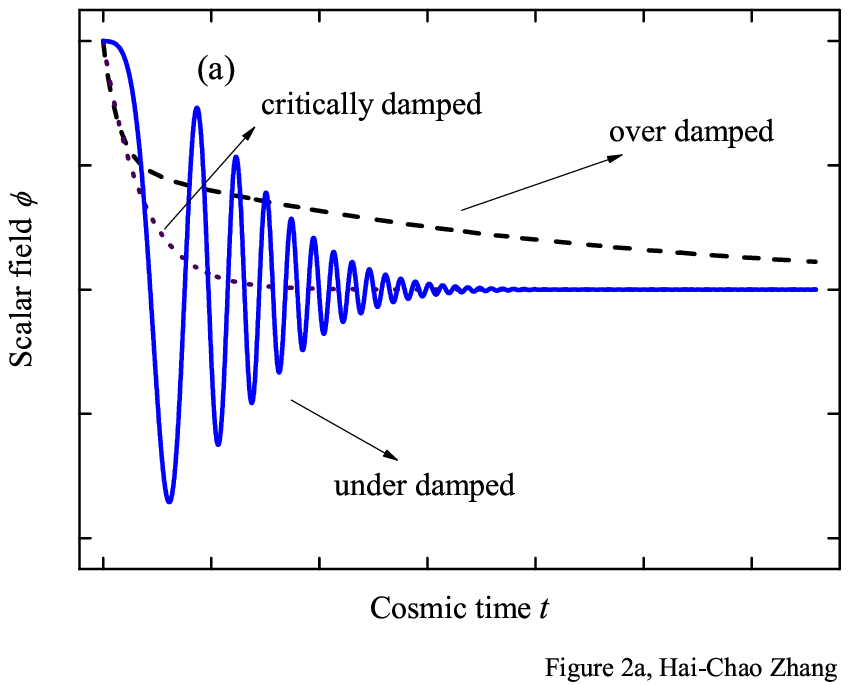}
\includegraphics[width=200pt]{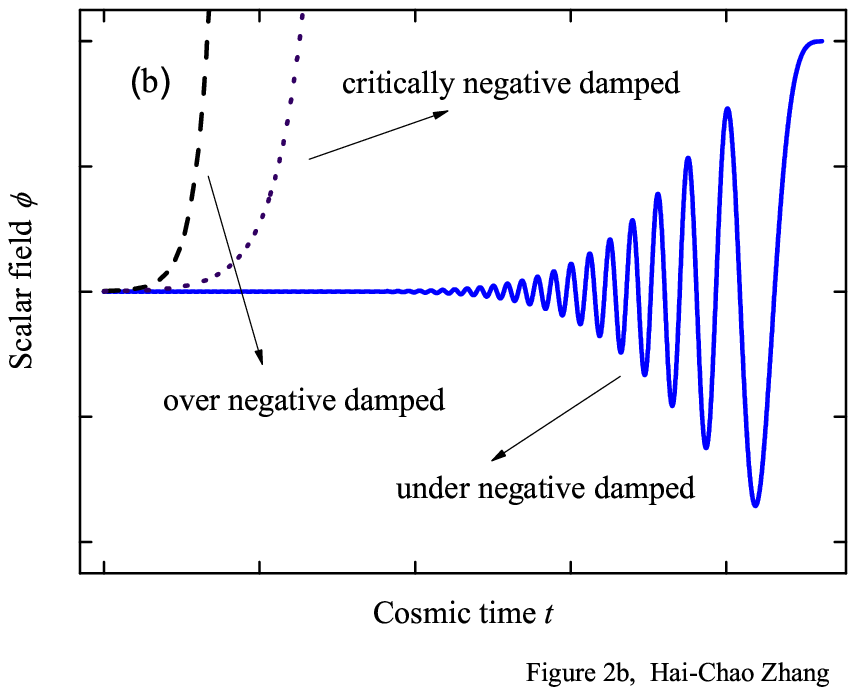}
\caption{The schematic sketch of the motion of the scalar field. (a) The curves of the motion of the scalar field correspond to the expansion of the Universe. The scalar field experiences damped oscillation in three cases: over-damping, critically-damping, and under-damping. The under-damped oscillation will last for a long time until the adiabatic condition shown in Eq. (\ref{equ18replus}) is satisfied. When the adiabatic condition is satisfied, the scalar field will quickly decay to the minimum and then stably sit at the minimum. (b) The curves of the motion of the scalar field correspond to the contraction of the Universe. The scalar field experiences negative damped oscillation in three cases: under-negative-damping, critically-negative-damping and over-negative-damping. In the case of under-negative-damping, the scalar field can initially sit and stably keep sitting at one of the minima of the effective potential until the adiabatic condition is broken, which will be discussed in section \ref{contraction}. When the temperature increases, the scalar field will finally decouple with matter and shift to the case of over-negative-damping. It is worth noting that neither graph shows the minimum moving over time. The moving rate of the minimum determines whether the adiabatic condition is satisfied. }\label{figure2}
\end{figure}

The symbol of absolute value is used because the damping coefficient $3H < 0$ in the contraction phase of the Universe. The oscillation frequency of the scalar field around the minimum is equal to $\sqrt{ {\omega _{\text{c}}^2 - 9{H^2}/4} }$, which is less than the Compton frequency ${\omega _c}$. Figure \ref{figure2} shows the schematic sketch of the curves of the motion of the scalar field in the cases of (a) the expansion and (b) the contraction of the Universe, respectively.

The negative-damping oscillation of the scalar field can absorb energy from the gravitational field by the negative damped manner during the contraction process of the Universe if the contraction really occurs. Thus, the negative-damping oscillation is different from forced oscillations. The negative-damping oscillation of the scalar field must result in exploding of the Universe due to the exponentially growing of the oscillation magnitude.

However, if the scalar field initially sits at the minimum, \emph{i.e.}, the oscillation magnitude is zero, the question is what thing activates the oscillation. When the adiabatic condition of Eq. (\ref{equ18replus}) is satisfied, the scalar field will keep sitting at the minimum. With the temperature of the Universe becoming hotter and hotter, the adiabatic condition does no longer hold and then the oscillation is triggered by the quick movement of ${\phi _{\min }}$ , which will be discussed in section \ref{contraction}.

\subsubsection{\label{cosmologicalconstant} The cosmological constant in quintessence form pinned by the broken-symmetry couple}

Since the adiabatic tracking always holds for $w_i=0$, the cosmological constant can be finely defined by Eq. (\ref{equ15}) and can be obtained in our scheme by Eq. (\ref{equ16a}) as follows
\begin{equation}
\Lambda _{\rm{E}}^4 \equiv V({\phi _{\min }}) = \frac{\lambda}{{{\rm{4}}}}\phi _{\min }^4.\label{equ21}
\end{equation}
If the symmetry is not broken, \emph{i.e.}, ${M_2} = 0$ in Eq. (\ref{equ16b}), one immediately has ${\phi _{\min }} = 0$ and $m_{{\rm{eff}}}=0$ by using Eq. (\ref{equ18}), and then the cosmological constant goes to zero. In this sense, the non-zero cosmological constant seems to stem from a broken-symmetry interaction between the scalar field and matter. However, if $\lambda = 0$, from Eq. (\ref{equ21}) one also obtain a zero. Indeed, our setting is essentially a quintessence model, in which the quintessence is trapped at the bottom of the interaction potential well. When the matter density large enough, the value of the self-interaction potential pinned by the broken-symmetry interaction approaches to a constant. That if the localization works well or not is left to next section \ref{expansion}.

\subsection{Summary}\label{suforsmb}

Since the adiabatic condition is satisfied in our setting, the value of the quartic self-interaction potential at the minimum of the effective potential can be localized stably by the broken-symmetry interaction potential. It has been proven that the value acts the cosmological constant. When the density of matter is large enough, the value of the self-interaction potential indeed approaches to a constant.

In the scheme, the effective mass of the scalar field has two important characters: 1). The mass in general is large enough so that the adiabatic instability can be suppressed; 2). The mass is unrelated to the self-interaction potential so that the zero-point-energy can be cancel out, and then the fine-tuning is avoided in deriving the cosmological constant, which will be discussed in section \ref{zeropoint}.  Although it has nothing to do with the effective mass, the self-interaction potential moves the position of the minimum of the effective potential. Thus, the observed cosmic acceleration can be ascribed entirely to the scalar field rather than any static vacuum energy.

Besides, the negative-damping oscillation of the scalar field in this section is introduced for the contraction process of the Universe.

\section{Application of the model to the expansion of the Universe (${H^+}> 0$)}\label{expansion}

We now test our setting how to explain the current astronomical observations, how to extrapolate backward in the early time and how to predict the future trends of the Universe.

\subsection{Quantitative comparison with some important astronomical observations}\label{quancom}

One can see from Eq. (\ref{equ17}) that only one parameter of ${M_2}$ needs to be determined by experimental data. The parameter of ${M_2}$ is chosen so as to satisfy both the values of the current cosmological constant and the current ratio of the energy density of matter to the total energy density of the Universe. After the determination of ${M_2}$, we will test the setting if works well or not.

\subsubsection{\label{determ2} Determining the only one adjustable parameter $M_{2}$ }

The ratio of matter density to the total mass density is
\begin{equation}
{\Omega _{\rm{m}}} \equiv \frac{{{\rho _{\rm{m}}}}}{{{\rho _{{\rm{tot}}}}}},\label{equ22minus1}
\end{equation}
where the physics matter density $ \rho _{\rm{m}} = \sum\nolimits_i \rho _{{\rm{m}} i}$ with $\rho _{{\rm{m}}i}=\rho _i A^{1 - 3{w_i}}\left( \phi  \right)$, the total mass density is a sum of the physics matter density and the mass density of the scalar field. It can be seen from the Friedmann equation (\ref{equ4}) why the real physics matter density is $ \rho _{\rm{m}}$ rather than ${\rho _i}$ or $\sum\nolimits_i {{\rho _{i}}} $. For pressureless matter sources ${w_i} = 0$, the physics matter density can be written as ${\rho _{\rm{m}}} = \rho A\left( \phi  \right)$ with ${\rho} = \sum\nolimits_i {{\rho _{i}}} $. The scalar field mass density is defined by ${\rho _\phi } = V\left( \phi  \right)/\left( {{\hbar ^3}{c^5}} \right) + {{\dot \phi }^2}/\left( {2\hbar {c^5}} \right)$ \cite{z9}, which can also be seen from the Friedmann equation (\ref{equ4}). When the scalar field adiabatically follows the minimum of the effective potential, the kinetic energy of the scalar field can be neglected and the scalar field mass density becomes ${\rho _\phi } = V\left( \phi_{\min }  \right)/\left( {{\hbar ^3}{c^5}} \right) $. Thus, Eq. (\ref{equ22minus1}) becomes
\begin{equation}
{\Omega _{\rm{m}}} = \frac{{\rho {\hbar ^3}{c^5}A\left( {{\phi _{\min }}} \right)}}{{\rho {\hbar ^3}{c^5}A\left( {{\phi _{\min }}} \right) + V\left( {{\phi _{\min }}} \right)}}. \label{equ22minus2}
\end{equation}
Substituting the current astronomical observation dada \cite{z1}
\begin{subequations}\label{equ22minus3re}
\begin{eqnarray}
{\Omega _{\rm{m0}}} &=& 31.58\%, \label{equ22minus3}\\
{\Lambda _{\rm{E0}}} &=& 2.239\,{\rm{ meV}}\label{equ22minus3reb}\\
({\Lambda_{0}}  &=& 4.24 \times {10^{ - 66}}\,{\rm{ e}}{{\rm{V}}^2} = 1.089 \times {10^{ - 52}}\,{{\rm{m}}^{ - 2}}),\nonumber
\end{eqnarray}
\end{subequations}
into Eqs. (\ref{equ22minus2}) and (\ref{equ21}) together with (\ref{equ18-a}), we obtain simultaneous equations. Noticing the expressions shown in Eq. (\ref{equ17}) and now regarding $M_2$ as a undetermined parameter, we solve the simultaneous equations to derive $M_2$ and the current matter (including CDM ) density of the Universe as follows:
\begin{subequations}\label{equ22minus4rezong}
\begin{eqnarray}
{M_2}& = & {4.96168 \; {\rm{meV/}}{c^2}= 8.845 \times {{10}^{ - 39}}\,\rm{kg};}\label{equ22minus4a}\\
{\rho _{{\rm{m}}0}}& \equiv & { {\rho _0}A_0 ({{\phi _{\min }}} )= 2.69271 \times {10^{-27}}\, \rm{kg/{m^3}}.}\label{equ22minus4}
\end{eqnarray}
\end{subequations}
Here subscript `0' marks the current time. The corresponding scalar-field-independent matter density is
\begin{equation}
{\rho _0} = 2.68026 \times {10^{ - 27}}\,{\rm{ kg/}}{{\rm{m}}^3},\label{equ22re}
\end{equation}
which is smaller than the real physics matter density $\rho _{{\rm{m}}0}$. The reason is that the physics matter density includes the interaction energy between matter and the scaler field.

The total energy density of the Universe is then obtained as follows:
\begin{equation}
{\rho _{{\rm{tot0}}}} = {\rho _0}A_0\left( {{\phi _{\min }}} \right) + \frac{{V_{0}\left( {{\phi _{\min }}} \right)}}{{{\hbar ^3}{c^5}}} = 8.52665 \times {10^{ - 27}}\;{\rm{kg}} \cdot {{\rm{m}}^{ - 3}}.\label{equ23}
\end{equation}
Therefore, by using the current values of ${\Omega _{\rm{m0}}}$ and ${\Lambda_{0}}$, not only the free parameter $M_2$ is determined, but also all the forms of the current energy density of the Universe are derived.

\subsubsection{\label{effectiequation } The effective equation of state for the scalar field in the present era}

The effective equation of state for the coupled scalar field in the present era is estimated by Eq. (\ref{s26plusere1}) in Appendix \ref{appB} to be
\begin{equation}
{w_{{\rm{eff0}}}} \equiv \frac{{{p_{{\rm{eff0}}}}}}{{{\rho _{{\rm{eff0}}}}{c^2}}} = \frac{{ - V_{0}\left( {{\phi _{\min }}} \right)}}{{{V_{{\rm{eff0}}}}\left( {{\phi _{\min }}} \right)}} =  - 0.998.\label{equ25s26pluse}
\end{equation}
This value is slightly larger than the result of $ w_{0}= -1.03 \pm {0.03} $ shown in \cite{z1}, but is slightly smaller than the result of $w = {-0.80^{+0.09} _{-0.11}} $ shown in \cite{z78}. The small differences may result from that the models used in the literature \cite{z1,z78} are not as the same as the model in this paper.

\subsubsection{The two transition redshifts in the past and the future}\label{theredshift}

We can now calculate the transition redshifts by letting the acceleration in Eq. (\ref{equ12}) equal to zero. Although the geometry curvature $K$ appears in the Friedmann equation, it disappears in the acceleration equation of the Universe. Consequently, the transition redshifts associated the zero-acceleration are independent of the curvature of the Universe. Since redshift is defined by $1+z=a_0/a(t)$ with $a(t)$ the scale factor of the Universe at cosmic time $t$ and $a_0$ the current value \cite{z26}, the scalar-field-independent matter density in the pressureless case can be expressed via Eq. (\ref{equ6}) as follows:
\begin{equation}
\rho  = \rho _0\left( {1 + z} \right)^3.\label{equ25plus0}
\end{equation}
The physical significance of Eq. (\ref{equ25plus0}) is that the particle number of the Universe is not altered during its expansion. However, it is worth noting that, in general, $ \rho _{\rm{m}}\neq \rho _{{\rm{m}}0}{\left( {1 + z} \right)^3}$ due to the interaction energy between matter and the scalar field (see also Appendix \ref{appB}). Substituting both $\ddot a = 0$ and Eq. $(\ref{equ25plus0})$ into Eq. (\ref{equ12}), we derive two solutions for transition redshift which mark the transition time of the Universe expansion from deceleration to acceleration and vice-versa.

The transition redshift corresponding to the deceleration-acceleration transition in the past of the Universe is
\begin{equation}
{z_{{\rm{past}}}} = 0.634478,\label{equ25plus1}
\end{equation}
which is consistent with \cite{z1,z14,z64,z87,z65}. At this transition time, the scalar-field-independent matter density $\rho_{\rm{past}}  = 1.1703 \times {10^{ - 26}}\,{\rm{ kg/}}{{\rm{m}}^3}$, which is slightly smaller than the corresponding physics matter density ${\rho _{\rm{mpast}}} = 1.1706 \times {10^{ - 26}}\, \rm{kg/{m^3}}$. The corresponding cosmological constant is obtained by Eqs. (\ref{equ18-a}) and (\ref{equ21}) as ${\Lambda _{{\rm{E past}}}} =2.241\,{\rm{ meV}}$ or equivalently, ${\Lambda _{{\rm{past}}}} = 1.1092 \times {10^{ - 52}}\, {{\rm{m}}^{ - 2}}$. The effective equation of state for the coupled scalar field is estimated by Eq. (\ref{s26plusere1}) to be ${w_{{\rm{eff}}}} =  - 1$.

Another transition redshift that corresponds to the next transition of acceleration-deceleration is obtained as follows
\begin{equation}
{z_{{\rm{future}}}} =  - 0.8977287,\label{equ25plus4}
\end{equation}
which will occur in the future. The scalar-field-independent matter density $\rho_{\rm{future}}  = 2.867 \times {10^{ - 29}} \, {\rm{ kg/}}{{\rm{m}}^3}$ is considerably smaller than the corresponding physics matter density ${\rho _{\rm{mfuture}}} = 1.3058 \times {10^{ - 27}}\,\rm{kg/{m^3}}$. This means that, with the density decreasing, the interaction potential energy between matter and the scalar field will increase due to the symmetry-breaking coupling function. Of course, one also finds that the cosmological constant will decrease with the matter-density decreasing. The corresponding cosmological constant is ${\Lambda _{{\rm{Efuture}}}} = 1.295\,{\rm{ meV}}$ (${\Lambda _{{\rm{future}}}} = 1.219 \times {10^{ - 53}}\, {{\rm{m}}^{ - 2}}$). The corresponding effective equation of state for the coupled scalar field is estimated by Eq. (\ref{s26plusere1}) to be ${w_{{\rm{eff}}}} =  - 0.334$.

\subsubsection{The nearly fixed cosmological constant before the present era}\label{thecosmologicalcons}

\begin{figure}
\centering
\includegraphics[width=250pt]{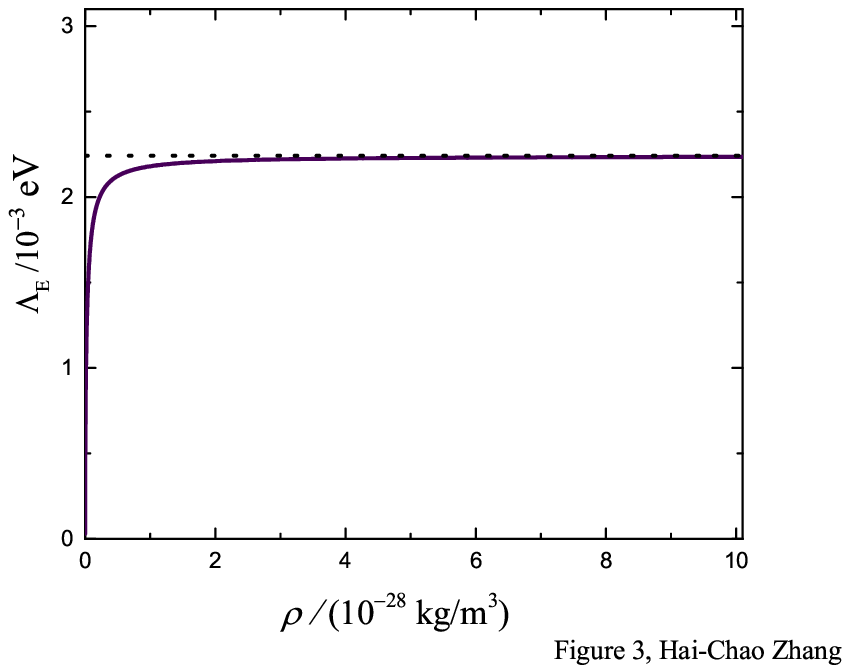}
\caption{The cosmological constant versus the density of matter. The density-dependent cosmological constant obtained by Eq. (\ref{equ21}) becomes nearly density-independent when the density large enough regardless of the Universe being flat, open or closed. However, for flat and open spaces, the cosmological constant approaches to zero when the density of matter approaches to zero. For closed space, the minimum cosmological constant is ${\Lambda _{{\rm{Emin}}}} = 8.664\times{10^{-8}}\,{\rm{ meV}}$ corresponding to the maximum radius of the Universe shown in Eq. (\ref{equ26-a}). }\label{figure3}
\end{figure}

If matter density increases in the pressure-less case, the interaction potential energy between matter and the scalar field will decrease and will finally approach to zero. According to Eqs. (\ref{equ16b}) and (\ref{equ18-a}), when $\rho$ approaches to infinity, the interaction potential Eq. (\ref{equ8plusa}) will vanish due to $A({\phi _{\min }})=1$. But the density-dependent cosmological constant will increase and finally approach to a constant. In other words, when the density large enough the cosmological constant obtained by Eq. (\ref{equ21}) together with Eq. (\ref{equ18-a}) is nearly density-independent, which is a desired result. In this sense, the cosmological constant really becomes a constant. For example, when $\rho  \to \infty $, one has
\begin{subequations}\label{equ25re}
\begin{eqnarray}
{\Lambda _{\rm{E}}} &=& 2.242\,{\rm{ meV}} \label{equ25rea} \\
(\Lambda &=& {\rm{1.093}} \times {{\rm{10}}^{{\rm{ - 52}}}}\,{{\rm{m}}^{ - 2}}), \nonumber \\
{w_{{\rm{eff}}}} &=&  - 1.\label{equ25reb}
\end{eqnarray}
\end{subequations}
The limit towards infinity does not represent any physical process: it is a mathematical construct to depict a nearly fixed value of $\Lambda $ in the past of the Universe \cite{z65replus}. In fact, that ${\Lambda _{\rm{E}}} \simeq 2.242\,{\rm{ meV}}$ always holds as long as $\rho \gg {\lambda {M_1}^{\rm{4}}{c^3}}/{{\hbar ^3}} \sim 10^{-30}\, \rm{kg/m^3}$. This implies that an actual time-variable of the cosmological `constant' in the matter-coupled scalar field model behaves a real constant before the present era. Figure \ref{figure3} shows the cosmological constant versus the matter density.

Due to the broken-symmetry of the interaction between the scalar field and matter, the value of the self-interaction potential at the minimum of the effective potential is pinned at a nearly fixed value if the density of matter is large enough. The minimum is mainly determined by three factors: the self-interaction potential, the shape of the coupling function and the density of matter. These remind us to have chosen the appropriate coupling shape so that the minimum is insensitive to the change of matter density for the large density of matter. The requirements of the adiabatic stability also implies that the effective mass of the scalar field should be large enough, which results in a very short range fifth force in wide region of ambient matter density (see section \ref{force}). It is the broken-symmetry coupling function that play a pivotal part in suppressing the scalar gradient force effect and in promoting the dark energy role of the scalar field in a matter environment.

\subsection{The space-curvature-dependent future of the expansion Universe }\label{thefuture}

The nearly fixed cosmological constant before the present era has been proven above. We now return to the case of matter density decreasing due to the Universe expanding. The Universe will switch into a deceleration expansion status according to the acceleration equation of the Universe. However, in order to judge whether the expansion in the distant future will stop or not, one has to consider the curvature of the Universe.

\subsubsection{$\Lambda  \propto {\rho ^2}$ long after the present era}\label{thecosmologicalconsafter}

With the density decreasing further in the future, for example, when $\rho \ll {\lambda {M_1}^{\rm{4}}{c^3}}/{{\hbar ^3}}$, it can be easily obtained from Eqs. (\ref{equ14}), (\ref{equ18-a}) and (\ref{equ21}) that
\begin{equation}
\Lambda  \simeq \left( {\frac{{2\pi G{\hbar ^3}{M_2}^4}}{{\lambda {c^5}{M_1}^8}}} \right){\rho ^2}.\label{equ26plusre1}
\end{equation}
Therefore, $\Lambda $ will reach a region where it decreases faster than matter density does in the future. In other words, $\ddot a < 0$ will occur according to Eq. (\ref{equ12}) as long as
\begin{equation}
 2V\left( {{\phi _{\min }}} \right)  < \rho _{\rm{m}} {\hbar ^3}{c^5}.\label{equ26plus}
\end{equation}
Consequently, the Universe will switch into a deceleration expansion status. It is clear that the self-interaction potential of the scalar field causes the expansion acceleration while matter density ${\rho _{\rm{m}}} $ decreases the acceleration. Although both $\Lambda  \to 0$ and ${\rho _{\rm{m}}} \to 0$ when $\rho  \to 0$, the different convergence rates result in the next decelerating expansion after the transition redshift of $ {z_{{\rm{future}}}} $.

What would happen next? Does the Universe keep expanding forever or switch into contracting? This cannot be solved by the acceleration equation (\ref{equ12}) alone. Applying the current Hubble constant $H_0$ to the Friedmann equation (\ref{equ11}), the question may be answered. Due to the Hubble tension \cite{z103,z109}, however, another criterion is needed, which will be shown in section \ref{contraction}.

\subsubsection{\label{flats}Flat space}

Let's discuss the flat space first,\emph{ i.e.} $ K=0 $. In this case, since only the combination $\dot a /a$ appears in the Friedmann Eq. (\ref{equ11}), one is free to rescale $a(t)$ as one chooses. For example, one can choose $a_0=1$ at the present time, which means that the physical coordinate system coincides with the comoving one at the present. The scalar-field-independent matter density in the pressure-less case can be expressed via Eq. (\ref{equ6}) as follows:
\begin{equation}
\rho  = \frac{{a_0^3{\rho _0}}}{{{a^3}}},\label{equ25re1}
\end{equation}
where the value of the current scalar-field-independent matter density ${\rho _0}$ has been estimated as shown by Eq.(\ref{equ22re}). Substituting (\ref{equ25re1}) into Eqs. (\ref {equ18-a}) and (\ref{equ21}), then substituting both into the Friedmann Eq. (\ref{equ11}), we obtain a little complicated differential equation about the scale factor in time.

However, we can study the asymptotic behavior for the late-time evolution of the Universe. When the matter density becomes smaller and smaller with the expansion so that $\phi _{\min }^2 \ll M_2^2{c^4}$, one will find the solution is $a(t) \propto {t^{2/3}}$ and then $\rho (t) \propto 1/{t^2}$, which is the same as the traditional matter-dominated solution in flat space. The Universe expands forever, but the Hubble parameter decreases with time as $H = 2/(3t)$ meaning that it will infinitely approach to zero as the time approaches to infinity. Substituting $K=0$ and the total density Eq. (\ref{equ23}) into the Friedmann Eq. (\ref{equ11}), one can calculate the current Hubble constant for the flat space as follows:
\begin{equation}
H_0(K=0)=67.376\,{\rm{ km }}{{\rm{s}}^{ - 1}}{\rm{ Mp}}{{\rm{c}}^{ - 1}}.\label{equ25re2}
\end{equation}
This value is close to the measured value of the current Hubble constant \cite{z1,z14}, such as, ${H_0} = 67.36\,{\rm{ km }}{{\rm{s}}^{ - 1}}{\rm{ Mp}}{{\rm{c}}^{ - 1}}$ \cite{z1}, or ${H_0} = 67.8\,{\rm{ km }}{{\rm{s}}^{ - 1}}{\rm{ Mp}}{{\rm{c}}^{ - 1}}$ \cite{z14}. The Hubble tension is not completely resolved \cite{z103,z109}. Although the Universe is close to a flat space at the present time, it still has possibility being the positive or negative curvature space. However, a closed space is favored by the observed feature of a concave potential \cite{z1}, which will be discussed in section \ref{concavevex}. In addition, since the measured Hubble constant corresponds to the Jordan frame, it should be transformed into the Einstein frame due to the calculation here performing in the Einstein frame. The relation of the two frames will be discussed in section \ref{selfacc}.

\subsubsection{\label{open}Open space}

In the cases of $K \neq 0$, one often rescales the scale factor $ a(t) $ by setting $|K|=1$ as shown in Eq (\ref{equ3}). When this rescaling is used, the freedom to set $a_0=1$ in flat space is lost, and the scale factor has definitely meaning in physical scale, e.g. curvature radius.

For the negative curve space $K=-1$ \cite{z14}, one can deduce from the Friedmann Eq. (\ref{equ11}) that, the Universe expands forever, but the Hubble parameter decreases with time and will infinitely approach to zero as the time approaches to infinity. These conclusions are the same as that in flat space. The asymptotic behavior of the Universe evolution in the future is also the same as the traditional matter-dominated solution in open space, \emph{i.e.,} $a \propto t$.

Besides, it is very clearly from Eq. (\ref{equ11}) that, $\dot a $, the expansion speed of the Universe exceeds the speed of light forever in this situation. This means that there is an infinite space part of the Universe cannot be observed. Thus, it is should be stressed that, essentially, the expansion based on the cosmological principle results from the entire homogeneous energy density and the corresponding pressure in the Universe. Due to the homogeneity, there is no gradient force driving the expansion. Due to the entirety, the density and pressure everywhere contribute their effect to the expansion. More specifically, the expansion does not involve any gradient force and then is not related to the Compton wave length of the scalar field.

\subsubsection{\label{closen}Closed space}

We now discuss the case of $K=1$. If we use the current Hubble constant \cite{z1} ${H_0} = 67.36\,{\rm{ km }}{{\rm{s}}^{ - 1}}{\rm{ Mp}}{{\rm{c}}^{ - 1}}$, the critical density can be estimated as
\begin{equation}
 {\rho _{\rm{c}}} \equiv \frac{{3{H_0}^2}}{{8\pi G}} = 8.52261 \times {10^{-27}}\;{\rm{kg}} \cdot {{\rm{m}}^{ - 3}}.\label{equ24}
\end{equation}
Since the total density ${\rho _{{\rm{tot}}}}$ shown by Eq. (\ref{equ23}) is slightly larger than the critical density ${\rho _{\rm{c}}}$, the Universe might be closed, \emph{i. e}., $K = 1$, and the current radius of the Universe is derived from Eq. (\ref{equ11}) as
\begin{equation}
{a_0} =6.305 \times {10^{27}}\rm{ m } =  204.3\;{\rm{ Gpc}}.\label{equ25}
\end{equation}
This value satisfies the constraint condition $a_0>\rm{ 81 Gpc}$ shown in the Planck 2018 results (X. constraints on inflation) \cite{z1i}.

From the Friedmann Eq. (\ref{equ11}), one can deduce easily that the Universe must go through infinite cycles of two stages: contraction and expansion. At the end of the decelerating expansion, the Universe will reach its maximum radius ${a_{\max }}$ where the expansion speed $\dot a = 0$. Substituting $\dot a = 0$ into Eq. (\ref{equ11}) (noticing the equation is only valid in the pressureless case), we get a redshift
\begin{equation}
z =  - 0.9999985281\label{equ26minus}
\end{equation}
corresponding to the maximum radius of the Universe. The maximum radius, the corresponding matter density, and the cosmological constant are given, respectively, as follows:
\begin{subequations}\label{equ26reabcd}
\begin{eqnarray}
{a_{\max }} &\equiv & {\frac{{{a_0}}}{{1 + z}} = 4.283 \times {10^{33}}\,{\rm{ m }}},\label{equ26-a}\\
{\rho_{\rm{min}}}  &\equiv & {{\rho _0}{\left( {1 + z} \right)^3}=8.547 \times {10}^{ - 45}\,{\rm{kg}} \cdot {{\rm{m}}^{ - 3}}}.\label{equ26}\\
{\Lambda _{{\rm{Emin}}}} &=& {8.664 \times{10^{-8}}\,{\rm{ meV}}}, \label{equ26rec}
\end{eqnarray}
\end{subequations}
The corresponding physics matter density is ${\rho _{\rm{mmin}}} = 8.761 \times {10^{ - 42 }}\, \rm {kg/{m^3}}$, which is about 3 orders of magnitude larger than the scalar-field-independent matter density shown in Eq. (\ref{equ26}). The cosmological constant corresponding to the maximum radius should be the minimum value in all of permissible values of the cosmological constant, that is, ${\Lambda _{{\min}}} = 2.44 \times {10^{ - 82}}\, {{\rm{m}}^{ - 2}}$. Therefore, there does not exist the situation of $\Lambda  = 0$. The effective equation of state for the coupled scalar field in this case is estimated by Eq. (\ref{s26plusere1}) to be ${w_{{\rm{eff}}}} =  - 1.494 \times {10^{ - 15}}$.

\begin{figure}
\centering
\includegraphics[width=200pt]{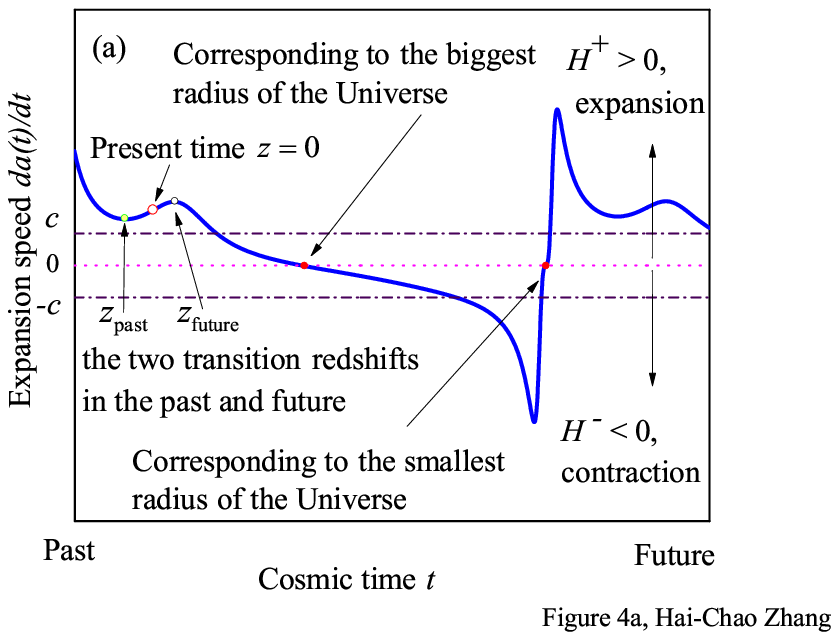}
\includegraphics[width=200pt]{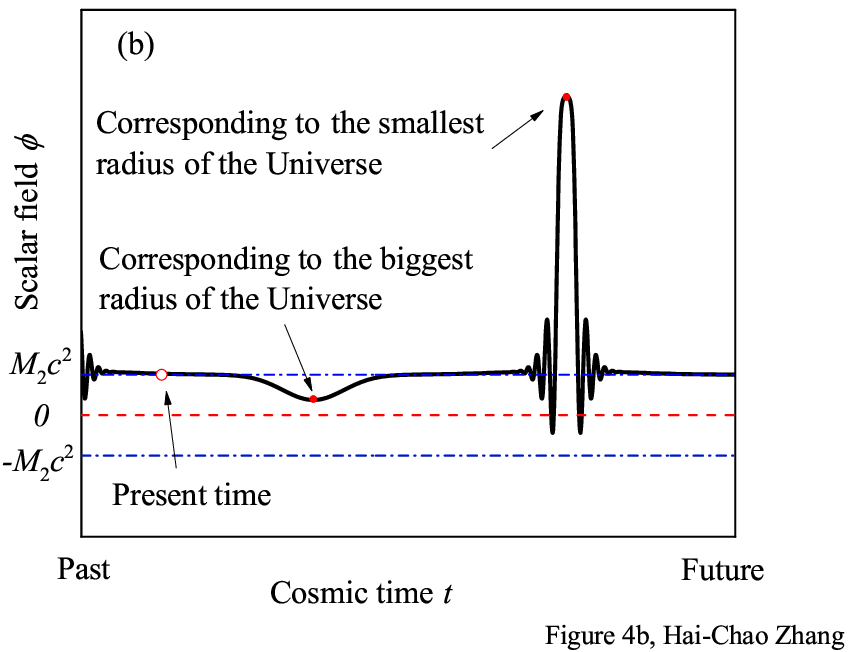}
\includegraphics[width=200pt]{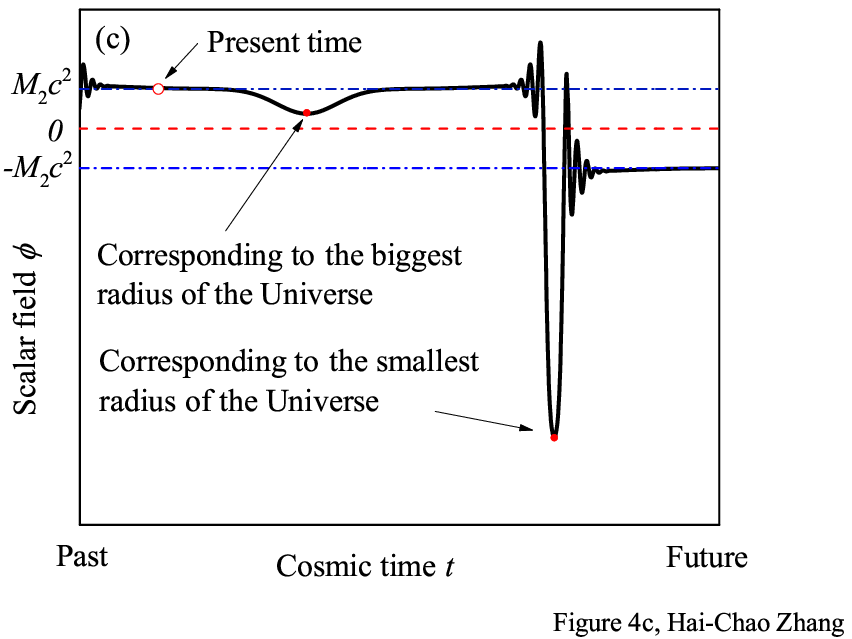}
\includegraphics[width=200pt]{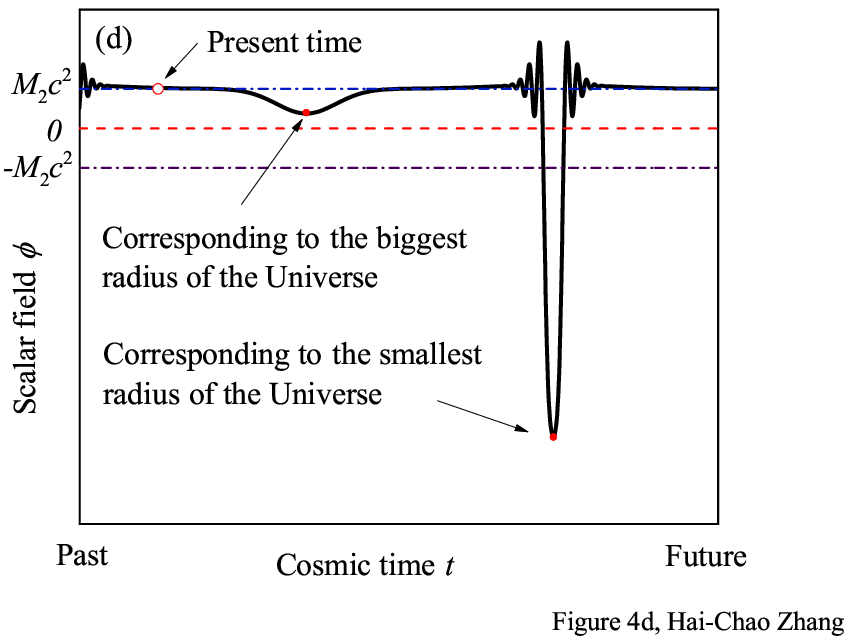}
\caption{The schematic sketch of the Universe evolution with time. (a) The sketch of the expansion speed of the Universe versus the cosmic time from the present to the future. $c$ is the speed of light. For acceleration expansion: $\ddot a > 0$ and $\dot a > 0$; for deceleration expansion: $\ddot a < 0$ and $\dot a > 0$; for acceleration contraction: $\ddot a < 0$ and $\dot a < 0$; for deceleration contraction: $\ddot a > 0$ and $\dot a < 0$; for the biggest radius of the Universe: $\ddot a < 0$ and $\dot a = 0$; for the smallest radius of the Universe: $\ddot a > 0$ and $\dot a = 0$; for the transitions between deceleration and acceleration expansion (contraction): $\ddot a = 0$; and such a pattern repeats itself. (b) The sketch of the scalar field versus the cosmic time during the evolution of the Universe for both the next new inflation and the second accelerating expansion occurring at $\phi>0$. (c) The sketch of the scalar field versus the cosmic time during the evolution of the Universe for both the next new inflation and the second accelerating expansion occurring at $\phi<0$. (d) The sketch of the scalar field versus the cosmic time during the evolution of the Universe for the next new inflation $\phi<0$ but the second accelerating expansion still occurring at the same as the current choice of $\phi>0$. The current value of the scalar field is chosen to be positive in all the situations of (b), (c) and (d). Since the potential density of the scalar field is equal to ${\phi}^4/4!$, all the cases present the same results of the Universe evolution. }\label{figure4}
\end{figure}

After the expansion end of  $\dot a = 0$, the Universe will begin its contraction \cite{z65replus}, which will be discussed in section \ref{contraction}. However, for completeness, the sketch of the whole evolution including both the expansion and the contraction of the Universe is presented here by Fig. \ref{figure4}. Figure \ref{figure4}(a) shows the sketch of the expansion speed versus the cosmic time. Correspondingly, the sketch of the scalar field versus the cosmic time during the evolution of the Universe is plotted schematically in Figs. \ref{figure4}(b), \ref{figure4}(c) and \ref{figure4}(d), respectively. The current value of the scalar field is chosen to be positive in all the Figs. \ref{figure4}(b), \ref{figure4}(c) and \ref{figure4}(d). However, the next new inflation due to the contraction may occur at either the same sign or opposite sign with the current choice of the sign of the scalar field. Figure \ref{figure4}(b) denotes both the next inflation and the second acceleration occurring at $\phi>0$ . Figures \ref{figure4}(c) and \ref{figure4}(d) denotes the next inflation occurring at $\phi<0$ with the second acceleration occurring $\phi<0$ and $\phi > 0$, respectively. Since the potential density of the scalar field is equal to ${\phi}^4/4!$, the three cases of Figures \ref{figure4}(b), \ref{figure4}(c) and \ref{figure4}(d) present the same results of the Universe evolution.

\subsection{Summary}\label{suforexp}

Using the only free parameter in the symmetry-breaking interaction model, the nearly fixed cosmological constant is obtained. The cosmological constant is estimated to be ${\Lambda _{\rm{E}}} \simeq 2.242\,{\rm{ meV}}$ before the present era, which is slightly larger than the current value ${\Lambda _{\rm{E}}} = 2.239\,{\rm{ meV}}$. Long after the present era, the cosmological constant is proven to be proportional to square of the density of matter. Therefore, the energy density of the scalar field will decrease faster than that of matter and then the Universe will shift into a deceleration expansion. Based on the current observation precision of the Hubble constant, one can deduce that the Universe is a nearly flat space in the present era. Due to the Hubble tension, it is necessary to invoke another criterion to judge the curvature of the Universe. The criterion will be shown in section \ref{contraction} and the positive curvature is preferred. For a closed Universe, the expansion will stop and the Universe will then contract.

The problem of how to reconcile our scheme with chameleon no-go theorems will be discussed in next section \ref{Nogo}.

\section{Evading the physical corollary of Chameleon No-Go theorems }\label{Nogo}

The main purpose of this paper is to find how to obtain a scalar fifth force under the requirement that the model should reflect as much as possible the measured value of the cosmological constant. The model should not add a cosmological constant to drive cosmic acceleration, since any force involves spatial gradient and then the effect of the added constant disappears in the differentiation. This reminds us that it is the energy density of the scalar field that drives cosmic acceleration rather than the spatial gradient fifth force does.

It is well known that the cosmological constant model provides the simplest explanation of cosmic acceleration. Apparently, there is no local fifth force in this model, albeit the nature of its negative pressure guarantees acceleration of the Universe. One does not regard this as a satisfactory solution based on the viewpoint of quantum field theory. When the vacuum expectation value of a conventional quantum field theory is used to mimic the cosmological constant, however, Weinberg's no-go theorem occurs \cite{z11}, which states that no tuning of the corresponding energy density can be naturally achieved.

Scalar field theories of dark energy, such as quintessence with a time-dependent energy density but without coupling to matter \cite{z34,z35,z104,z106}, may be able to circumvent Weinberg's no-go theorem. If the scalar field does not couple to matter, \emph{i.e}., $A(\phi)=1$, our model mentioned above becomes to one of the typical quintessence forms. To mimic a cosmological constant, the scalar field must be in a very slow roll state so that its kinetic energy can be negligible. This requires that the Compton frequency of the scalar field is smaller than the Hubble rate, \emph{i.e}. $m_{\phi}c^2 <{\hbar} H_0\sim 10^{-33} \, \rm{eV}$ with $m_{\phi}$ being the mass of the scalar field. In this case, the scalar fifth force does not appear but the scalar field is still used to act the dark energy field. The drawback of the models is that the mass of the scalar field is too small. In any case, one can see that the Universe accelerating expansion is independent of any gradient force of the scalar field. Thus, the interaction range of the scalar field does not play an important role in the observed cosmic acceleration.

Scalar-tensor theories, such as chameleon \cite{z3} and symmetron \cite{z6} model, have a coupling of the scalar field to matter and then scalar fifth forces appear. Unfortunately, a corollary based on chameleon no-go theorems states that the chameleon-like fields cannot drive the observed cosmic acceleration \cite{z101,z13}. The detailed analysis of the conclusion is given as follows.

\subsection{Self-acceleration problem}\label{selfacc}

Acceleration Eq. (\ref{equ12}) clearly shows that acceleration will take place when $2V\left( {{\phi _{\min }}} \right)  > \rho _{\rm{m}} {\hbar ^3}{c^5}$. The acceleration is caused by the stable value of $2V\left( {{\phi _{\min }}} \right)$ with $\phi _{\min }$ being the minimum of the effective potential. The effect of the coupling function to the acceleration is indirectly through the matter-density-dependent interaction potential.

Let us check the possibility of self-acceleration in our scheme. We need introduce the Jordan-frame which, indeed, has been implied in the second term on the right-hand side of the action Eq. (\ref{equ1}). The Jordan-frame metric $g_{\mu \nu }^J $ is related to the Einstein-frame metric $g_{\mu \nu }$ by the positive coupling function $ A( \phi )$ as following \cite{z101,z13}:
\begin{equation}
g_{\mu \nu }^J = {A^2}(\phi ){g_{\mu \nu }}.\label{equ26replus}
\end{equation}

Self-accelerating theories attempt to attribute the observed (Jordan-frame) cosmic acceleration to the self-acceleration \cite{z101}. That is, the cosmic acceleration stems entirely from the conformal transformation shown in Eq. (\ref{equ26replus}). Literature \cite{z101} has proven that this is impossible. This is one of the chameleon no-go theorems. It will be seen later in this subsection that our model does not conflict with this no-go theorem, albeit the scalar field can account for the observed cosmic acceleration as has been shown in section \ref{expansion}.

To obtain the observable quantities in the Jordan frame, the following translation between the Einstein and Jordan frames need to be used,
\begin{subequations}\label{equ26replus2}
\begin{eqnarray}
{{a^J}({t^J})} & =& { A(\phi )a(t)}, \label{equ26replus2a}\\
{d{t^J}} &=& {A(\phi )dt}, \label{equ26replus2b}
\end{eqnarray}
\end{subequations}
where $a^J$ and $t^J$ are the scale factor and the cosmic time in the Jordan frame, respectively.

Suppose that the scalar field has been sitting stable at the minimum of the effective potential, then the coupling function is completely determined by the density of matter due to $A(\phi )=A{\left[\phi_{\min}(\rho )\right]}$. In our scheme when $w_i =0$, the function $\phi_{\min}(\rho )$ has been shown as Eq. (\ref{equ18-a}). Following \cite{z101}, in the case of pressureless matter source it can be easily obtained that,
\begin{subequations}\label{equ26replus3ab}
\begin{eqnarray}
{{\dot a}^J} &=& {\dot a - 3\dot a\frac{{d\ln A}}{{d\ln \rho }}},\label{equ26replus3a}\\
{{\ddot a}^J} &=& {\frac{{\ddot a}}{A}\left( {1 - 3\frac{{d\ln A}}{{d\ln \rho }}} \right) + \frac{{9{{\dot a}^2}}}{{Aa}}\frac{{{d^2}\ln A}}{{{{\left( {d\ln \rho } \right)}^2}}}},\label{equ26replus3b}
\end{eqnarray}
\end{subequations}
where ${\rho }$ is scalar-field-independent matter density, ${{\dot a}^J} \equiv d{a^J}/d{t^J}$ and ${{\ddot a}^J} \equiv {d^2}{a^J}/({dt^J})^2 $, respectively. Apparently, if the expansion speed in the Einstein frame is equal to zero,\emph{ i.e.}, $\dot a = 0$, the expansion speed in the Jordan frame is also equal to zero, \emph{i.e.}, ${{\dot a}^J} = 0$. However, if the acceleration in the Einstein frame is equal to zero, \emph{i.e.}, $\ddot a = 0$,  the acceleration in the Jordan frame is no longer equal to zero, \emph{i.e.}, ${{\ddot a}^J} \ne 0$ . This no-zero acceleration in the Jordan frame can be regarded as self-acceleration, \emph{i.e}., a genuine modified gravity effect.

\begin{figure}
\centering
\includegraphics[width=250pt]{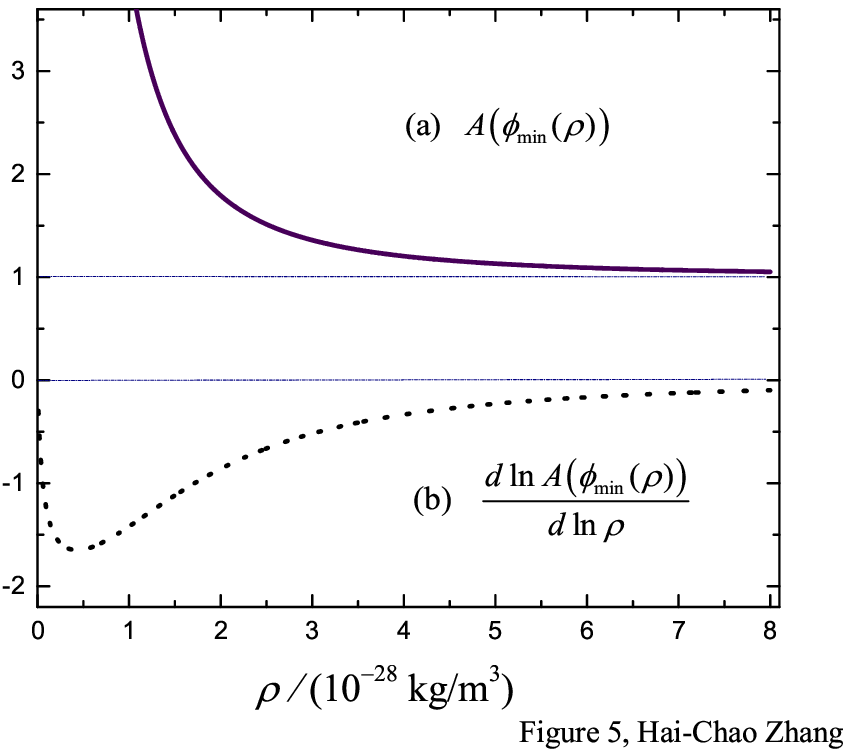}
\caption{(a) The matter-density-dependent value of the coupling function at the minimum of the effective potential. When the density of matter is large enough, e.g., larger than the current matter density of the Universe, the value of the coupling function is almost equal to 1 and then the corresponding self-acceleration is nearly equal to zero. (b) The coefficient $(d\ln A)/(d\ln \rho ) $ varies with the density of matter in the pressure-less case. Since the coefficient is always negative, the cosmological observation value of the expansion speed (performed in the Jordan-frame) is larger than the corresponding calculation value obtained in the Einstein-frame. The derivative of the coefficient with respect to the density of matter can roughly reflect the effect of self-acceleration. If the derivative is positive, the self-acceleration is positive in the Jordan-frame even if the corresponding acceleration in the Einstein-frame is zero, and vice versa. When the density of matter is large enough, e.g., larger than the current matter density of the Universe, the coefficient and its derivative with respect to the density of matter is nearly equal to zero. Therefore, Jordan- and Einstein-frame metrics are indistinguishable before the current era for the pressure-less fluid of matter sources. }\label{figure5}
\end{figure}

The existence of the self-acceleration implies that the transition redshifts calculated in section \ref{expansion} should be corrected, because cosmological observations are implicitly performed in the Jordan frame. To obtain the correction to the transition redshifts, one should use ${{\ddot a}^J} =0$ rather than $\ddot a = 0$. The calculation of the transition redshifts can still perform in the Einstein frame. After obtaining the transition redshifts $z$ in the Einstein frame, one can use the following translation to obtain the transition redshift $z^J$ in the Jordan frame,
\begin{equation}
1 + {z^J} = \frac{{{A_0}}}{A}\left( {1 + z} \right).\label{equ26replus4}
\end{equation}
where $1 + {z^J} = {{a_0}^J}/{a^J}$ and subscript `0' marks the current time. The Hubble parameter should also be converted into the appropriate frame. It is worth noting that, the previous calculation in the last section is indeed in the Einstein frame. When the measured Hubble parameter (in the Jordan frame $H^J= {\dot a}^J/a^J$) is used to the calculation, it should be transformed into the Einstein frame as $H= \dot a/a$, which is ignored in the last section. According to Eqs. (\ref{equ16b}) and (\ref{equ18-a}), the coupling function $A( {\phi }_{\min}  )$ is nearly fixed to the value of 1 in the past of the Universe due to $\rho  \gg  {\lambda {M_1}^{\rm{4}}{c^3}}/{{\hbar ^3}} \sim  10^{-30}\,\rm{kg/m^3}$. Therefore, the correction to the Hubble parameter related to the different frames can be approximately neglected \cite{z115}. We may introduce a notation $C(\rho)$ to mark the part of self-acceleration in Eq. (\ref{equ26replus3b}) as follows:
\begin{equation}
C(\rho ) = \frac{{9{{\dot a}^2}}}{{Aa}}\frac{{{d^2}\ln A}}{{{{(d\ln \rho )}^2}}}.\label{nogo6}
\end{equation}
Of course, the second term in the parentheses on the right-hand of Eq. (\ref{equ26replus3b}) can also contribute the so-called self-acceleration. In any case, when the density of matter is larger than the current matter density, from Eq. (\ref{equ16b}) together with Eq. (\ref{equ18-a}), one can easily obtain that $A({\phi _{\min }}) \cong 1$. Therefore, self-acceleration is approximated to zero and then Jordan- and Einstein-frame metric are indistinguishable. Substituting Eqs. (\ref{equ16b}) and (\ref{equ18-a}) into Eq. (\ref{nogo6}), one also obtain that $C(\rho  = 4.5 \times {10^{ - 29}}{\text{ kg}} \cdot {{\text{m}}^{ - 3}}) = 0$, and $C(\rho ) > 0$ ($C(\rho ) < 0$) when $\rho  > 4.5 \times {10^{ - 29}}{\text{ kg}} \cdot {{\text{m}}^{ - 3}}$ ($\rho  < 4.5 \times {10^{ - 29}}{\text{ kg}} \cdot {{\text{m}}^{ - 3}}$). Only if matter density is small enough, the effect of self-acceleration (self-deceleration) occurs.  Figure \ref{figure5}(a) shows that the value of the coupling function at the minimum varies with the density of matter in the pressure-less case. Figure \ref{figure5}(b) describes the trend of self-acceleration varying with the density of matter.

From Eq. (\ref{equ26replus3a}) together with Figure \ref{figure5}(b), one sees that ${{\dot a}^J} > \dot a $ due to $(d\ln A)/(d\ln \rho ) < 0 $.  From Figure \ref{figure5}, with the Universe expanding the self-acceleration marked by Eq. (\ref{nogo6}) will occur in the future and then turn to self-deceleration with the matter-density further decreasing. Before the present time, since the coupling function nearly keeps a constant of value 1 (that is equivalent to the statement of $\Delta A/A \ll 1$ in \cite{z101}), the self-acceleration then vanishes. This clearly concludes that our scheme coincides with the second chameleon no-go theorem \cite{z101}. Unfortunately, based on the chameleon no-go theorem, a misleading corollary is deduced, which states that, the chameleon-like scalar field cannot impact cosmological observations \cite{z101}. Apparently, the corollary conflicts with our scheme. By adopting a broken-symmetry coupling function, the cosmological constant has been obtained in section \ref{expansion} since the chameleon-like scalar field in our proposal indeed is also a quintessence field. The vanishing of self-acceleration does not mean that the quintessence effect of the scalar field must vanish. One cannot deduce the above corollary from the almost zero-self-acceleration. There is no convincing no-go theorem hinders the establishment of a chameleon-like model to mimic the cosmological constant.

\subsection{Overshooting problem}\label{overshoot}

Since the minimum of the effective potential changes with time, a characteristic time has been introduced in \ref{symetrybreaking} to describe whether the minimum moves quickly or slowly. The changing rate of the minimum position has been naturally defined by ${{\dot \phi }_{{\text{min}}}}/{\phi _{{\text{min}}}} $. If the rate is smaller than the damping rate of $|3H(t)/2|$ with $H(t)$ being the Hubble parameter, the scalar field can adiabatically follow the minimum of the effective potential; On the contrary, overshooting must occur. Overshooting problem in the chameleon-like model is sometimes regarded as another no-go theorem, although it is not explicitly mentioned in \cite{z101,z13}.

In our scheme, it has been demonstrated that $|{{\dot \phi }_{{\text{min}}}}/{\phi _{{\text{min}}}}| \leq |3H/2|$ [see Eq. (\ref{s8plus2}) in Appendix \ref{appC}]. The higher the density of matter, the smaller the changing rate of ${\phi _{{\text{min}}}}$. This means that the scalar field sits more stably at the minimum in a higher density of matter. The transition redshift denoting the deceleration-acceleration transition has been estimated in section \ref{expansion}. At the transition redshift, the matter density $\rho_{\rm{past}}  = 1.17 \times {10^{ - 26}}\,{\rm{ kg/}}{{\rm{m}}^3}$. That is, when matter density $\rho > \rho_{\rm{past}} $ the Universe is in the phase of decelerating expansion, and vice versa. Near the transition density, the scalar field sits very stably at the time-dependent minimum due to $\rho_{\rm{past}}   \gg  {\lambda {M_1}^{\rm{4}}{c^3}}/{{\hbar ^3}} \sim  10^{-30}\;\rm{kg/m^3}$. We conclude that the large effective mass around the minimum results in the small changing rate of the minimum position and then suppresses the possibility of overshooting.

Now, we explain why the overshooting occurs in the symmetron \cite{z5}, one of the chameleon-like models, in which the self-interaction is a symmetry-breaking potential and the coupling function is not broken-symmetry, \emph{i.e}.,
\begin{subequations}\label{symm}
\begin{eqnarray}
{V(\phi )} &=&  {- \frac{{{\mu ^2}{c^4}}}{2}{\phi ^2} + \frac{\lambda }{4}{\phi ^4}}, \label{symm1} \\
{A(\phi )} &=& {1 + \frac{1}{{2{M^2}{c^4}}}{\phi ^2}}.\label{symm7}
\end{eqnarray}
\end{subequations}
One can easily obtain the minima of their effective potential as follows:
\begin{equation}
{\phi _{{\text{min}}}} =
\begin{cases}
0, & {\rho  \ge \frac{{{\mu ^2}{M^2}{c^3}}}{{{\hbar ^3}}}}\\
{ \pm {{[\frac{1}{\lambda }({\mu ^2}{c^4} - \frac{{\rho {\hbar ^3}c}}{{{M^2}}})]}^{1/2}}}, & {\rho  < \frac{{{\mu ^2}{M^2}{c^3}}}{{{\hbar ^3}}}}.\label{symm7+1}
\end{cases}
\end{equation}
The corresponding effective mass around the minima are
\begin{equation}
m_{{\text{eff}}}^2 =
\begin{cases}
{\frac{{\rho {\hbar ^3}}}{{{M^2}{c^3}}} - {\mu ^2}}, & {\rho  \ge \frac{{{\mu ^2}{M^2}{c^3}}}{{{\hbar ^3}}}} \\
{2({\mu ^2} - \frac{{\rho {\hbar ^3}}}{{{M^2}{c^3}}})},& {\rho  < \frac{{{\mu ^2}{M^2}{c^3}}}{{{\hbar ^3}}}}.\label{symm7+2}
\end{cases}
\end{equation}
It should be emphasized that the effective mass equals to zero at the critical value of matter density.

Literature \cite{z5} attempts to mimic the deceleration-acceleration transition redshift in the recent past by a phase transition when the effective mass vanishes. Apparently, the vanishing undoubtedly causes a serious overshooting. When the effective mass far smaller than the Hubble rate in the mass scale, the stability of the minimum is too fragile to resist perturbations. A little energy of the perturbations can result in a huge moving speed of the minimum.

Substituting Eqs (\ref{symm}) and (\ref{symm7+1}) into Eq. (\ref{s8plus}) in Appendix \ref{appC}, one has

\begin{equation}
\frac{{{{\dot \phi }_{{\text{min}}}}}}{{{\phi _{{\text{min}}}}}} = H \cdot \frac{{3{\hbar ^3}\rho }}{{{M^2}m_{{\text{eff}}}^2{c^3}}}.\label{symm7+3}
\end{equation}
Since the effective mass $m_{{\text{eff}}}^2 \to 0$ at the critical value of matter density, Eq. (\ref{symm7+3}) gives a result that the changing rate ${{\dot \phi }_{{\text{min}}}}/{\phi _{{\text{min}}}} \to \infty $. Thus, the minimum of the effective potential moves too quickly for the scalar field to follow it adiabatically during the expansion of the Universe. In no-adiabatic tracking cases, when the Compton frequency of the scalar field is smaller than damping rate of $|3H/2|$, it undergoes an over-damped evolution; when the Compton frequency is larger than the damping rate, the scalar field undergoes under-damped oscillations. The two phenomena just are what have been shown in the literature \cite{z5} near its transition redshit. In addition, when the symmetry-breaking potential (\ref{symm1}) is chosen as a quintessence field, its value at the minima is too small to drive cosmic acceleration.

Symmetron-like generalized potential and coupling function shown in \cite{z5} cannot also avoid the overshooting problem, because the effective mass still vanishes at the critical value of matter density.

\subsection{Compton wavelength problem}\label{comlength}

At the beginning of this section, it has been shown by way of example that the observed cosmic acceleration is ascribed to the self-interaction potential density of the scalar field rather than any form of the spatial gradient of the scalar field. Although the gradient may contribute its effect locally to the Universe, its resulting contribution to the cosmic acceleration should be zero. The reason is that, the part of positive values of the gradient has to be offset by the corresponding part of negative values, otherwise the cosmological principle would be violated.

In our scheme, as long as the matter density is large enough, the minimum of the effective potential of the scalar field becomes almost matter-density independent [see Eq. (\ref{s6minus}) in Appendix \ref{appC}]. In this case of the large matter-density, even if the matter-density varies with space, the minimum almost keeps the same value. Thus, dark energy behaves like the cosmological `constant' not only in temporal scale but also in spatial scale. Unlike the minimum, however, the corresponding effective mass of the scalar field is strongly dependent on the density of matter and is very large in general case. The very large mass guarantees the stability of the minimum to perturbations. The large effective mass means that the Compton wavelength of the scalar field is short ranged and then the effects of the corresponding fifth force is considerably suppressed, which will be discussed in section \ref{force}.

Although the broken-symmetry coupling function has been used to explain the cosmological constant and will be used to explain inflation of the Universe in section \ref{contraction}, a seriously misleading problem to the short interaction range of the scalar field needs to be presented. Unfortunately, the problem cannot be clearly resolved by some mathematical equations as we have done above, because it essentially results from the misunderstanding on the physical concept related to the accelerating expansion. It has always been considered that the scalar field must be light if it is to address the cosmological-constant problem \cite{z80}. This questionable viewpoint stems at least from the two requirements as follows: One results from the case that the scalar field does not couple with matter, \emph{i.e}., the traditional quintessence situation. The shallow potential is required so that the evolution of the scalar field can satisfy the slow roll condition. This requirement is unreasonably extrapolated to the coupling case \cite{z80}. We call this one as the shallow-potential requirement; The other is that the scalar field should mediate a long range interaction so as to explain the acceleration expansion of the Universe \cite{z21,z46,z47}. We call the second one as the long-range-interaction requirement.

The long-range-interaction requirement is not the same as the shallow-potential one. In the quintessence situation where the light scalar field does not interact directly with matter, the shallow potential guarantees that the quintessence field can roll down slowly and then its kinetic energy can be neglected.  Thus, as long as the self-interaction potential large enough, the quintessence model can mimic the cosmological constant. In the quintessence model, there is no long-range interaction. This also implies that a long-range interaction is not a necessary condition to explain the cosmological constant by a scalar field. Apparently, the first requirement of the shallow potential is not necessary in our scheme because the value of the self-interaction potential can now be localized by the symmetry-breaking coupling. Therefore, in the main rest of the section we will focus on the topic that the second requirement of long-range interaction is not necessary.

\subsubsection{\label{scalarcw}The Compton wavelength of the scalar field}

We now need to review the second chameleon no-go theorem. The theorem is an upper bound on the chameleon Compton wavelength at present cosmological density \cite{z101,z13}, which is given as follows:
\begin{equation}
{\mathchar'26\mkern-10mu\lambda _c} \equiv \frac{\hbar }{{{m_{{\text{eff}}}}c}} \lesssim {\text{Mpc}}.\label{nogo8}
\end{equation}
In our model, the Compton wavelength is estimated to be about $\sim 5 \,\mu \rm{m}$ at present cosmological density, which of course satisfies the constraint denoted by Eq. (\ref{nogo8}). According to \cite{z101}, any cosmological observable probing liner scales should see no deviation from general relativity in our model due to the short range. From the mathematical point of view, our model does not conflict with the chameleon theorems. Based on chameleon theorems, however, literature \cite{z101} claims that, chameleons have negligible effect on the linear growth of structure, and cannot account for the observed cosmic acceleration except as some form of dark energy. This creates a paradox since the cosmological constant is obtained in our chameleon-like setting.

For what reason can one deduce a wrong physical corollary using the correct chameleon theorems? The reason is the misunderstanding on the concept of the energy density and pressure. One confuses the effect of the energy density and pressure with that of their gradient. Lots of literature thinks that it is the long-range interaction that drive the accelerating expansion. However, as pointed by \cite{z100}, there are no forces in a homogeneous universe, because the density and pressure are everywhere the same. To supply a force some gradient is required. Energy density and pressure do not contribute any force helping the expansion along. It is the density and pressure that drive the Universe accelerating expansion. One should not confuse the acceleration of the Universe expansion with an acceleration of a test particle in a scalar field, where the force originates from the spatial gradient and which will be discussed in section \ref{force}. It is worth noting that the fifth force is not the same as the pressure gradient force, albeit the fifth force is also a gradient force. A concrete example of pressure gradient force is buoyant force, while the scalar fifth force is a fundamental force.

The acceleration of the Universe is determined by all the everywhere density and pressure including observable and no-observable part of the Universe, which is has been discussed in section \ref{open}. The observable Universe is defined by a region with radius that light can travel during the lifetime of the Universe. Even the fastest light cannot establish the causal relationship to all the part of the Universe, so does a light scalar field. That using a light scalar field mediate a long-range interaction not only is unreasonable and unrealistic, but also is in disagreement with the current precision tests of gravity that there is no evidence of the long-range fifth force. The requirement that the scalar field should be light is not necessary, because the acceleration of the Universe in essential is related to the density and pressure of matter, the potential energy density and the kinetic energy density of the scalar field. Only the potential energy density of the scalar field drive the Universe accelerating expansion and the others not. There is no evidence that interaction-range play an important role in the expansion. This conclusion can be derived directly from the acceleration equation (\ref{equ9}) of the Universe, in which there is no term related to the Compton wavelength of the scalar field. The Compton wavelength of the scalar field is related to the second derivative of the effective potential with respect to the scalar field, while the acceleration of the Universe expansion is related to the self-interaction potential itself.

However, another long Compton-wavelength is needed in our scheme, which will be discussed immediately in the following.

\subsubsection{\label{dmrcw}The Compton wavelength of dark matter}

The matter-density-dependent cosmological constant in the symmetry-breaking coupling model requires matter to permeate all of the Universe space according to Eqs. (\ref{equ18}) and (\ref{equ21}). Due to the nature of the asymptote shown in Figure \ref{figure3}, a locally concentrative distribution of matter cannot enhance the cosmological constant further, but the presence of voids that are completely empty of matter may lower the cosmological constant. The observed spatial independent cosmological constant implies that the distribution region of the complete voids is smaller if they exist and dark matter should be cold or/and fuzzy in the present time.

Since the rapid motion of relativity particles may destroy the inhomogeneous seed structures generated in inflation, the current dark matter model assumes dark matter gas to be cold \cite{z26,z63,z72,z73}. That is to say, its thermal velocity is negligible with respect to the Hubble flow \cite{z73}. But the mass $m$ of its particles is not determined \cite{z73,z74,z75}, which is ranged widely, for example, from the so-called axion $\sim 10^{-6}-10^{-4}\,\rm{eV}$ \cite{z74} to $\rm{ WIMPs}$ (weakly interacting massive particles) $\sim 10^{2}-10^{3}\,\rm{GeV}$ \cite{z76}. The requirement of matter permeating all space is not incompatible with the CDM model. When dark matter gas is cold, the thermal DeBroglie wavelength $h/{\sqrt{3m k_{\rm{B}}T}}$, with Planck's constant $h$ and Boltzmann's constant $k_{\rm{B}}$, of the particles can get large values due to the small values of the root-mean-square speed in the low temperature $T$, and there will be a large extension of the wavefunctions for the particles. The colder the dark matter gas, the more notable the quantum effect of the gas. Thus, the dark matter wave can permeate everywhere in the Universe.

Fuzzy dark matter is also permissible \cite{z67,z110}. Fuzzy dark matter gas corresponds to ultralight particles, such as, the mass of dark matter particles $\sim 10^{-22}\,\rm{eV}$ \cite{z67,z110}. If the ultralight particles are fermions, the Fermi-energy of the Fermi gas is ultrahigh and then the gas is in a state of complete quantum degeneracy although it may also be a relativistic gas \cite{z25}. Thus, when a structure forms, it will be stable and nearly not be disturbed by another particle collision due to the Pauli principle. If they are bosons, since the characteristic temperature is inversely proportional to the particle mass \cite{z25}, the characteristic temperature is then extremely high. Thus, when a structure forms, it will be a stable Bose-Einstein condensation. The lighter the particles, the more notable the quantum effect of the gas.

Of course, as time goes on, the nearly void distribution region may occur and even exceed the concentrative distribution region of matter, which will decrease the cosmological constant. Consequently, the Universe will decelerate its expansion rate and then shift to a contraction phase. A detailed dark matter model that matches the scalar fifth force model needs a further investigation.

\subsubsection{\label{relacd}The relationship between the cosmological constant and dark matter}

Like the isotropic microwave background possessing the same temperature, there exists light dark matter which distributes the space all most homogeneous and isotropic at least in the Universe scale. As we have demonstrated, the cosmological constant can be obtained by the scalar field via the broken-symmetry couple to matter.  Apparently, like the explanation to the isotropic temperature of the microwave background, the current cosmological constant still needs a inflation era.

The reason is given in the following. Although the late-time acceleration is ascribed to the self-interaction potential of the scalar field in our setting, only light dark matter can localize the value of the self-interaction potential through the symmetry-breaking couple. If most of the space is completely empty and dark matter is cluster distribution in the space, the cosmological constant will be so small since the larger density of matter cannot enhance further the cosmological constant while the empty can reduce it. To acquire a cosmological constant, a homogeneous background of dark matter is needed. The homogeneous distribution of dark matter can establish during the inflationary era because different regions of the Universe is able to interact and move towards thermodynamics equilibrium. That is to say, temperature, pressure and density of mater are all the same values everywhere in the inflationary era.

If there is no light dark matter but only agglomerate ordinary matter, the current cosmological cannot be obtained from our setting. Our model is dependent on both the scalar field and dark matter with light masses. It is light dark matter that helps the Universe generating the homogeneous cosmological constant through the scalar field and the symmetry-breaking couple to matter. The property of light mass of dark matter \cite{z105,z108,z113} guarantees that dark matter fills space everywhere. Even if dark matter itself is not very homogeneous, the cosmological constant is still spatial homogeneous and time-independent, as long as the density of dark matter is large enough, \emph{i.e.}, $\rho  \gg  {\lambda {M_1}^{\rm{4}}{c^3}}/{{\hbar ^3}} \sim  10^{-30}\;\rm{kg/m^3}$.

If one wants a light mass field as a medium to permeate all the space of the Universe, it should be dark matter. One can say that, the scalar field, though its Compton wavelength is short at present cosmological density, makes a cosmological impact mediated through light dark matter. In this sense, the light dark matter seems to play a role to mediate `a long-range interaction' that does not really exist. In our setting both the Compton wavelengths vary indeed with time: The Compton wavelength of the scalar field will become longer and longer with the Universe expanding, while the Compton wavelength of dark matter will increase to some extent limited by the value of the symmetry-breaking coupling function at $\phi =0$.

For other reasons, fuzzy dark matter has become an interesting topic recently \cite{z102,z111}. This means that the requirement of the ultralight particles for dark matter is not a disadvantage of our setting.

\subsection{The zero-point energy problem}\label{zeropoint}

Essentially, Weinberg's no-go theorem results from the problem of the zero-point energy of quantum field theory \cite{z11,z12}. In order to evade Weinberg's no-go theorem, the possibility is to use a new dynamical scalar field \cite{z12,z34,z9}(e.g., quintessence), instead of the assumption of the vacuum energy density to mimic the cosmological constant. Therefore, we have introduced a space-time-dependent scalar field that preserves Lorentz invariance to mimic a material-free dynamical vacuum instead of the traditional vacuum in quantum field theory. When the kinetic energy density of the scalar field is much less its self-interaction potential density, the potential density is defined as a cosmological constant.

Of course, even if the new degree of freedom is introduced, one still faces the problem of the traditional zero-point energy. Since we require that the new degree of freedom should account entirely for the cosmic acceleration, the zero-energy problem must be avoided in our scheme. The reason is that the lowest energy density of the system in question is always divergence whether the new degree of freedom is invoked or not. We may call the scalar field here as a dynamical vacuum to distinguish it from the lowest energy state of quantum field theory. Although the scalar field sits at the minimum of the effective potential in the case of pressureless matter fluid, the definition of the cosmological constant [see Eq. (\ref{equ14})] makes the dynamical vacuum different from the vacuum of quantum field theory. Besides, in next section \ref{contraction} we will discuss the case that the scalar field departs from the minimum. Therefore, the dynamical vacuum is not always corresponding to the lowest energy state of the system. Even so, the zero-energy problem has to be faced since the current cosmological constant is related to the minimum of the effective potential.

The absence of the value of the static vacuum energy density in our scheme indicates that some counteracting mechanism is hidden and used in the model. The counteracting mechanism requires that the cosmological constant is just related to the difference of the energy density values of the system in some parameter space rather than the zero-point-energy-density itself. The parameter can be naturally defined by a so-called self-coupling coefficient which marks the strength of the self-interaction of the scalar field. $\lambda$ in the self-interaction potential shown in Eq. (\ref{equ16a}) can just act as the parameter role, in spite of the fact that its value has been determined in our scheme.

We now discuss how the scheme can avoid the zero-point energy problem. We assume that the scalar field has been sitting at the minimum of effective potential. Without the self-interaction potential, the minimum is completely fixed as a constant. When the self-interaction is considered, the minimum is shifted but the effective mass around the minimum keeps the same value as in the absence of the self-interaction. We can see from Eq. (\ref{equ18-b}) that $m_{{\text{eff}}}$ does not depend on $\lambda$, which is a very important character to guarantee the validity of the following discussion. $\lambda=0$ denotes the absence of the self-interaction of the scalar field. Thus, if we apply second quantization and discuss the lowest energy states of the coupled scalar field in both the cases of $\lambda=0$ and $\lambda \neq 0$, the biggest difference between the two cases is the $\lambda$-dependent lowest states. The $\lambda$-dependence of ${\phi _{\min }}$ has been clearly shown by Eq. (\ref{equ18-a}). Since both the equation of motion for the scalar field and the effective mass around the minimum are just related to the partial derivative of potentials with respect to the scalar field, one naturally deduces that the choice of the zero potential energy should be arbitrary. However, in the traditional quintessence model, it requires that the self-interaction potential is small or zero at the value of $\phi$ where ${V_{,\phi }}\left( \phi  \right)=0$ \cite{z116}. Weinberg argues that theories of quintessence offer no explanation why this should be the case \cite{z116}.

Let us give a brief explanation in our model. Assuming that the first derivative of the self-interaction potential has a $\lambda$-dependent form as follows:
\begin{equation}\label{nogo9}
{V_{,\phi }}\left( \phi  \right) = \lambda {f_{,\phi }}\left( \phi  \right).
\end{equation}
If one wants to discuss the case without the self-interaction potential, one can let $\lambda=0$, which is then corresponding to a completely conformal transformation theory. When the scalar field is added to the physical system, the minimum of the effective potential should be shifted. And then the value of the self-interaction potential corresponding to the minimum appears, which indeed is the definition of the cosmological constant when the scalar field sits at the minimum, \emph{i.e}.,
\begin{equation}\label{nogo10}
\Lambda _{\text{E}}^4 = \int_{{\phi _{\min }}(\lambda  = 0)}^{{\phi _{\min }}(\lambda  \ne 0)} {{V_{,\phi }}\left( {{\phi _{\min }}} \right)d{\phi _{\min }}}.
\end{equation}
Thus, the self-interaction potential has a general form as follows:
\begin{equation}\label{nogo11}
V\left( \phi  \right) = {V_0} + \lambda f\left( \phi  \right),
\end{equation}
where ${V_0}$ is an integration constant. If ${V_0}$ has any observable effect, it must be the other form of energy. Consequently, if one genuinely wants to using the coupled scalar field to describe entirely the cosmological constant without any other form of energy to drive the cosmic acceleration, he must choose $V_{0}=0$. That is, the self-interaction potential should have a form of ${V}\left( \phi  \right) = \lambda {f}\left( \phi  \right)$. Otherwise, the other form of energy will contribute to dark energy. This is a main reason that we have chosen $V\left( \phi  \right) = \lambda {\phi ^4}/4$ without $V_{0}$.

In brief, the strong constraint, that the coupled scalar field should account entirely for the observed cosmic acceleration, result in the following mathematical requirements: the cosmological constant should vanish if there does not exist the scalar field, \emph{i.e.}, $\phi=0$; the cosmological constant should also vanish if there is no self-interaction potential in the presence of the scalar field, \emph{i.e}., $\lambda=0$.

Of course, the scalar field needs not to keep in the lowest energy state according to its equation of motion. Due to the expansion, however, the scalar field can be damped to sit at the lowest energy state. In this situation, the cosmological constant is defined by a part of the energy-density-difference between the two cases of the lowest energy state: one with the self-interaction potential and the other without the self-interaction potential. Since the effective masses are the same in the both cases, the part of zero-point-energy-density cancels exactly each other out in the subtraction. And the left part in the subtraction is a sum of the two expectation values in the lowest energy state of the system with the self-interaction. The two expectation values include both the expectation value of the interaction potential energy density with matter and the expectation value of the self-interaction potential energy density of the scalar field. It is the expectation value of the self-interaction potential density that acts as the cosmological constant due to the scalar field sitting at the minimum stably.

\subsection{Summary}\label{nogosummary}

The scalar field can entirely account for the cosmic acceleration. Our model does not conflict with chameleon no-go theorems, at least mathematically. But our model conflicts with the corollary that a chameleon-like scalar field cannot account for the observed cosmic acceleration. The corollary is a misleading conclusion. The reasons are given in the following. Firstly, the negligible self-acceleration shown by one of the no-go theorems \emph{only} implies that a (narrowly defined) appropriate value of self-interaction potential of the scalar field is needed to drive the Universe accelerating expansion. It does not imply that all the chameleon-like scalar fields cannot entirely explain the cosmological constant. Secondly, the short chameleon Compton wavelength shown by the other no-go theorem also misleadingly implies that chameleon-like fields have a negligible effect on the large-scale of the Universe. This misleading corollary results from the assumption that only a long-range force mediated by the scalar field can impact the cosmic acceleration. However, it is the pressure of the scalar field that drives the cosmic acceleration rather than any form of force either from the pressure gradient or from the field gradient. Thirdly, the overshooting problem in the symmetron model can be avoided in our setting due to the heavy effective mass of the coupled scalar filed.

In the end of this section, it is shown briefly that the zero-point energy density cancels exactly out in our scheme and the expectation value of the self-interaction potential in the lowest energy state acts as the cosmological constant. Thus, Weinberg's no-go theorem is circumvented.

\section{Application of the model to the contraction of the Universe (${H^ - } < 0$) }\label{contraction}

We have obtained the cosmological constant, which is dependent on the adiabatic condition. In the pressureless fluid of matter source, the adiabatic condition always holds. In this section, we will discuss the unavailability of the adiabatic condition of the scalar field when matter source becomes hotter and hotter due to the contraction of the Universe.

In section \ref{expansion}, it has been deduced that the Universe might be closed, \emph{i.e.}, $K = 1$. In the closed space, the Universe must pass through both $\dot \phi =0$ and $H=0$ which are schematically shown in Figure \ref{figure4}. Consequently, if the scalar field that account for the late-time acceleration is used as inflation field, one of the slow-roll conditions $| {\ddot \phi } | \ll 3| {H\dot \phi } |$ is apparently not satisfied. In fact, in the slow-roll approximation in inflation, one often requires that both $\dot \phi $ and the Hubble parameter do not pass through zero \cite{z22}. Therefore, it is the right time to discuss the unavailability of the slow-roll approximation in this section.

\subsection{The mechanism of the quasi-cyclic Universe}\label{conberela}

The end of the expansion that has been discussed in section \ref{expansion} is also the beginning of the contraction. After the expansion stops ($\dot a = 0$, $\ddot a < 0$) in the closed Universe, the contraction process will start and the non-relativistic matter density will increase. The scalar field can adiabatically follow the minimum of the effective potential at the initial stage of the contraction period. In this stage, the value of the self-interaction potential at ${\phi _{\min }}$ still acts as the cosmological constant. With the density of non-relativistic matter increasing, the cosmological constant approaches a fixed value which has been estimated by using $\rho  \to \infty $ to be ${\Lambda _{\rm{E}}} = 2.242\,{\rm{ meV}}$ ($\Lambda = {\rm{1.093}} \times {{\rm{10}}^{{\rm{ - 52}}}}\,{{\rm{m}}^{ - 2}}$). However, $\rho  \to \infty $ for ${w_{i}}=0$ is just a mathematics construct. In the contraction process the parameter $ w_i $ of the equation of matter state must depart from zero due to the collision of matter particles.

\subsubsection{The start of the negative-damping evolution of the scalar field \label{startneda}}

With the further contracting the violent particle collisions occurs and the temperature becomes higher and higher. As a result, the parameter of the equation of state for matter fluid is no longer equal to zero. The better stability of the minimum of the scalar field at the higher matter density gradually becomes weaker and weaker until entering an unstable stage, which can be seen by Eq. (\ref{s8plusre8zong}) in Appendix \ref{appC}.

In this non-adiabatic tracking case, with the density of matter increases further, the kinetic energy density of the scalar field has to increase much more quickly to store the redundant part of the scalar field energy increased during the contraction process. The field can no longer sit stable at the minimum, which can be described by the negative damping oscillation shown in Figures \ref{figure2}(b), \ref{figure4}(b), \ref{figure4}(c) and \ref{figure4}(d). The negative damping, which grows exponentially the magnitude of the oscillation rather than attenuates the magnitude, is one of the most prominent characteristics during the contraction process. It is worth noting that, although the adiabatic condition of $\left| {{{\dot \phi }_{{\text{min}}}}/{\phi _{{\text{min}}}}} \right| \leq \left| {3H/2} \right|$ does no longer hold, the oscillation condition of $\left| {3H/2} \right| \leq {\omega _{\text{c}}}$ still holds when the oscillation is activated by the quick movement of ${\phi _{\min }}$. Thus, the start of the under-negative-damping oscillation stems from the adiabatic instability. With matter fluid approaching to the extremely relativistic case of $w_i \to 1/3$, the effective mass of the scalar field approaches to zero which can be seen by Eq. (\ref{s8plusre5}) in Appendix \ref{appC}. Since both the adiabatic condition and the oscillation condition are no longer satisfied in the relativistic case, the over-negative-damping occurs.

\subsubsection{The energy exchange between the scalar field and matter \label{energyexchange}}

Let's now analyze the energy exchange among the scalar field, matter and the gravitational field more mathematically. In the absence of any couplings to the scalar field, both the density and the temperature of matter would grow due to the contraction of the Universe, and it can be regarded as adiabatic compression for matter system according to thermodynamics. However, the presence of interactions between the scalar field and matter allows for the conversion of the scalar field energy density and matter energy density. By using Eqs. (\ref{equ6})- (\ref{equ8}), the coupled equations are easily obtained as follows (the detail derivation is shown in Appendix \ref{appB}):

\begin{subequations}\label{equ27}
\begin{eqnarray}
{{\dot \rho }_\phi } + 3{H^ - }\left( {{\rho _\phi } + \frac{{{p_\phi }}}{{{c^2}}}} \right) =  - \sum\limits_i {{\rho _i}\frac{{d{A^{1 - 3{w_i}}}\left( \phi  \right)}}{{dt}}},\label{equ27a}\\
{{\dot \rho }_{\rm{m}}} + 3{H^ - }\left( {{\rho _{\rm{m}}} + \frac{{{p_{\rm{m}}}}}{{{c^2}}}} \right) = \sum\limits_i {{\rho _i}\frac{{d{A^{1 - 3{w_i}}}\left( \phi  \right)}}{{dt}}}.\label{equ27b}
\end{eqnarray}
\end{subequations}
In Eq. (\ref{equ27}) above,

\begin{subequations}\label{equ27plus0}
\begin{eqnarray}
{\rho _\phi } = \frac{{V\left( \phi  \right)}}{{{\hbar ^3}{c^5}}} + \frac{{{{\dot \phi }^2}}}{{2\hbar {c^5}}},\label{equ27plus0a} \\
{p_\phi } =  - \frac{{V\left( \phi  \right)}}{{{\hbar ^3}{c^3}}} + \frac{{{{\dot \phi }^2}}}{{2\hbar {c^3}}}, \label{equ27plus0b}
\end{eqnarray}
\end{subequations}
are the scalar field energy density and the pressure of the scalar field, respectively \cite{z9}. And
\begin{subequations}\label{equ27plus}
\begin{eqnarray}
{\rho _{\rm{m}}} = \sum\limits_i {{\rho _i}{A^{1 - 3{w_i}}}\left( \phi  \right)},\label{equ27plusa} \\
{p_{\rm{m}}} = \sum\limits_i {{p_i}{A^{1 - 3{w_i}}}\left( \phi  \right)},\label{equ27plusb}
\end{eqnarray}
\end{subequations}
are the mass density and pressure of matter fluid, respectively. If only pressure-less matter sources ${w_i} = 0$ are considered, matter energy density and pressure can be simplified as ${\rho _{\rm{m}}} = \rho A\left( \phi  \right)$ and ${p_{\rm{m}}} = pA\left( \phi  \right)$, respectively. Superscript`$-$' on the right-side of Hubble parameter $H$ in Eq. (\ref{equ27}) is just used to emphasize that the Hubble parameter describes a contraction process of the Universe. If the adiabatic condition is not satisfied due to the increase of temperature, the scalar field will undergo a negative-damping motion in the contraction process. It is the contraction in the adiabatic instability case that can drive the negative-damping motion of the scalar field. Eq. (\ref{equ27}) is different from th equation (4.88) in literature \cite {z12} because the total energy density is defined by $\rho+\rho_{\phi}$ and is considered being conserved shown in Eqs. (4.82) and (4.83) in the literature. The total energy should be $\rho_{\rm{m}}+\rho_{\phi}$ rather than $\rho+\rho_{\phi}$, which is discussed in Appendix \ref{appB}.

In the contraction process, beside the increase of the energy densities of both matter and the scalar field due to the contracting, there also exists a complicated energy exchange between matter and the scalar field shown as Eq. (\ref{equ27}). In the adiabatic instability case, the energy exchange is more complicated that can be roughly classified into two types: one corresponds to the high-frequency oscillations of the scalar field in the case of under-negative-damping, which can cause the rapid energy exchanges between matter and the scalar field, and accordingly obtain much more energy from matter that heated by the contraction and the oscillating field; the other corresponds to the exponential growth magnitude in the cases of negative-damping oscillations (such as, over-negative-damping evolution, critically-negative-damping oscillations and the exponential growth profile of the curve of under-negative-damping oscillations).

During high-frequency oscillations of the scalar field, the attractive gravitational effect grows rapidly. If we average both the kinetic energy and the self-interaction potential of the scalar field in Eqs. (\ref{equ7}) and (\ref{equ9}) over a period of time large compared with the period of oscillation, it can be easily proven that the high frequency oscillating field corresponds to a nearly pressure-less fluid \cite{z26}. Thus, with the magnitude growing exponentially, a rapidly contracting occurs. Consequently, not only the adiabatic condition is not satisfied but also the oscillation condition is no longer satisfied due to the huge Hubble parameter and the extremely small mass of the scalar field. In this case, the field will experience an over-negative damping motion with huge kinetic energy density. It is well known that, when the kinetic energy density far larger than the potential energy, the parameter of the equation of state for the scalar field approaches to 1 [see Eq. (\ref{s26plusere3})]. The scalar field now will generate attractive gravitational effect more effective than matter, and then more dramatic contraction will occur, which may be called deflation in contrast with inflation.

The scalar field can also be used to drive inflation of the Universe, which will be discussed in the next part \ref{concavevex} of this section.

\subsubsection{The minimum radius of the Universe corresponding to the maximum of the scalar field \label{theminiradius}}

The contraction heats matter and causes energy to transfer from matter to the scalar field. When ${w_i}\left( T \right) \to 1/3$, matter particles become relativistic and matter fluid decouples from the scalar field. After the decoupling completely, the Universe continues contracting. The vigorous scalar field will climb up along its self-interaction potential to the maximum value $V\left( {{\phi _{\max }}} \right)$ with ${\phi _{\max }} > 0$ (One could equivalently consider the case ${\phi _{\max }} < 0$ as shown in Figure \ref{figure4}c and \ref{figure4}d) and then roll down from the maximum. It should be emphasized that the maximum value $V\left( {{\phi _{\max }}} \right)$ exceeds the initial maximum of the kinetic energy density that corresponds to the zero value of the self-interaction potential of the scalar field. The reason is that the scalar field can also acquires additional energy from gravitational field due to the contraction. If the scalar field passes through its zero value, overshooting would occur due to the zero mass of the scalar field. Since the Hubble rate at this time has approached to huge values, this very light mass marks a over-negative-damped evolution of the scalar filed. The higher the Hubble parameter, the stronger the scalar field absorbing energy from the gravitational field.

During the contracting process, there exists a transition from acceleration contraction ($\dot a < 0$, $\ddot a < 0$) to deceleration contraction $\dot a < 0$, $\ddot a > 0$ due to the increase of the potential of the scalar field. And then the Universe gradually approaches to the end of the decelerating contraction. At the end of the decelerating contraction, the Universe will reach its minimum radius corresponding $\dot a = 0$ and $\ddot a > 0$. Substituting $\dot a = 0$ and ${w_i} = 1/3$ into Eq. (\ref{equ4}) and assuming that the scalar field is close to the maximum of its self-interaction potential, \emph{i.e}., $V\left( {{\phi _{\max }}} \right)$ with $\dot \phi=0$, the minimum radius of the Universe can be estimated by:

\begin{equation}
{a_{\min }} = {\left( {\frac{{3{\hbar ^3}{c^7}}}{{8\pi G\left( {V({\phi _{\max }}) + {\hbar ^3}{c^5}{\rho _{\max }}} \right)}}} \right)^{1/2}}.\label{equ28reminise}
\end{equation}
The scalar-field-independent matter density ${\rho _{\max }} = \sum\nolimits_i {{\rho _i}} $ will also reach its highest value due to the minimum volume of the Universe. However, when the Universe departs from its minimum volume, the energy density of matter decreases more rapidly than the potential energy density of the scalar field. Thus, to roughly estimate the minimum radius of the Universe, one may neglect the matter density in Eq. (\ref{equ28reminise}), which gives
\begin{equation}
{a_{\min }} = {\left( {\frac{{3{\hbar ^3}{c^7}}}{{8\pi GV({\phi _{\max }})}}} \right)^{1/2}}.\label{equ28}
\end{equation}
Substituting $\dot \phi=0$ and ${w_i} = 1/3$ into Eq. (\ref{equ9}), the expansion acceleration of the Universe at its minimum radius is given by:
\begin{equation}\label{equ28replus}
\frac{{{{\ddot a}_{\min }}}}{{{a_{\min }}}} = \frac{{8\pi G}}{{3{\hbar ^3}{c^5}}}\left( {V({\phi _{\max }}) - {\hbar ^3}{c^5}{\rho _{\max }}} \right).
\end{equation}

According to the astronomical observation and the Big Bang Nucleosynthesis calculation, we can deduce the decoupling temperature to be $T\ge{\rm{0.1 \, MeV}}$ \cite{z16}. Therefore, the scalar field will approach to its maximum of $|\phi _{\max }| \ge {\rm{0.1 \,MeV}}$ and correspondingly climb up to the maximum of its self-interaction potential density of $V\left( {{\phi _{\max }}} \right) \ge ({\rm{0.1 \,MeV}})^4/4!$ because of $V(\phi)={\phi}^{4}/4!$ shown in Eq. (\ref{equ16a}). The upper bound value of the scalar field is plausibly assumed as the reduced Planck energy ${M_{{\rm{pl}}}}{c^2} = 2.4 \times {10^{18}}{\rm{\, GeV}}$ with ${M_{{\rm{Pl}}}}  \equiv {(\hbar c/8\pi G)^{1/2}}$ and the corresponding  self-interaction potential density is $V\left( {{\phi _{\max }}} \right) = (2.4 \times {10^{18}}{\rm{\, GeV}})^4/4!$. Correspondingly, the minimum radius of the Universe is estimated from Eq. (\ref{equ28}) to be
\begin{equation}
{10^{ - 33}}\,{\rm{ m}} \le {a_{\min }} \le {10^{12}}\,{\rm{ m}}. \label{equ28pluse2}
\end{equation}
When the scalar field achieves its maximum value, the equation of state ${w_\phi } \equiv p_\phi  /\left({\rho _\phi }{c^2}\right)$ for the scalar field equals to $-1$ due to ${\dot \phi ^2} = 0$.

\subsubsection{The maximum of the cosmological constant for relativistic matter fluid\label{maxicc}}

Comparing Eq. (\ref{equ9}) with the acceleration of the Universe in the $\Lambda$CDM model, the maximum of the self-interaction potential acts as the maximum of the cosmological constant due to the zero kinetic energy of the scalar field, \emph{i.e.},
\begin{equation}
\Lambda _{{\rm{Emax}}}^4 \equiv V({\phi _{\max }}) = \frac{1}{{{\rm{4!}}}}\phi _{\max }^4.\label{equ28pluse3}
\end{equation}
And $\rho_{\rm {m}}=\rho$ is also obtained due to $w_i =1/3$, which indicates that matter particle masses are no longer affected by the scalar field. This conclusion is natural because $w_i =1/3$ describes the decoupling case about the scalar field and matter. It is interesting to get one little sentence out here for the coupling case of $w_i = 0 $. According to Eq. (\ref{equ18-a}), when matter density is so large that $|\phi_{\rm min}|$ achieves to $M_{2}c^2$, $A(\phi_{\rm {min}})$ is then approaches to 1. As a result, $\rho_{\rm {m}}=\rho$ also almost completely holds in the coupling case, which implies that matter particle masses are hardly affected by the scalar field. The larger the matter density, the less the influence of the scalar field to the particle mass. This does not mean that the interaction does not play an important role in this dense density case. It localizes the scalar field around $|\phi_{\rm min}|$. The confinement strength can be described by the curvature of the effective potential shown in Eq. (\ref{equ18-b}). Therefore, the confinement strength depends on the density of matter. The larger the matter density, the stronger the confinement. The curvature of the effective potential around the minimum is not related to the self-interaction potential as mentioned in section \ref{masseff}. However, in the case of $w_i =1/3$ here, when the scalar field climbs up to the highest value along its self-interaction potential, the unstable point also corresponds to a curvature which is completely dependent on the self-interaction potential. One cannot confuse the two types of the masse of the scalar field.

Only in the case that the kinetic energy is considerably smaller than the potential energy of the scalar field can we introduce a cosmological constant. In the decoupling case of $w_i=1/3$, when the scalar field climbs up on its highest potential value, it must return into a rolling down phase along the self-interaction potential curve. Then the maximum cosmological constant is not a stable value (see Figure \ref{figure4}). In the coupling case of $w_i = 0 $, however, the scalar field is trapped into the minimum of the effective potential. Since the cosmological constant is acted by the value of the self-interaction potential at the minimum, which is shown by Eq. (\ref{equ15}), it becomes stable and appears as a constant during the Universe evolution.

Unlike the stable cosmological constant for $w_i = 0 $, the maximum cosmological constant for $w_i=1/3$ is not able to determine completely due to the lack of the decoupling parameter.

\subsubsection{The quasi-cyclic Universe}\label{qusicyc}

When the constraints of the cosmological constant are satisfied, the symmetry-breaking coupling function between matter and the scalar field has been determined in obtaining the fifth force. For the sake of completeness, the scheme has also been extended to the cosmic evolution, which leads to a quasi-cyclic universe. There are many alternative scenarios to the standard inflationary paradigm, such as bouncing cosmologies \cite{z17} and cyclic universes \cite{z18,z19}. However, our quasi-cyclic model corresponding to a regular scalar field differs from not only the bouncing cosmologies but also the cyclic universes. To realize a contracting phase, many bouncing and cyclic models violate the Null Energy condition ${\rho _\phi }{c^2} + {p_\phi } > 0$ \cite{z17,z18,z19}. For example, a negative potential energy is introduced in \cite{z18} while a ghost field with negative kinetic energy is introduced in \cite{z19}.

\subsection{The biggest challenge to the scheme }\label{concavevex}

Until now, there are no real challenges specific to the the scheme. However, when the scalar field is used to drive inflation of the Universe, the biggest challenge is encountered. On the one hand, the quartic form of the self-interaction potential of the scalar field in our scheme is a convex function,\emph{ i.e.}, ${{V_{,\phi \phi }}\left( \phi  \right)}>0$ shown in Eq. (\ref{equ16a}). On the other hand, according to \cite{z1i}, in the framework of standard single-field inflationary models with Einstein gravity, the most probable candidate shape of the self-interaction might be a concave potential, \emph{i.e.}, ${{V_{,\phi \phi }}\left( \phi  \right)}<0$. In this sense, our scheme seems impossible to be used to drive inflation of the Universe. The observed data are strongly dependent on the assumption of the slow-roll parameters. However, the slow-roll parameters are not always valid in our scheme. We analyse the possibility that the `observed concave behavior' can be explained by the self-interaction potential with a convex property.

\subsubsection{The solutions to equations of motion near the neighborhood of the minimum radius  }\label{concavevex1}

\begin{figure}
\centering
\includegraphics[width=250pt]{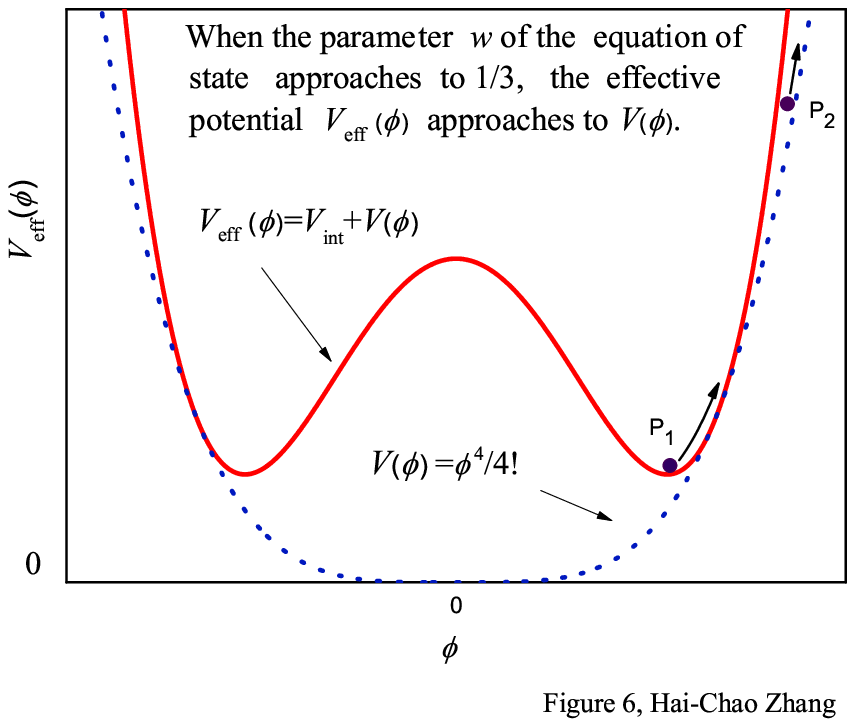}
\caption{The sketch of the dynamical vacuum energy that appears in the climbing-up process of the scalar field along the self-interaction potential. With the temperature increasing during the contraction process of the Universe, the scalar field becomes vigorous and ultimately decouples from matter. In the coupling case, the scalar field can oscillate around one of the minima ${{\rm{P}}_1}$ of the effective potential. With the Universe contracting further, the vigorous scalar field decouples from matter and climbs up along its self-interaction potential (see also ${{\rm{P}}_2}$), which results in the repulsive gravitational effect.}\label{figure6}
\end{figure}

It is worth noting that, the method that using the scalar field as a time variable \cite{z22,z23,z26} will not work here due to its assumption of $\dot \phi $ not passing through zero. In our models, the repulsive gravitational effect even starts at the beginning of the decelerating contraction process of the Universe. When the potential energy of the scalar field gradually grows and then exceeds the sum of matter energy and the kinetic energy of the field in the climbing-up process along the self-interaction potential curve, the Universe shifts to decelerating contraction from accelerating contraction. Unfortunately, there has been no investigation on this climbing-up model at the present time, to our knowledge. With the temperature increasing, from under-negative-damping oscillation around one of the minima of the effective potential, the scalar field shifts to a climbing-up phase. After reaching at the highest value along the self-interaction potential, the scale field will roll down. Near the minimum radius of the Universe, there does not exist a stable minimum of the effective potential. Thus, the stable `vacuum expectation value' of the scalar field in the effective potential minimum is then meaningless. `Dynamical vacuum energy' has been used in the previous section and can be used to characterize the time-variable value of the self-interaction potential of the scalar field which account for both the late-time acceleration and inflation. Figure \ref{figure6} shows the sketch of the dynamical vacuum energy appearing in the climbing-up and rolling-down model of the scalar field.

Near the neighborhood of the minimum radius of the Universe, since  $V\left( {{\phi _{\max }}} \right) \gg {\hbar ^2}{\dot \phi ^2}/2{\rm{ + }}\sum\limits_i {{\rho _i}{\hbar ^3}{c^5}} $, Eqs. (\ref{equ7}) and (\ref{equ8}) become as follows:
\begin{subequations}
\begin{eqnarray}
{\frac{{\dot a}}{a}  \equiv {H^ \pm }} = { \pm {\left( {\frac{{{c^2}}}{{a_{\min }^2}} - \frac{{{c^2}}}{{{a^2}}}} \right)^{1/2}}},\label{equ28pluse5}\\
{{\hbar ^2}\ddot \phi  + \frac{1}{{3!}}\phi _{\max }^3}  = { 0 }.\label{equ28pluse6}
\end{eqnarray}
\end{subequations}
Therefore, the evolutions of the radius and the scalar field with the cosmic time are respectively obtained as
\begin{subequations}\label{equation28p}
\begin{eqnarray}
{a(t)} & =& { {a_{\min }}\cosh \left[ {\frac{{c\left( {t - {t_{\rm{c}}}} \right)}}{{{a_{\min }}}}} \right]},\label{equ28pluse7}\\
{\phi (t)} & = & { {\phi _{\max }} - \frac{1}{{12}}\phi _{\max }^3{\left( {\frac{{t - {t_{\rm{c}}}}}{\hbar }} \right)^2}},\label{equ28pluse8}
\end{eqnarray}
\end{subequations}
where $t_{\rm{c}}$ stands for the shift time from contracting to expanding and then corresponding to the minimum radius of the Universe. It is the self-interaction potential of the scalar field that drives the Universe evolution rather than the Hubble parameter since the Hubble parameter equals to zero at the minimum radius of the Universe.

Apparently, one of the slow-roll conditions $| {\ddot \phi } | \ll 3| {H\dot \phi } |$ in literature, such as \cite{z55, z26}, is not satisfied and is not necessary due to the zero Hubble rate. Because the next rapid expansion always follows the last rapid contraction of the Universe, one needs not to worry about the amount of inflation being sufficient or not. The slow-roll condition, however, is satisfied at the time when the kinetic energy becomes to the nearly same value as the potential energy of the scalar field. Thus, whether the slow-roll condition is valid or not is determined by what time is in the expansion process.

Near the shift time $t_{\rm{c}}$, both the scale factor and the scalar field have time reversal symmetry. The characteristic time of the scalar field near its maximum is $\tau_{\phi} \sim \hbar/|\phi _{\max }|$, while the characteristic time of the cosmic scale factor near its minimum is $\tau_{a} \sim a_{\min }/c $. Thus, $\tau_{a}/\tau_{\phi} \sim {M_{{\rm{pl}}}}{c^2}/|{\phi _{\max }}|$. If $|{\phi _{\max }}|={M_{{\rm{pl}}}}{c^2}$, both the characteristic times are the same as the Planck time. If $|\phi _{\max }|$ equals to the calculation temperature of the Big Bang Nucleosynthesis $ {\rm{0.1 \,MeV}}$ \cite{z16}, the two characteristic times become larger. But, the characteristic time of the cosmic scale factor near its minimum value is considerably larger than that of the scalar field. On the face of it, this seems to mean that the Universe will remain longer time in the phase of the minimum radius compared to the staying time in the maximum of the scalar field. However, noticing that the Universe radius near the minimum increases or decreases in the form of exponential growth, one may conclude that a slow evolution of the scalar field corresponds to a fast evolution of the Universe radius in this case. Therefore, the total time period near the minimum radius of the Universe is not very different for the both physical quantities. In the language of inflation, it is the slow evolution of the scalar field that drives the fast inflation of the Universe. Unfortunately, the property of the slow evolution of the scalar field is always extrapolated to such an extent that $|{\ddot \phi }|$ is required to be neglected, \emph{i.e}., $|{\ddot \phi }| \ll 3| {H\dot \phi }|$, which is apparently not the same as the case near the maximum value of the scalar field, \emph{i.e.,} $|{\ddot \phi }| \gg 3| {H\dot \phi }|=0$. Thus, one should be carefully in applying the traditional slow-roll conditions to the close space.

The solutions described by Eqs. (\ref{equ28pluse7}) and (\ref{equ28pluse8}) can be rewritten as useful alternative forms by using the scalar field as an independent variable. Since the traditional slow-roll conditions are not always valid due to the zero Hubble rate, the slow-roll parameters is not appropriate for describing the inflation for the close Universe. Notice that the method shown in \cite{z22} cannot be used because it assumes that $\dot \phi$ not change sign during inflation. We really face the case that both $\dot \phi$ and $H$ pass through zero. Substituting Eq. (\ref{equ28pluse8}) into Eq. (\ref{equ28pluse7}) to eliminate the time $t$, the $\phi $-dependent $a(\phi)$ near the minimum radius of the Universe can be obtained as follows:
\begin{equation}
a\left( \phi  \right) = {a_{\min }}\cosh \left[ {\frac{{\hbar c}}{{{a_{\min }}}}{{\left( {\frac{{12\left( {{\phi _{\max }} - \phi } \right)}}{{\phi _{\max }^3}}} \right)}^{1/2}}} \right],\; \rm{for}\, \left| \phi  \right| \le \left| {{\phi _{\max }}} \right|.\label{equ28pluse9}
\end{equation}
Substituting Eq. (\ref{equ28pluse9}) into Eq. (\ref{equ28pluse5}) and squaring it, one has
\begin{equation}
H^2(\phi)=\frac{{{c^2}}}{{a_{\min }^2}}\tanh^2 \left[ {\frac{{\hbar c}}{{{a_{\min }}}}{{\left( {\frac{{12\left( {{\phi _{\max }} - \phi } \right)}}{{\phi _{\max }^3}}} \right)}^{1/2}}} \right],\; \rm{for}\, \left| \phi  \right| \le \left| {{\phi _{\max }}} \right|.\label{equ28pluse10}
\end{equation}
Apparently, ${H^2}\left( \phi  \right)$ is a concave function of  $\phi$. It worth noting that, these solutions are valid if $V\left( {{\phi _{\max }}} \right) \gg {\hbar ^2}{\dot \phi ^2}/2{\rm{ + }}\sum\limits_i {{\rho _i}{\hbar ^3}{c^5}} \approx  {\hbar ^2}{\dot \phi ^2}/2$. Since the value of the self-interaction potential corresponding to the minimum radius of the Universe is extremely large, they are indeed good approximation solutions around the minimum.

\subsubsection{The pseudo-potential density}\label{psedo}

To investigate inflation by our scheme, we neglect the density of matter and introduce a pseudo-potential in Friedmann equation. Eq. (\ref{equ4}) is then rewritten as follows:
\begin{equation}\label{equ28pluse3revied1}
{H^2} = \frac{{8\pi G}}{{3{\hbar ^3}{c^5}}}\left[ {\left( {{V_{{\text{pse}}}}\left( \phi  \right) + \frac{{{\hbar ^2}}}{2}{{\dot \phi }^2}} \right)} \right],
\end{equation}
where the pseudo-potential
\begin{equation}\label{equ28vpseu}
{V_{{\text{pse}}}}\left( \phi  \right) \equiv V\left( \phi  \right) - \frac{{3{\hbar ^3}{c^7}}}{{8\pi G}}\frac{1}{{{a^2}\left( \phi  \right)}}
\end{equation}
with $K = 1$ having been used. Therefore the pseudo-potential acts the role of the real potential in the literature's \cite{z22} equation (1) in inflation of the Universe. In the traditional treatment for inflation, however, it is always assume that the space is flat  \cite{z22}.

\subsubsection{A short review to slow-roll parameters}\label{reviewtosrp}

First Hubble slow-roll parameter and ($n+1$)st Hubble slow-roll parameter are given respectively as follows \cite{z22,z23,z26}: ${\epsilon}_{1}=- \dot{H}/H^{2}$ and ${\epsilon _{n + 1}} ={{\dot \epsilon }_n}/{H{\epsilon _n}} $ with $n\geq 1$. Apparently, the Hubble slow-roll parameters are just valid when the radius of the Universe large enough to render the curvature term small but the term cannot be neglected completely [see Eq. (\ref{s16}) in Appendix \ref{appA}]. Noticing Eq. (\ref{s16}) and neglecting the effect of matter in the equation, one can easily see that only the positive curvature of the Universe can give a positive value of $\dot {H}$ as long as ${\hbar ^2}{{\dot \phi }^2}/2 < {\hbar ^3}{c^7}/\left[ {8\pi G{a^2}\left( \phi  \right)} \right]$. However, this does not mean that inflation can only occur in the closed space since inflation is described by $\ddot {a}$ rather than by $\dot {H}$, although the closed space is interested in this section only. If enough inflation has occurred so that $3{\hbar ^3}{c^7}/\left[ {4\pi G{a^2}\left( \phi  \right)} \right] - 2{\hbar ^2}{{\dot \phi }^2} \ll V\left( \phi  \right)$ (in fact, this means that the minimum radius is large enough so as to hold the inequality in small kinetic energies), the Hubble slow-roll parameters are good approximations. However, the curvature term must not vanish. If the curvature term completely vanishes, there is no difference between the pseudo-potential and the self-interaction potential.

When the pseudo-potential is introduced, the traditional slow-roll parameters for the real potential of the scalar field should be replaced by the pseudo-potential. Thus, first pseudo-potential slow-roll parameter and second pseudo-potential slow-roll parameter are given respectively as follows:
\begin{subequations}\label{28pluse3re2}
\begin{eqnarray}
{\epsilon _V} &=& {  \frac{{M_{{\text{Pl}}}^2{c^4}}}{2}{\left( {\frac{{{V_{{\text{pse,}}\phi }}}}{{{V_{{\text{pse}}}}}}} \right)^2}},\label{28pluse3re2a} \\
{\eta _V} &=& { \frac{{M_{{\text{Pl}}}^2{c^4}{V_{{\text{pse,}}\phi \phi }}}}{{{V_{{\text{pse}}}}}}} ,\label{28pluse3re2b}
\end{eqnarray}
\end{subequations}

\subsubsection{The pseudo-potential having a concave property }\label{pseuconcave}

We now demonstrate how the convex potential we adopted can behavior a concave property. Substituting Eq. (\ref{equ28pluse10}) into Eq. (\ref{equ28pluse3revied1}) and neglecting the kinetic energy density of the scalar field, one has
\begin{equation}
{V_{{\text{pse}}}}\left( \phi  \right) \simeq \frac{{3{\hbar ^3}{c^5}}}{{8\pi G}}{H^2}\left( \phi  \right).\label{equ28pluse11}
\end{equation}
Thus, ${V_{{\text{pse}}}}\left( \phi  \right) $ is really a concave function of  $\phi$ due to the concave property of ${H^2}\left( \phi  \right)$, although $V(\phi)$ is a convex function. A further detailed investigation that using the convex potential to analyze the concave behavior is needed.

In fact, the conclusion can be directly obtained from the definition of the pseudo-potential marked by Eq. (\ref{equ28vpseu}). The minimum value of the pseudo-potential corresponds to the minimum radius of the Universe, which can be demonstrated by substituting Eq. (\ref{equ28pluse9}) into Eq. (\ref{equ28vpseu}). In this sense, the observed concave property implies that the Universe is a closed space. Planck evidence for a closed Universe has also shown very recently in \cite{z114} from the presence of an enhanced lensing amplitude in cosmic microwave background power spectra.

\begin{figure}
\centering
\includegraphics[width=250pt]{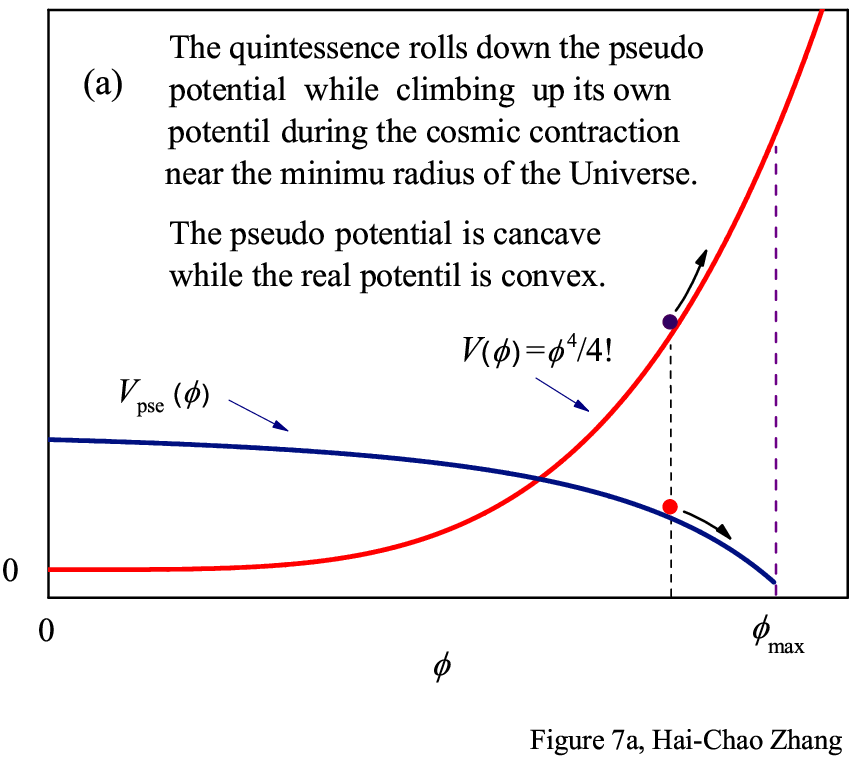}
\includegraphics[width=250pt]{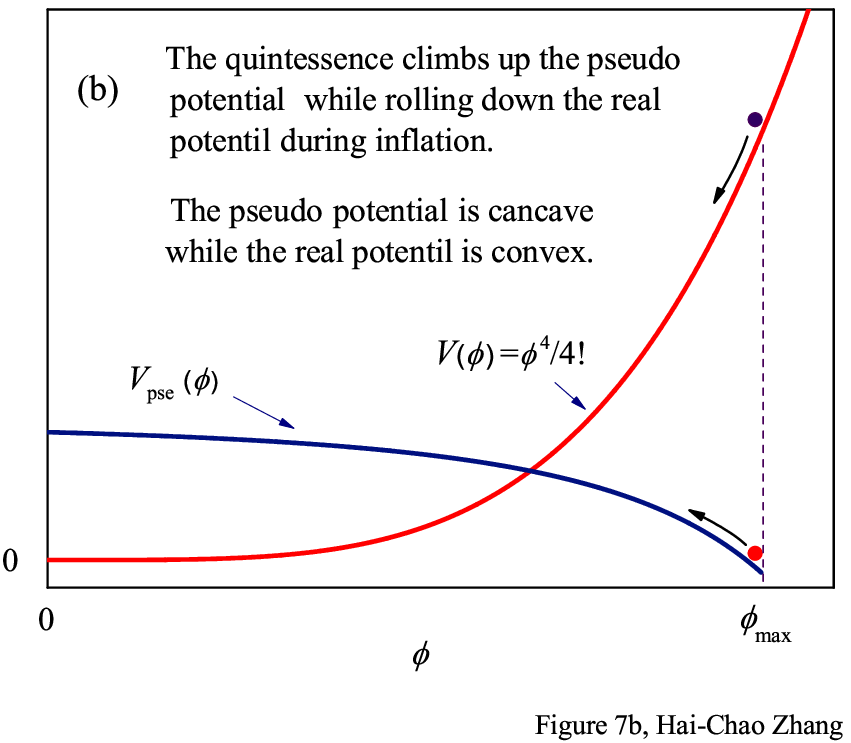}
\caption{The self-interaction potential and the corresponding pseudo-potential near the maximum value of the scalar field. (a) For the case of the contraction of the Universe near the minimum radius of the Universe; (b) for the case of the expansion near the minimum radius of the Universe, \emph{i.e.}, inflation.}\label{figure7}
\end{figure}

Fig. \ref{figure7} shows the rolling down and climbing up of the scalar field along the pseudo-potential and the self-interaction potential, respectively, near the maximum value of the scalar field.

\subsection{Summary}\label{suforcont}

Using the quartic self-interaction potential and the symmetry-breaking interaction potential, the quasi-cyclic Universe model is obtained. We have introduced the pseudo-potential to describe the evolution of the Universe near its minimum radius. The pseudo-potential is the sum of the self-interaction potential density and the energy density scale of the positive curvature of the Universe. The observed potential concave property is explained by the pseudo-potential rather than the self-interaction potential only. Thus, the prediction of the closed space is favored by the observation data.

\section{The scalar fifth force }\label{force}

We have determined the parameters of the symmetry-breaking interaction model under the constrains of the cosmological constant. Since the cosmological constant is based on the homogeneous and isotropic assumption, the gradient terms in the scalar equation of motion is ignored. If ordinary agglomerate matter couples to the scalar field in the same way as dark matter doing, the gradient force cannot be ignored and might be tested in laboratory, provided that experiments are designed properly. In this section, we will discuss the characters of the fifth force, such as, its strength, interaction range, thin shell and saturation effects.

In order to obtain the modified geodesic equation for non-relativistic matter particles in the case of the inhomogeneity and anisotropy, instead of the line element of Eq. (\ref{equ3}), we use a linearly perturbed form Eq. (\ref{equ2minus}) with two potentials $\Phi $ and $\Psi $ in the weak-gravitational-field limit. Then one gets the scalar field fifth force shown in Eq. (\ref{equ2}). Since the fifth force occurs in the extremely small scale in comparing with the scale factor of the Universe, the time-variable scale factor is not necessary to be introduced in Eq. (\ref{equ2minus}). In the equation of motion for the scalar field, the generally covariant d'Alembertian operator ${D^\mu }{D_\mu }$, where ${D_\mu }$ is the covariant derivative with respect to the Einstein metric, can be approximated to the common form as ${D^\mu }{D_\mu } = {\nabla ^2} - {c^{ - 2}}\partial _t^2$ in the weak gravitational field \cite{z20}. Thus, to compute the scalar field in the case of weak gravitational field, it is sufficient to use the Minkowskian line element as $d{s^2} =  - {c^2}d{t^2} + d{\vec {x}}^2$. For a static, space-variable density of non-relativistic matter source, the scalar equation of motion that results from the action (\ref{equ1}) is given as follows:

\begin{equation}
{\hbar ^2}{c^2}{\nabla ^2}\phi  = {V_{{\rm{eff,}}\phi }}(\phi ).\label{equ29}
\end{equation}

\subsection{The Klein-Gordon equation for a massive scalar field }\label{hydrogenic}

In this subsection, we will solve one of the two biggest challenges that the upper bound of the linear coupling coefficient is about ${10}^{14}$ \cite{z15}. The other challenge of the observed potential concave feature has been solved in section \ref{concavevex}.

In order to distinguish the roles of a homogeneous ambient density and a space-variable source density that generates the spatial gradient of the scalar field, we may imagine a scalar field profile induced by the source embedded in the medium of background density ${\rho _{\rm{b}}}$. Since the scalar field-independent matter mass density is frequently used in this section, it will be often called as matter density or density afterwards for conciseness. In the homogeneous background, ${\nabla ^2}\phi {\rm{  = 0}}$ and the equilibrium value of the scalar field is ${\phi _{\rm{b}}}{\rm{ = }}{\phi _{\min }}\left( {{\rho _{\rm{b}}}} \right)$ corresponding to a minimum of the effective potential. Assuming the density of the source $\rho  = {\rho _{\rm{b}}} + \delta \rho $, but not assuming $ \delta \rho \ll {\rho _{\rm{b}}}$, we can still expand the field around the background value $\phi  = {\phi _{\rm{b}}} + \delta \phi $. The reason is that matter density in laboratory experiments is always larger than the current density of the Universe. Correspondingly, the scalar field is almost fixed and then $\delta \phi \ll {\phi _{\rm{b}}}$ [see also Figure \ref{figure3}, Eqs. (\ref{equ18-a}) and (\ref{equa34})]. An equation of motion for a massive scalar field from Eq. (\ref{equ29}) is then obtained as follows:
\begin{equation}
\left( {{\nabla ^2} - \frac{{m_{{\rm{eff}}}^{\rm{2}}{c^2}}}{{{\hbar ^2}}}} \right)\delta \phi  = {A_{,\phi }}\left( {{\phi _{\rm{b}}}} \right)\hbar {c^{\rm{3}}}\delta \rho ,\label{30}
\end{equation}
where the effective mass is:
\begin{equation}
m_{{\rm{eff}}}^{\rm{2}} \equiv m_{{\rm{eff}}}^{\rm{2}}\left( {{\phi _{\rm{b}}}} \right) + \frac{{{A_{,\phi \phi }}({\phi _{\rm{b}}})\delta \rho {\hbar ^3}{c^5}}}{{{c^4}}}\label{31}
\end{equation}
with $m_{{\rm{eff}}}^{\rm{2}}\left( {{\phi _{\rm{b}}}} \right) = {V_{{\rm{eff}}}}_{,\phi \phi }\left( {{\phi _{\rm{b}}}} \right)/{c^4}$. If $\delta \rho \ll {\rho _{\rm{b}}}$, then $m_{{\rm{eff}}}^{\rm{2}} = m_{{\rm{eff}}}^{\rm{2}}\left( {{\phi _{\rm{b}}}} \right)$. However, in general the density of source is always larger than that of background, \emph{i.e,} $\delta \rho  \gg {\rho _{\rm{b}}}$. Consequently, $m_{{\rm{eff}}}^{\rm{2}} \gg m_{{\rm{eff}}}^{\rm{2}}\left( {{\phi _{\rm{b}}}} \right)$ in source region. This means that the equation of motion is different from the usual Klein-Gordon equation in which the value of the mass is the same everywhere. If the masse in the equation of motion for the scalar field varies in space, the linear superposition principle is not valid.

\subsubsection{Matter-density-dependent interaction range }\label{mddir}

From the Klein-Gordon equation (\ref{30}), the interaction range is naturally defined by ${\mathchar'26\mkern-10mu\lambda _{\rm{c}}} \equiv \hbar /\left( {{m_{{\rm{eff}}}}c} \right)$. The density-dependent interaction-rang has been estimated by Eq. (\ref{equ19}), which can be rewritten with the background density ${\rho _{\rm{b}}}$ as an approximate expression in the following:
\begin{equation}
{\mathchar'26\mkern-10mu\lambda _{\rm{cb}}}\left[ {\rm{m}} \right]  \approx \frac{{1.648 \times {{10}^{ - 19}}}}{{{{\left( {\rho_{\rm{b}} \left[ {{\rm{kg/}}{{\rm{m}}^{\rm{3}}}} \right]} \right)}^{1/2}}}}.\label{equ19and33}
\end{equation}
Apparently, in the common density, the interaction range is so short that the fifth force is suppressed from local tests of gravity \cite{z21,z46,z47}.

We have known well that the four fundamental interactions: the gravitational, electromagnetic, strong and weak interactions. The gravitational and electromagnetic ones produce long-range forces. The strong and weak ones produce forces at subatomic distances and govern nuclear interactions. The fifth force discussed here has a density-dependent interaction range. Even in the extremely vacuum space, such as, in the case of the current matter density of the Universe, the interaction range is estimated to be about $\sim 5 \,\mu \rm{m}$. It is worth noting that the short-range force is not responsible for the Universe acceleration. The force results from the space inhomogeneity of matter distribution. It is the pressure of the scalar field that drives the Universe acceleration.

\subsubsection{Matter-density-dependent coupling coefficient $\beta$  }\label{lpcbeita}

Substituting $\phi  = {\phi _{\rm{b}}} + \delta \phi $ into Eq. (\ref{equ2}), the acceleration on a test particle due to the scalar field becomes
\begin{equation}
\vec {a} =  - {c^2}\frac{{{A_{,\phi }}\left( \phi  \right)}}{{A\left( \phi  \right)}}\nabla \delta \phi  \approx  - {c^2}\frac{{{A_{,\phi }}\left( {{\phi _{\rm{b}}}} \right)}}{{A\left( {{\phi _{\rm{b}}}} \right)}}\nabla \delta \phi .\label{32}
\end{equation}
The coefficient ${A_{,\phi }}\left( \phi  \right){\rm{/}}A\left( \phi  \right)$ defines a coupling coefficient $\beta(\phi)$ to characterize the strength of the fifth force \cite{z2,z7,z15,z57}, \emph{i.e.},
\begin{equation}
\beta \left( \phi  \right) \equiv {M_{{\rm{Pl}}}}{c^2}\frac{{{A_{,\phi }}\left( \phi  \right)}}{{A\left( \phi  \right)}} \approx {M_{{\rm{pl}}}}{c^2}\frac{{{A_{,\phi }}\left( {{\phi _{\rm{b}}}} \right)}}{{A\left( {{\phi _{\rm{b}}}} \right)}}.\label{32plus0}
\end{equation}
Since $\beta \left( \phi  \right)$ corresponds to the slope of the coupling function $A(\phi)$, it describes only the linear part of the nonlinear coupling function \cite{z15}. Due to the nonlinear property of $A(\phi)$, the coupling coefficient $\beta (\phi)$ is not a constant and varies with $\phi$. For the symmetry-breaking coupling function of Eq. (\ref{equ16b}), the coupling coefficient is derived in Appendix \ref{appC} as Eq. (\ref{s9}). For density far larger than the current matter density of the Universe, the density-dependent coupling coefficient is estimated from Eqs. (\ref{s9}), (\ref{equ16b}) and (\ref{equ18-a}) to be as:
\begin{equation}
|\beta | \approx \frac{{1.15 \times {{10}^4}}}{{{\rho _{\rm{b}}}\left[ {{\rm{kg/}}{{\rm{m}}^3}} \right]}}.\label{33}
\end{equation}

For density smaller the current matter density of the Universe, Eq. (\ref{33}) above is not valid. A complete expression of $\beta$ is shown by Eq. (\ref{s9}). Besides, an mass scale $M_{\rm{m}}$ is often used to describe the coupling strength \cite{z15}, which is defined by
\begin{equation}
M_{\rm{m}}  \equiv \frac{{{M_{{\rm{pl}}}}}}{{{|\beta|}}}.\label{32plus1}
\end{equation}
${M_{\rm{m}}}$ is also not a constant due to the nonlinear property of $A(\phi)$.

\begin{figure}
\centering
\includegraphics[width=250pt]{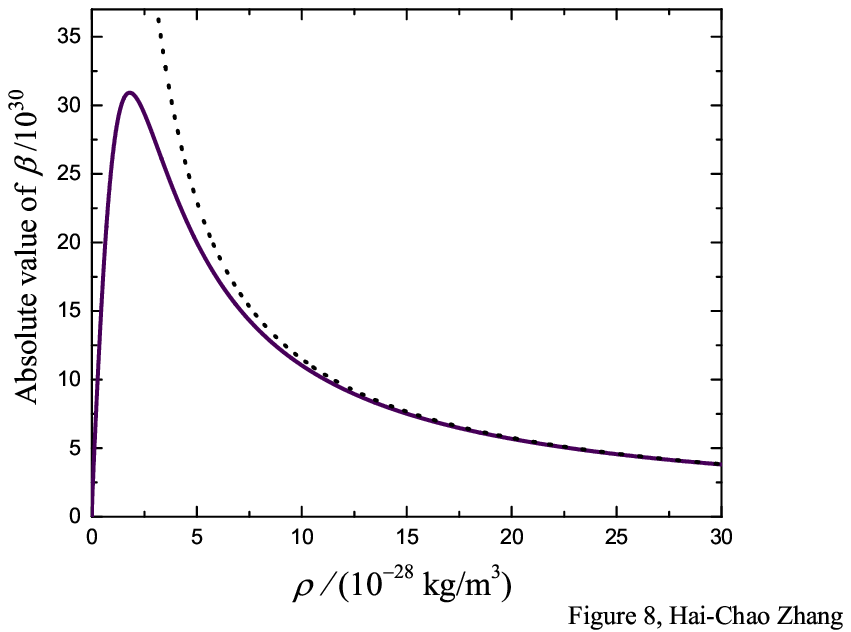}
\caption{The coupling $|\beta|$ versus the ambient density of matter. The solid curve corresponds to a complete and a little complicated expression of Eq. (\ref{s9}), while the dotted curve corresponds to an approximation of Eq. (\ref{33}) to estimate the order of magnitude of the coupling. When the density of matter approaches to zero, the approximate expression is no longer valid.}\label{figure8}
\end{figure}

Obviously, the absolute value of $\beta $ can be far bigger than $O\left( 1 \right)$ in a wide density region. Strong coupling to matter does not means that it must dissatisfy experimental constraints, which has been discussed in \cite{z43}. For lower density in the local environment, the magnitude of the fifth force is much larger than gravity. Figure \ref{figure8} shows the coupling $\beta$ versus the ambient density of matter.

The huge values of $\beta $ maybe mislead one into an illusion that the interaction should be easily experienced and then contradicts the absence of observable interaction. The coupling $\beta $ has a shortcoming that just describes the linear property of the nonlinear coupling function. The shortcoming can be partly remedied by the parameter of the Compton wavelength of the scalar field. The Compton wavelength estimated by Eq. (\ref{equ19and33}) is, in general, very short, which leads to the absence of observable interaction.

\subsubsection{The constraint to $\beta$ from the precision measurements }\label{upperbound}

The `upper' bound of $|\beta|$ is claimed having been derived in \cite{z15}. In fact, \cite{z15} reports the equivalent `lower' bound of ${M_{\rm{m}}} > {10^4}\,{\rm{ GeV}}/{c^2}$ from the precision measurements of hydrogenic energy levels. It is a strong constraint. But it should be emphasized that, since the coupling coefficient is dependent on ambient density, the statement of `lower' is not appropriate, especially in the case of low ambient density.

We now explain the constraint by our scheme. For hydrogen atom, the mass density of the electron cloud around the atomic nucleus can be estimated to be as $\rho _{\rm{b}} \sim {m_{\rm{e}}}/(4\pi r_{\rm{0}}^3/3) \approx 1.468\,{\rm{kg/}}{{\rm{m}}^3}$, where ${r_0} = 5.29 \times {10^{ - 11}}\,{\rm{ m}}$ and ${m_{\rm{e}}}$ are the Bohr radius and electron mass, respectively. The mass density of the electron cloud should screen the scalar field perturbation by the point-like density of the atomic nucleus. Substituting the background density of the electron cloud into Eq. (\ref{33}), the coupling coefficient is estimated to be $|\beta |\approx 7.836 \times {10^{3}}$. By using Eq. (\ref{32plus1}) the mass scale of the coupling is obtained to be ${M_{\rm{m}}} \approx 3 \times {10^{14}}\,{\rm{GeV/}}{c^2}$, which satisfies the constraint of ${M_{\rm{m}}} > {10^4}\,{\rm{GeV/}}{c^2}$ \cite{z15}. The satisfactory results from the large ambient density of $1.468\,{\rm{kg/}}{{\rm{m}}^3}$.

\subsection{Screening effects to the scalar fifth force}\label{Saturation }

To detect the fifth force, one must uncover the shield on the scalar field. Of course, all the screening effects to shield the fifth force mainly and essentially originate from the symmetry-breaking interaction between the scalar field and matter. We now discuss the two important screening mechanisms.

\subsubsection{Saturation effect in a high density}\label{Saturation1 }

Besides the short interaction-range, the fifth force is also suppressed by the saturation effect discussed in the following. The broken-symmetry interaction potential acts like a trap which confines the value of the scalar field falling in the range either $(0, M_2 {c^2} )$ or $(- M_2{c^2} ,0)$ in the static situation, where ${M_2} = 4.96168 \;{\rm{meV/}}{c^2}$. As an example we choose the case of $\rm{VEV}>0$ (one can also discuss the case of $\rm{VEV}<0$ ). From Eq. (\ref{equ18}), one can see that ${\phi_{\rm {min}}}$ is almost independent on the density of matter and approaches $M_2 {c^2}$ as long as matter density large enough, \emph{i.e.},
\begin{equation}
 \rho \gg {\frac{\lambda {M_1}^{\rm{4}}{c^3}}{{\hbar ^3}}} \simeq {5.722 \times { 10^{-30}}} \, \rm{kg/m^3}. \label{equa34}
\end{equation}
No matter how violently matter-density varies in the space, as long as matter-density is large enough, the fifth force denoted by Eq. (\ref{32}) vanishes due to the scalar field almost keeping the same value. This may be called saturation effect. Therefore, the strength of the fifth force cannot be further enhanced by continuously increasing the density of sources. It is unnecessary to use high density metal as the sources to induce the scalar field. But for detecting the fifth force in laboratory, ultrahigh vacuum is necessary.

\subsubsection{Thin shell of the scalar fifth force}\label{thinshell }

In an unbounded homogeneous background, the scalar field always equals to $\phi _{\min }$ at any space point and then the acceleration on a test particle due to the scalar field interaction is zero. When a test particle traveling in one medium with matter density $\rho _{\rm{1}}$ impinges on another medium with a different matter density $\rho _{\rm{2}}$, it experiences a scalar fifth force. For simplicity we assume that the boundary surface is the plane $z=0$ and the test particle travels in the $+z$ direction from the left region 1 of the boundary surface to the right region 2. It is helpful to gain intuition on how the fifth force is localized near the interface between the two mediums. The region is called thin shell \cite{z44,z45,z50,z58}. Since inside medium 1 the scalar field $\phi=\phi_{\rm{min1}}$ and inside medium 2 the scalar field $\phi=\phi_{\rm{min2}}$, and the gradient along $z$ direction near the interface can be roughly estimated as

\begin{equation}
\frac{{d\phi }}{{dz}} \sim \frac{{{\phi _{\min 2}} - {\phi _{\min 1}}}}{{{\mathchar'26\mkern-10mu\lambda _{{\rm{c1}}}} + {\mathchar'26\mkern-10mu\lambda _{{\rm{c2}}}}}}.\label{equa35minise-1}
\end{equation}
Thus, an asymptotic solution to the scalar field equation of motion can be guessed as follows:
\begin{equation}
\phi \left( z \right) = \frac{{{\phi _{\min 2}}}}{2}\left( {1 + \tanh \frac{z}{{{\zeta}}}} \right) + \frac{{{\phi _{\min 1}}}}{2}\left( {1 - \tanh \frac{z}{{{\zeta}}}} \right).\label{equa35-1}
\end{equation}
Then, the gradient of $\phi$ along the $+z$ direction is
\begin{equation}
\nabla \phi \equiv \frac{{d\phi }}{{dz}}\hat z= \frac{{{\phi _{\min 2}} - {\phi _{\min 1}}}}{{2{\zeta}}}{{\mathop{\rm sech}\nolimits} ^2}\left( {\frac{z}{{{\zeta}}}} \right)\hat z, \label{equa36}
\end{equation}
where $\hat z$ is the unit vector of the coordinate $z$ and $\zeta$ denotes a effective range of the fifth force. If the effective fifth force length $\zeta$ defined by
\begin{equation}
\zeta = \mathchar'26\mkern-10mu\lambda _{\rm{c2}} + \mathchar'26\mkern-10mu\lambda _{\rm{c1}},\label{equa37-1}
\end{equation}
with $\mathchar'26\mkern-10mu\lambda _{\rm{c2}}$ and $ \mathchar'26\mkern-10mu\lambda _{\rm{c1}}$ corresponding to the Compton wavelengths for $\rho _{\rm{2}}$  and $\rho _{\rm{1}}$, respectively, then near the surface, $d\phi /dz \sim  \left( {{\phi _{\min 2}} - {\phi _{\min 1}}} \right)/\left( 2{{\zeta} } \right)$, as expected in Eq. (\ref{equa35minise-1}) except the factor 2.

\begin{figure}
\centering
\includegraphics[width=250pt]{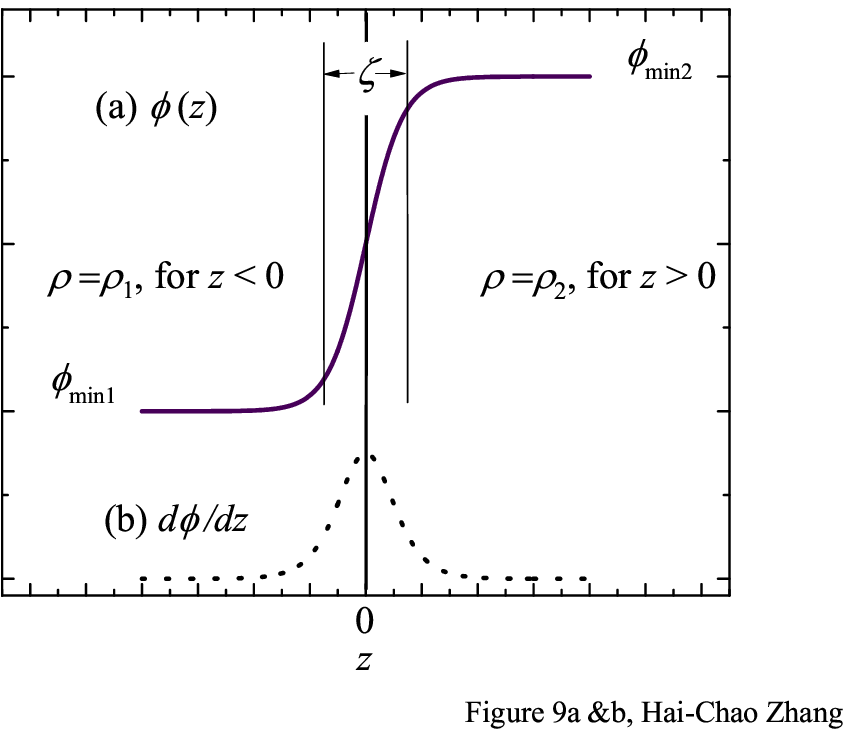}
\caption{The thin shell near the boundary surface $z=0$. (a) $\phi (z)$ versus $z$. (b) $ d\phi(z)/dz$ versus $z$. $\zeta $ denotes the thin shell length. The fifth force is strongly localized in space with a maximum on the boundary surface $z=0$ and falling rapidly to zero for $|z|>\zeta$ }.\label{figure9}
\end{figure}

The fifth force mainly appears at a thin shell region around the boundary surface. The thin-shell effect suppresses the scalar fifth force away from the boundary surface. Figure \ref{figure9} depicts the thin shell region through both (a) $\phi (z)$ and (b) $ d\phi(z)/dz$. The thin shell length can be defined by the effective range $\zeta$ of the fifth force.

All in all, the peak magnitude of the gradient shown in Eq. (\ref{equa36}) is proportional to the difference of $\left( {{\phi _{\min 2}} - {\phi _{\min 1}}} \right)$. Due to the saturation effect, the largest difference $\left( {{\phi _{\min 2}} - {\phi _{\min 1}}} \right)$ is $M_2 {c^2}$, which cannot increase further. Although the peak magnitude of the gradient is inversely proportional to the thin shell length $\zeta$, a narrow thin shell makes a sharp decay of the fifth force away from the boundary surface. However, if experiments are properly designed in which test objects are able to pass through the thin-shell \cite{z8} and performed in the ultrahigh vacuum \cite{z82}, the fifth force might be detected significantly due to the bigger value of the coupling $\beta$.

\subsection{Approximate solution in one-dimensional case}\label{magnitude}

Now, we estimate the magnitude order of the scalar fifth force more quantitatively in the case that the density difference between a source and background is much smaller than the homogeneous background, that is, $\delta \rho  \ll {\rho _{\rm{b}}}$. In this case, the effective mass of the coupled scalar field can be seen as the same value in all of space,\emph{ i. e.,} $m_{{\rm{eff}}}^{\rm{2}} = m_{{\rm{eff}}}^{\rm{2}}\left( {{\phi _{\rm{b}}}} \right)$ and then Eq. (\ref{30}) is indeed a Klein-Gorden equation. Therefore, the linear superposition principle is valid. Supposing that the source is homogeneously filled in the region between the two planes $z=-z_{0}$ and $z=z_{0}$,
\begin{equation}
\delta \rho(z) =
\begin{cases}
\delta {\rho _0}& |z|\le {z_0}\\
0& |z|>z_{0}
\end{cases}
\end{equation}
with $z_{0}$ and $\delta {\rho _0}$ are positive values, the solution to Eq. (\ref{30}) can be easily obtain as follows:
\begin{eqnarray}
{\delta \phi(z)} &  = & { {A_{,\phi }}\left( {{\phi _{\rm{b}}}} \right)\hbar {c^{\rm{3}}}\int_{ - \infty }^{ + \infty } {\delta \rho (\eta )g(z;\eta )} d\eta} \nonumber \\
 & = & {{A_{,\phi }}\left( {{\phi _{\rm{b}}}} \right)\hbar {c^{\rm{3}}}{\delta \rho_{0}\int_{ - z_{0} }^{ + z_{0} }g(z;\eta )} d\eta,\;\;\rm{for}\; \delta \rho_{0}  << {\rho _{\rm{b}}}},\label{equa39}
\end{eqnarray}
where the Green function
\begin{equation}
g(z;\eta ) =  - \frac{{{\mathchar'26\mkern-10mu\lambda _{{\rm{cb}}}}}}{2}\exp \left( { - \frac{{\left| {z - \eta } \right|}}{{{\mathchar'26\mkern-10mu\lambda _{{\rm{cb}}}}}}} \right)\label{equa40}
\end{equation}
with the Compton wavelength $\mathchar'26\mkern-10mu\lambda _{{\rm{cb}}}$ corresponding to the ambient density $\rho _{\rm{b}}$. The scalar fifth force along the $+z$ direction can be obtained by Eq. (\ref{32}) as follows:
\begin{equation}
a\left( z \right) = -\frac{{\beta \left( {{\phi _{\rm{b}}}} \right)}}{{{M_{{\rm{Pl}}}}}}\frac{{d\delta \phi \left( z \right)}}{{dz}}. \label{equa41}
\end{equation}

\subsubsection{Estimation of the value of thin shell}\label{thinshellsca }

Since the interaction range calculated by Eq. (\ref{equ19and33}) is extremely small in the common background density of laboratory case, we consider another extreme case where the interaction range can achieve about the order of magnitude of $10\, \mu \rm{m}$ so that one can gain a quantitative picture. In addition, although the extreme background vacuum might not be achieved in laboratory, it is reasonable that in the dilute gas of atoms the space among the atoms must be extreme vacuums, which can be used to design the test experiment for the scalar fifth force.

\begin{figure}
\centering
\includegraphics[width=250pt]{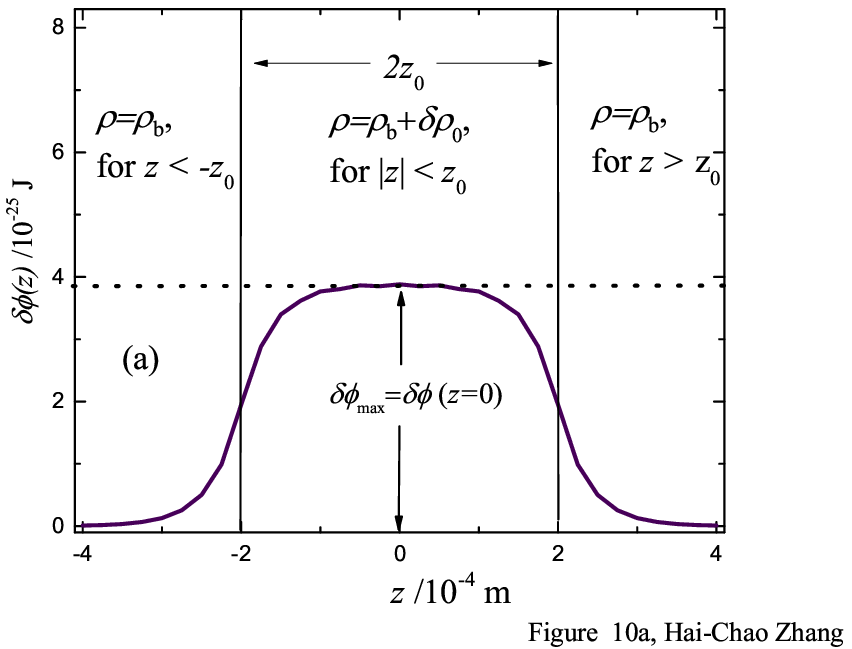}
\includegraphics[width=250pt]{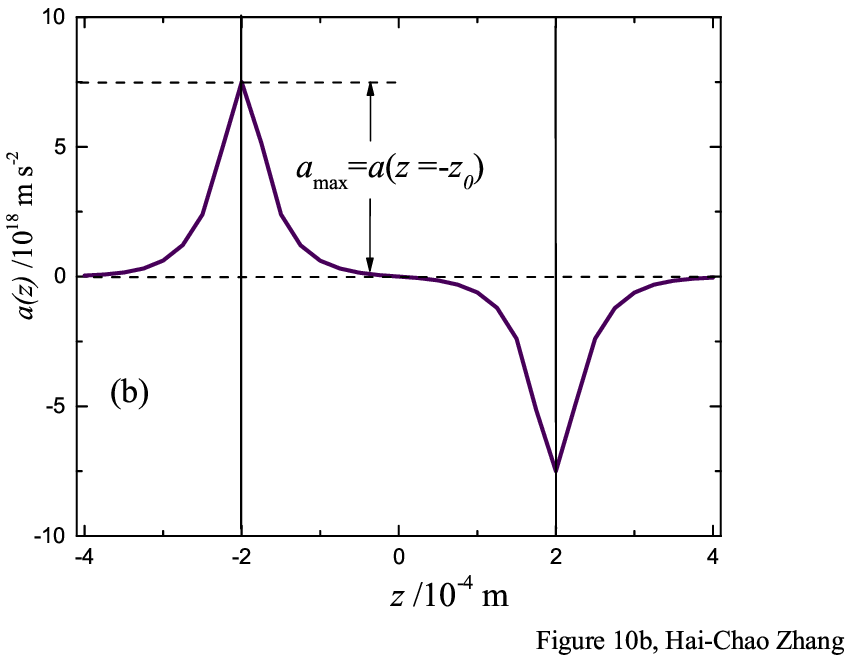}
\caption{(a) The scalar field difference $\delta \phi(z)$ along the $z$ direction. The maximum difference appears at the $z=0$ plane. (b) The scalar fifth force along the $z$ direction. The two maxima of the absolute value of the fifth force are at the two boundary surfaces $z=-2 \times {10^{ - 4}}\,\rm{m}$ and $z=2 \times {10^{ - 4}}\,\rm{m}$, respectively,  but the directions of the two forces are opposite. Since the Compton wavelength is too short in the case of large matter density, we plot the pictures in the very smaller density of matter. That is, ${\rho _{\rm{b}}} = 2 \times {10^{ - 29}}\,\rm{kg \cdot {m^{ - 3}}}$ and $\delta {\rho _0} = 1 \times {10^{ - 31}}\, \rm{kg \cdot {m^{ - 3}}} \ll {\rho _{\rm{b}}}$, respectively. }\label{figure10}
\end{figure}

Of course, if dark matter permeates everywhere, the current minimum density should be $ \simeq 2.7 \times {10^{ - 27}}\,\rm{kg \cdot {m^{ - 3}}}$ and the corresponding interaction range is $3 \times {10^{ - 6}}\,\rm{m}$, which is too small to draw a clearly illustrated graph. To clearly display the character of the fifth force, we choose a lower value of the background density of ${\rho _{\rm{b}}} = 2 \times {10^{ - 29}}\,\rm{kg \cdot {m^{ - 3}}}$ (the corresponding Compton wave length ${\mathchar'26\mkern-10mu\lambda _{\rm{cb}}}= 3.684 \times {10^{ - 5}}\,\rm{m}$). We also choose that $\delta {\rho _0} = 1 \times {10^{ - 31}}\, \rm{kg \cdot {m^{ - 3}}} \ll {\rho _{\rm{b}}}$, and ${z_0} = 2 \times {10^{ - 4}}\,\rm{m}$ to calculate $\delta \phi(z)$ and $a\left( z \right)$. The calculated results of $\delta \phi(z)$ and $a(z)$ are shown in Figures \ref{figure10}(a) and (b), respectively.

One can see that two thin shell regions appear on the two boundary surfaces in Figure \ref{figure10}. The maximum value of $\delta \phi(z)$ in Figure \ref{figure10}(a) appears at the center plane $z=0$ of the source, which is denoted by $\delta\phi_{\max}$; The maxima of $|a( z )|$ in Figure \ref{figure10}(b) appear at the two boundary surfaces of $|z|=2 \times {10^{ - 4}}\,\rm{m}$. The positive (negative) value of the force describes its direction along the $+z$ ($-z$) direction, and the maximum value at $z=-z_{0}$ is denoted by $a_{\max}$ in Figure \ref{figure10}(b).

The source region acts as an attractive trap due to $\delta {\rho _0}>0$. Oppositely, if $\delta {\rho _0}<0$, the region acts like a repulsive barrier.  Whether the length of the source region is larger than the Compton wave length ${\mathchar'26\mkern-10mu\lambda _{\rm{cb}}}$ or not, the fifth force is indeed equal to zero on the center plane of $z=0$.

\subsubsection{The maximum value of fifth force at boundary surface}\label{forceboundary}

As long as $\delta \rho_{0} \ll {\rho _{\rm{b}}}$, both $\delta\phi_{\max} \equiv \delta \phi \left( z=0 \right)$ and $a_{\max}\equiv a\left( { z=-{z_0}} \right)$ are proportional to $\delta \rho_{0}$ described by Eqs. (\ref{equa39}) and (\ref{equa41}). When $\delta \rho_{0}  > {\rho _{\rm{b}}}$, Eqs. (\ref{equa39}) and (\ref{equa41}) are no longer valid. However, it can be deduced that $\delta\phi_{\max}$ cannot exceed $M_2 {c^2}$ since the value of the scalar field falls in the range of $(0, M_2 {c^2} )$. The gradient of $\delta \phi(z)$ can be estimated by $\delta\phi_{\max}/{\mathchar'26\mkern-10mu\lambda _{\rm{cb}}}$. Then the gradient cannot exceed $M_2 {c^2}/{\mathchar'26\mkern-10mu\lambda _{\rm{cb}}}$. Consequently, one can deduce that $a_{\max}$ cannot exceed $M_2 {c^2}/(M_{{\rm{m}}}{\mathchar'26\mkern-10mu\lambda _{\rm{cb}}})$ even if $\delta {\rho _0} \to  + \infty $.

\begin{figure}
\centering
\includegraphics[width=250pt]{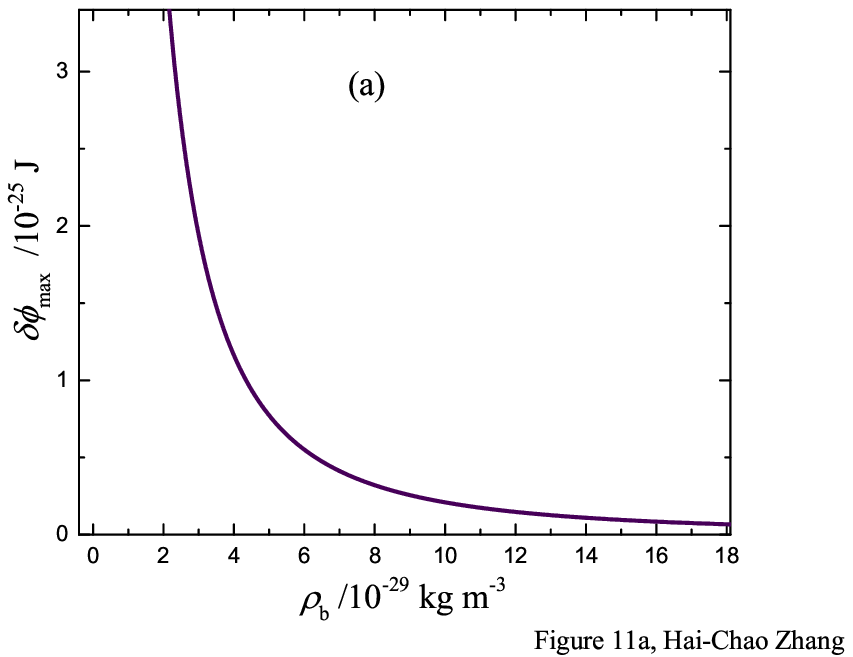}
\includegraphics[width=250pt]{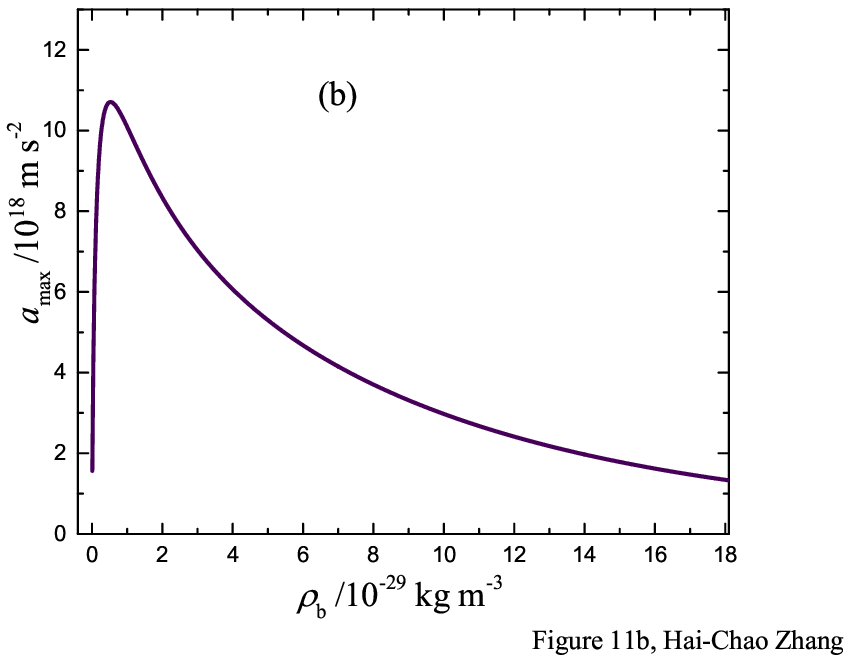}
\caption{(a) $\delta \phi_{\max}$ versus $\rho _{\rm{b}}$. This is a monotonically decreasing function. When $\delta {\rho _0}$ is fixed, the difference of the scalar field between the deep inside and outside of the source decreases rapidly with the ambient density increasing. (b) $a_{\max}$ versus $\rho _{\rm{b}}$. This is a concave function. When $\delta {\rho _0}$ is fixed, the acceleration on a test particle approaches to zero if the ambient density approaches either to zero or to infinity. }\label{figure11}
\end{figure}

We now return again to the case of $\delta \rho_{0}  \ll {\rho _{\rm{b}}}$. Both $\delta\phi_{\max} $ and $a_{\max}$ are dependent on the background density $\rho _{\rm{b}}$, but they do not increase with the increase of the density of background. To see the density-dependence trend , we fix $\delta {\rho _0} = 1 \times {10^{ - 31}}\, \rm{kg \cdot {m^{ - 3}}} \ll {\rho _{\rm{b}}}$ and plot the curves of $\delta\phi_{\max}$ and $a_{\max}$ versus $\rho _{\rm{b}}$. The background-density-dependence of $\delta\phi_{\max}$ and $a_{\max}$ are shown in Figure \ref{figure11}(a) and (b), respectively.

The curve of $\delta\phi_{\max}$ versus $\rho _{\rm{b}}$ is a monotonically decreasing function shown in Figure \ref{figure11}(a). However, the curve of $a_{\max}$ versus $\rho _{\rm{b}}$ is a concave function shown in Figure \ref{figure11}(b) and the maximum point appears at $ \rho _{\rm{b}} \approx 5 \times{ 10^{-30}} \,\rm{kg/m^3} $. When the background density $\rho _{\rm{b}}\to 0$ or $\rho _{\rm{b}}\to \infty$, the fifth force rapidly approaches to zero.

\subsubsection{Estimation expression of average fifth force in thin shell}\label{averagingforce}

To gain intuition picture on how the scalar fifth force is dependent on the background density in the case of $\delta \rho_{0}  \ll {\rho _{\rm{b}}}$, we will derive a concise approximation expression. Since the force is localized in the thin shell, we now consider an average acceleration in this region rather than the maximum value of the acceleration discussed above. The average gradient of the scalar field along the $z$ direction is roughly estimated as follows:
\begin{equation}
\left\langle {\frac{{d\phi }}{{dz}}} \right\rangle \sim \frac{{{\phi _{\min }}(\rho _{\rm{b}}+\delta \rho_{0}) - {\phi _{\min}}(\rho _{\rm{b}})}}{{{{\mathchar'26\mkern-10mu\lambda _{{\rm{cb}}}}}}}, \label{equa42}
\end{equation}
where the density-dependent $\phi _{\min }$ is described by Eq. (\ref{equ18-a}). From Eq. (\ref{equa41}), one gets the average acceleration in the thin shell as follows:
\begin{eqnarray}
\bar a \sim \frac{{\left| {\beta \left( {{\rho _{\rm{b}}}} \right)} \right|}}{{{M_{{\rm{Pl}}}}}}\left\langle {\frac{{d\phi }}{{dz}}} \right\rangle.\label{equa43minus}
\end{eqnarray}
For $\delta \rho_{0}  \ll {\rho _{\rm{b}}}$, Eq. (\ref{equa43minus}) becomes
\begin{eqnarray}
\bar a \sim\frac{{\left| {\beta \left( {{\rho _{\rm{b}}}} \right)} \right|}}{{{M_{{\rm{Pl}}}}}}{\left. {\frac{{d{\phi _{\min }}}}{{d\rho }}} \right|_{\rho  = {{\rm{\rho }}_{\rm{b}}}}}\frac{\delta {\rho _0}}{{{\mathchar'26\mkern-10mu\lambda _{{\rm{cb}}}}}}.\label{equa43}
\end{eqnarray}
For $\rho_{\rm{b}} \gg 10^{-30}\;\rm{kg/m^3}$ and $ \delta \rho_{0}  \ll {\rho _{\rm{b}}}$, using Eqs. (\ref{equ18-a}), (\ref{equ19and33}) and (\ref{33}), we get a concise expression as
\begin{eqnarray}
{\bar a} [\rm{m/s^2}] \sim \frac{{3.668 \times {{10}^{ - 20}}}}{{{(\rho _{\rm{b}}[\rm{kg/m^3}])}^{5/2}}}  \delta {\rho _0}[\rm{kg/m^3}].\label{equa44}
\end{eqnarray}
It can be seen that the fifth force decreases rapidly with the ambient density increasing. Although the condition of $\rho_{\rm{b}} \gg 10^{-30}\;\rm{kg/m^3}$ can always be easily satisfied in laboratory, $ \delta \rho_{0}  \ll {\rho _{\rm{b}}}$ cannot be easily achieved.

\subsubsection{Estimation expression of fifth force away from a solid surface}\label{forceaway}

In general, when $ \delta \rho_{0}  \ge {\rho _{\rm{b}}} $, Eqs. (\ref{equa43}) and (\ref{equa44}) are no longer valid. However, one can still use Eqs. (\ref{equa42}) and (\ref{equa43minus}) to roughly estimate the average acceleration in the thin shell. In the most experimental designs, the source objects are solid. The test objects cannot pass through the region of the thin-shell but can only sense the fifth force away from the surface. Supposing that the boundary surface is the plane $z=z_{0}$, and noticing Eq. (\ref{equa40}), the acceleration of a test object to linear order from the action (\ref{equ1}) is then given by
\begin{equation}
a\left( z \right) = a({z_0}){e^{ - \left| {z - {z_0}} \right|/{\mathchar'26\mkern-10mu\lambda _{{\rm{cb}}}}}}.\label{equa45}
\end{equation}
The average acceleration in the thin shell can be defined by
\begin{eqnarray}
{\bar a} & \equiv & {\frac{1}{{{\mathchar'26\mkern-10mu\lambda _{{\rm{cb}}}}}}\int_{{z_0}}^{{z_0} + {\mathchar'26\mkern-10mu\lambda _{{\rm{cb}}}}} {a\left( z \right)dz}} \nonumber \\
& = & {a({z_0})\left( {1 - {e^{ - 1}}} \right) = 0.632a({z_0})}.\label{equa46}
\end{eqnarray}
Noticing Eqs. (\ref{equa42}) and (\ref{equa43minus}), Eq. (\ref{equa45}) then becomes
\begin{equation}
a\left( z \right) \sim -1.582\frac{{ {\beta \left( {{\rho _{\rm{b}}}} \right)} }}{{{M_{{\rm{Pl}}}}}}\frac{{{\phi _{\min }}\left( {{\rho _{\rm{b}}} + \delta {\rho _0}} \right) - {\phi _{\min }}\left( {{\rho _{\rm{b}}}} \right)}}{{{\mathchar'26\mkern-10mu\lambda _{{\rm{cb}}}}}}{e^{ - \left| {z - {z_0}} \right|/{\mathchar'26\mkern-10mu\lambda _{{\rm{cb}}}}}}.\label{equa47}
\end{equation}
Both the saturation factor of $\left[{\phi _{\min }}\left( {{\rho _{\rm{b}}} + \delta {\rho _0}} \right) - {\phi _{\min }}\left( {{\rho _{\rm{b}}}} \right)\right]$ and the exponential attenuation factor of $\exp \left( { - \frac{{\left| {z - {z_0}} \right|}}{{{\mathchar'26\mkern-10mu\lambda _{{\rm{cb}}}}}}} \right)$ suppress the scalar fifth force. Thus, the fifth force is too small to be detected \cite{z80,z102}. Apparently, the main screening factor in detecting the fifth force is the short ranged interaction.

Since the fifth force mainly appears at the thin shell region, the fifth force might be detected if the designed experiments allow test objects passing through the region of the thin-shell. To obtain a huge value of the coupling coefficient $\beta ( \rho _{\rm{b}})$, an extreme-high background vacuum in laboratory is necessary. If dark matter permeates all of the Universe space, the current minimum density should be as small as $\sim {10^{ - 27}}\,\rm{kg \cdot {m^{ - 3}}}$, and then the upper bound of the force range is $\sim 5\,\mu \rm{m}$.

\subsection{Summary}\label{suforce}

The scalar fifth force is investigated in our model, especially, some concise approximate expressions of the fifth force are deduced. Under the requirements that the coupled scalar field must account for the observed cosmic acceleration, the interaction range of the fifth force is extremely short even in the current density of the Universe. Therefore, the local test of gravity is satisfied. To test the force in laboratory, experiments may be designed allowing that the test objects are able to pass the thin shell of the source.

We have also explained the so-called `upper' bound of $|\beta|\sim {10}^{14}$ from the precision measurements of hydrogenic energy levels. It is worth noting that, this value indeed is not an upper bound due to the matter-density dependence of $|\beta|$. $|\beta|$ can acquire very huge values in the extremely low density of matter. For the case of the precision measurements of hydrogenic energy levels, the calculated value by our model is $\sim {10}^{4}$, which is considerably smaller than the `upper' bound of $\sim {10}^{14}$. Because the mass density of the electron cloud around the atomic nucleus is used in our model, the satisfactory result implies that both ordinary matter (e.g., electron) and dark matter can couple to the scalar field in the same manner.

\section{Discussions and Conclusions}\label{discussion}

\subsection{The equivalence principle, Mach's principle, and the Copernican principle }\label{principle}

Since the scalar field couples universally to all matter fields as shown in the action (\ref{equ1}), the weak equivalence principle holds. In terms of Jordan frame variables, the action (\ref{equ1}) describes a Brans-Dicke-type-like scalar-tensor theory with a field-dependent Brans-Dicke parameter. In their original work \cite{z20}, the Brans-Dicke parameter is chosen as a constant. Since both the self-interaction potential and the symmetry-breaking interaction potential are not inserted specifically in their original action, the model describes a long-range interaction and cannot provide a cosmological constant to drive the Universe accelerating expansion in the present time.

Based on the equivalence principle, Brans and Dicke had introduced a scalar-tensor combination inducing their gravitational field to incorporate Mach's principle into general relativity \cite{z20}. According to Mach, the inertial forces observed in a laboratory may originate in distant matter accelerated relative to the laboratory \cite{z20,z20p1}. The action (\ref{equ1}) combining with Eq. (\ref{equ16revise}) may be seen as a Machian model. Since the equation of motion for the scalar field is determined by a sum of the self-interaction and the interaction potential with matter, the scalar field no longer appears a long-range field as in the original Brans-Dicke model. The interaction range is dependent on the ambient mass density and the range is very short in the common matter density. But the coupled scalar field can indirectly influence matter by a global effect of driving the Universe expansion since the Einstein tensor is determined by both the energy-momentum tensor and the scalar field. In addition, the scalar field can also directly influence matter by the short-range interaction if the local gradient of the scalar field is present.

It is well-known that the principle of a homogeneous and isotropic universe is a spatial embodiment of the Copernican principle. The extended Copernican principle indicates further that there is no special space-time position in the Universe. That is, every part or every era of the Universe is not the center or the origin. However, suitable initial conditions are required for inflation to start in the current cosmological theory \cite{z16,z42}. In general, a flat potential of the inflaton is also required so that sufficient amount of inflation will be obtained. This starting origin of time is at odds with the extended Copernican principle \cite{z56}. The quasi-cyclic model discussed in this paper may resolve the incompatibility with the extended Copernican principle and avoid fine tuning in choosing initial conditions in inflation.

When the coupled scalar field is invoked to drive the Universe acceleration successfully, a strong fifth force appears. However, the very short interaction ranges and the screening effects enable our Brans-Dicke-type-like model in the calculation to converge to the results of general relativity. Although the coupling to ordinary matter should not lead to observable long-range forces, it may result in observable short-range forces. The strength of the fifth forces is dependent on the ambient density, and the density-dependence is a concave curve. In both extremely opposite cases, that is, when the mass density approaches to zero or infinity, the fifth forces vanish.

\subsection{Conclusions and outlook}\label{outlook}

In order to obtain the scalar fifth force, we have required that the scalar field must entirely account for the observed cosmic acceleration. This requirement leads to the symmetry-breaking coupling that can localize the value of the self-interaction potential to act as the cosmological constant. Since the interaction potential between matter and the scalar field is directly proportional to the particle number density of matter, the localization gives the nearly fixed cosmological constant as long as the number density large enough in the non-relativistic-matter case.

Our scheme does not conflict with chameleon no-go theorems, at least mathematically. Just as that has been pointed out in the original literature \cite{z101}, any model that purports to explain the cosmic acceleration, and passes solar system tests, must be doing so using some form of quintessence or vacuum energy. Our model in fact uses a coupled scalar field as quintessence, just only the quintessence is pinned via a symmetry-breaking coupling.

As to Weinberg's no-go theorem, since our scheme uses dynamic quintessence rather than static vacuum energy, the no-go theorem is circumvented. In our model, the self-interaction potential of the scalar field has nothing to do with the effective mass around the minimum of the effective potential but can move the position of the minimum. These properties guarantee that the observed cosmic acceleration stems entirely from the coupled scalar field rather than any stable vacuum energy.

The scheme has also been applied to the relativistic-matter case (e.g., during inflation). In this case, matter decouples completely with the scalar field, and the evolution of the Universe is mainly dominated by the scalar field. The pseudo-potential density has been introduced as the sum of the self-interaction potential energy density of the scalar field and the energy density scale of the curvature of the Universe. It is the pseudo-potential that acts the role of the inflaton potential in the inflationary slow-roll approximation rather than the self-interaction potential itself only. Thus, the observed concave potential feature naturally belongs to the pseudo-potential, and then is in favor of the closed space of the Universe.

When the cosmic constraints are satisfied, the scheme predicts that the magnitude of the fifth force is considerably larger than gravity, especially for lower density in the local environment. However, the fifth force is suppressed into a very short interaction range. Tests of gravity are then satisfied. A typical interaction range is estimated to be about $\sim 5 \,\mu \rm{m}$ for the current matter density of the Universe.

The fifth force might be detectable significantly, provided that experiments are designed allowing test particles to pass through the thin-shell region of sources, at least test particles can approach the thin-shell as closely as possible.

\begin{acknowledgments}
We acknowledge discussions with J. N. Zhang and C. Liu. This work was supported by Science Challenge Project, TZ2018003 and the National Natural Science Foundation of China through Grant Nos. 11604348.
\end{acknowledgments}

\appendix

\section{Obtaining the acceleration equation of the Universe}\label{appA}
\setcounter{section}{1}

We now derive the acceleration equation of the Universe in detail from the Friedmann equation shown as Eq. (\ref{equ4}). For the simplification, the abbreviate notation $B_i(\phi)$ is introduced through $B_i(\phi)\equiv A^{1 - 3{w_i}}\left( \phi  \right)$ at the beginning in deriving equations. The final results are obtained by using ${A^{1 - 3{w_i}}}\left( \phi  \right)$ instead of $B_i(\phi)$. The Friedmann equation (\ref{equ4}) in a Friedmann-Robertson-walker (FRW) metric then becomes as follows:
\begin{equation}
{H^2} \equiv \frac{{{{\dot a}^2}}}{{{a^2}}} = \frac{{8\pi G}}{{3{\hbar ^3}{c^5}}}\left[ { \sum\limits_i\rho_i {\hbar ^3}{c^5}B_i(\phi) + \left( {V\left( \phi  \right) + \frac{{{\hbar ^2}}}{2}{{\dot \phi }^2}} \right)} \right] - \frac{{K{c^2}}}{{{a^2}}},\label{s10}
\end{equation}
where the values $K = 1,\, 0 ,\,{\rm{ or \, - 1}}$ correspond to closed, flat or open spaces, respectively. Differentiating Eq. (\ref{s10}) with respect to time, one has
\begin{eqnarray}
 2H\dot H & \equiv & 2H\left( {\frac{{\ddot a}}{a} - {H^2}} \right)\nonumber\\
 &= & \frac{{8\pi G}}{{3{\hbar ^3}{c^5}}}\left\{ { \sum\limits_i\rho_i {\hbar ^3}{c^5}{B_{i,\phi }}\left( \phi  \right)\dot \phi  +  \sum\limits_i\dot \rho_i {\hbar ^3}{c^5}B_{i}\left( \phi  \right) + \left[ {{V_{,\phi }}\left( \phi  \right)\dot \phi  + {\hbar ^2}\ddot \phi \dot \phi } \right]} \right\} + \frac{{2K{c^2}\dot a}}{{{a^3}}}.\label{s11}
\end{eqnarray}
Noticing Eq. (\ref{equ8}), \emph{i.e.},
\begin{equation}
{\hbar ^2}\ddot \phi  + 3H{\hbar ^2}\dot \phi  + {V_{,\phi }}\left( \phi  \right) +  \sum\limits_i\rho_i {\hbar ^3}{c^5}{B_{i,\phi }}\left( \phi  \right) = 0,\label{s12}
\end{equation}
one has
\begin{eqnarray}
2H\dot H = \frac{{8\pi G}}{{3{\hbar ^3}{c^5}}}\left[ { - 3H{\hbar ^2}{{\dot \phi }^2} +  \sum\limits_i\dot \rho_i {\hbar ^3}{c^5}B_i\left( \phi  \right)} \right]{\rm{ + 2}}H\frac{{K{c^2}}}{{{a^2}}}.\label{s13}
\end{eqnarray}
Substituting the conservation law of Eq. (\ref{equ6}), \emph{i.e.},
\begin{equation}
\dot \rho_i  =  - 3H\left( {\rho_i  + \frac{p_i}{{{c^2}}}} \right) \label{s14}
\end{equation}
into Eq. (\ref{s13}), one has
\begin{equation}
2H\dot H = \frac{{8\pi G}}{{3{\hbar ^3}{c^5}}}\left[ { - 3H{\hbar ^2}{{\dot \phi }^2} - 3H\sum\limits_i\left(  {\rho_i  + \frac{p_i}{{{c^2}}}} \right){\hbar ^3}{c^5}B_i\left( \phi  \right)} \right] + {\rm{2}}H\frac{{K{c^2}}}{{{a^2}}}.\label{s15}
\end{equation}
Again, we need to emphasize that $\rho_i $ and $p_i$ are independent of the scalar field and the satisfaction of the conservation law is a definition-like choice or constraint. This conservation law of Eq. (\ref{s14}) presents that the corresponding entropy is conserved. The number of the particles is not altered but the masses of the particles are shifted due to the coupling of the particles to the scalar field. In other words, $\rho_i$ and $p_i$ denote the matter mass density and pressure in the decoupled case with the scalar field, but ${\rho _i}{A^{1 - 3{w_i}}}\left( \phi  \right)$ describes the physics matter density. Thus, only expression ${p_i}{A^{1 - 3{w_i}}}\left( \phi  \right)$ can be used to describe the physics pressure of the perfect fluid, so that the parameter of the equation of state for matter is independent of the scalar field [see Eq. (\ref{s18}) later].

Removing the common factor $2H$ in Eq. (\ref{s15}), we have
\begin{eqnarray}
\dot H = \frac{{8\pi G}}{{3{\hbar ^3}{c^5}}}\left[ { - \frac{3}{2}{\hbar ^2}{{\dot \phi }^2} - \frac{3}{2}\sum\limits_i\left( {\rho_i  + \frac{p_i}{{{c^2}}}} \right){\hbar ^3}{c^5}B_i\left( \phi  \right)} \right]{\rm{ + }}\frac{{K{c^2}}}{{{a^2}}}.\label{s16}
\end{eqnarray}
Since $\ddot a/a \equiv \dot H + {H^2}$, from Eqs. (\ref{s10}) and (\ref{s16}), one has
\begin{eqnarray}
\frac{{\ddot a}}{a} = \frac{{4\pi G}}{{3{\hbar ^3}{c^5}}}\left[ {2V\left( \phi  \right) - 2{\hbar ^2}{{\dot \phi }^2} - \sum\limits_i\left( {\rho_i  + \frac{{3p_i}}{{{c^2}}}} \right){\hbar ^3}{c^5}B_i\left( \phi  \right)} \right].\label{s17}
\end{eqnarray}
Noticing the equation of state Eq. (\ref{equ5}), \emph{i.e}., ${w_i} \equiv {p_i}/\left( {{\rho _i}{c^2}} \right)$, and introducing a coupled matter mass density for each species matter as
\begin{equation}
{\rho _{{\rm{m}}i}} = {\rho _i}{A^{1 - 3{w_i}}}\left( \phi  \right)\equiv \rho _i B_i(\phi)\label{s17plus1}
\end{equation}
that includes the coupling energy with the scalar field, and a corresponding coupled pressure of the perfect fluid as
\begin{equation}
{p_{{\rm{m}}i}} = {p_i}{A^{1 - 3{w_i}}}\left( \phi  \right)\equiv p_i B_i(\phi)\label{s17plus2}
\end{equation}
that includes the coupling pressure with the scalar field, we immediately have
\begin{equation}
{w_i} \equiv \frac{{{p_i}}}{{{\rho _i}{c^2}}} = \frac{{{p_{{\rm{m}}i}}}}{{{\rho _{{\rm{m}}i}}{c^2}}}.\label{s18}
\end{equation}
We expect that $\rho _{{\rm{m}}i}\,(p_{{\rm{m}}i})$ denotes the real physics matter density (physics pressure). Although $\rho _{{\rm{m}}i}\,(p_{{\rm{m}}i})$ is $\phi$-dependent, from Eq. (\ref{s18}) we can see that the parameter ${w_i}$ of the equation of state for matter is really independent of the scalar field as we desired.

Replacing $B_i(\phi)$ by $A^{1 - 3{w_i}}\left( \phi  \right)$, Eq. (\ref{s17}) can be rewritten as Eq. (\ref{equ9}), \emph{i.e.,}
\begin{eqnarray}
\frac{{\ddot a}}{a} = \frac{{4\pi G}}{{3{\hbar ^3}{c^5}}}\left[ {2V\left( \phi  \right) - 2{\hbar ^2}{{\dot \phi }^2} - \sum\limits_i {{\rho _{\rm{m}}}_i\left( {1 + 3{w_i}} \right){\hbar ^3}{c^5}} } \right].\label{s19}
\end{eqnarray}
In addition, Eq. (\ref{s10}) or Eq. (\ref{equ4}) can also be rewritten as follows:	
\begin{eqnarray}
{H^2} = \frac{{8\pi G}}{3}\left[ {\sum\limits_i {{\rho _{{\rm{m}} i}}}  + \frac{1}{{{\hbar ^3}{c^5}}}\left( {V\left( \phi  \right) + \frac{{{\hbar ^2}}}{2}{{\dot \phi }^2}} \right)} \right] - \frac{{K{c^2}}}{{{a^2}}}\nonumber\\
=\frac{{8\pi G}}{3}\left[ { {{\rho _{{\rm{m}}}}}  + \frac{1}{{{\hbar ^3}{c^5}}}\left( {V\left( \phi  \right) + \frac{{{\hbar ^2}}}{2}{{\dot \phi }^2}} \right)} \right] - \frac{{K{c^2}}}{{{a^2}}},\label{s10plus}
\end{eqnarray}
where ${\rho _{\rm{m}}} = \sum\nolimits_i {{\rho _{{\rm{m}}i}}}$ denotes the total scalar-field-dependent matter density.

In this Appendix, it has also been shown through Eq. (\ref{s18}) that one can find the equation of state for matter being temperature dependent but free of the scalar field. The property of $\phi$-independence is very important to get Eq. (\ref{s8plusre6zong}) in Appendix \ref{prappc}.

\section{The conservation laws}\label{appB}
\setcounter{section}{2}

When the scalar field couples with matter, the energy exchange between the scalar field and matter would happen. Thus, the equation of state for the scalar field should be modified. In this Appendix, the average energy of an individual matter-particle will also be introduced as a supplement to the conservation law of Eq. (\ref{s14}).

\subsection{The coupled equations }\label{ceappb}

The basic assumption of the conservation law of Eq. (\ref{s14}) and its equivalent form of Eq. (\ref{equ6}), mean that both ${\rho _i}$ and ${w_i}$ are independent of the scalar field with $i$ denoting several species of non-interacting perfect fluids of matter sources. However, this does not mean that the energy density of matter is independent of the scalar field. The energy exchange between matter and the scalar field is discussed as follows: To avoid the complicatedness, we derive the rest of conservation laws by using $B_i(\phi)$ instead of $ A^{1 - 3{w_i}}\left( \phi  \right)$. We have introduced the scalar-field-dependent matter density and the scalar-field-dependent pressure shown as Eqs. (\ref{s17plus1}) and (\ref{s17plus2}), respectively. Since $i$-species matter density ${\rho _{{\rm{m}}i}} = \rho _i B_i(\phi)$, differentiating with respect to time reads
\begin{equation}
{\dot \rho _{{\rm{m}}i}} = \dot \rho_i B_i\left( \phi  \right) + \rho_i \dot {B_i}\left( \phi  \right).\label{B1}
\end{equation}
Substituting the conservation law of Eq. (\ref{s14}) into Eq. (\ref{B1}), we get
\begin{equation}
{\dot \rho _{{\rm{m}}i}}  =  - 3H\left( {\rho_i  + \frac{p_i}{{{c^2}}}} \right)B_i\left( \phi  \right) + \rho_i \dot B_i\left( \phi  \right).\label{B2}
\end{equation}
Noticing Eqs. (\ref{s17plus1}) and (\ref{s17plus2}), we then get
\begin{equation}
{\dot \rho _{{\rm{m}}i}} + 3H\left( {{\rho _{{\rm{m}}i}} + \frac{{{p_{{\rm{m}}i}}}}{{{c^2}}}} \right) = \rho_i \dot B_i\left( \phi  \right).\label{s21}
\end{equation}
Summing over all species matter $i$ to Eq. (\ref{s21}), we get
\begin{equation}
{\dot \rho _{{\rm{m}}}} + 3H\left( {{\rho _{{\rm{m}}}} + \frac{{{p_{\rm{m}}}}}{{{c^2}}}} \right) =\sum\limits_i \rho_i \dot B_i\left( \phi  \right),\label{s21summ}
\end{equation}
where ${\rho _{\rm{m}}} = \sum\nolimits_i {{\rho _{{\rm{m}}i}}} $ and ${p _{\rm{m}}} = \sum\nolimits_i {{p _{{\rm{m}}i}}} $, respectively. Eq. (\ref{s21summ}) is different from the conservation law of Eq. (\ref{s14}). The difference results from the interaction between matter and the scalar field.

Since the energy density of the scalar field is defined by ${\hbar ^3}{c^5}{\rho _\phi } = V\left( \phi  \right) + {{\dot \phi }^2}{\hbar ^2}/2$, differentiating this equation gives
\begin{equation}
{\dot \rho _\phi }{\hbar ^3}{c^5} = {V_{,\phi }}\left( \phi  \right)\dot \phi  + {\hbar ^2}\dot \phi \ddot \phi .\label{s22}
\end{equation}
Substituting Eq. (\ref{s12}) into Eq. (\ref{s22}) and noticing the pressure definition of the scalar field ${\hbar ^3}{c^3}{p_\phi } =  - V\left( \phi  \right) + {{\dot \phi }^2}{\hbar ^2}/2$,  it follows that
\begin{eqnarray}
{{\dot \rho }_\phi }{\hbar ^3}{c^5} =  - 3H\left( {{\hbar ^3}{c^5}{\rho _\phi } + {\hbar ^3}{c^3}{p_\phi }} \right) - \sum\limits_i\rho_i {\hbar ^3}{c^5}\dot B_i\left( \phi  \right).\label{s23}
\end{eqnarray}
Then, removing the common factor ${\hbar ^3}{c^5}$ in Eq. (\ref{s23}), we have
\begin{equation}
{\dot \rho _\phi } + 3H\left( {{\rho _\phi } + \frac{{{p_\phi }}}{{{c^2}}}} \right) =  - \sum\limits_i\rho_i \dot B_i\left( \phi  \right).\label{s24}
\end{equation}
Introducing the interaction energy density between $i$-species matter and the scalar field by ${\rho _{i{\mathop{\rm int}} }} = \rho_i \left[ {B_i\left( \phi  \right) - 1} \right]$, and the corresponding interaction pressure by ${p_{i{\mathop{\rm int}} }} = p_i\left[ {B_i\left( \phi  \right) - 1} \right]$, and noticing the conservation equation (\ref{s14}), one has
\begin{equation}{\dot \rho _{i{\mathop{\rm int}} }} + 3H\left( {{\rho _{i{\mathop{\rm int}} }} + \frac{{{p_{i{\mathop{\rm int}} }}}}{{{c^2}}}} \right) = \rho_i \dot B_i\left( \phi  \right).\label{s25}
\end{equation}
Summing over all species matter $i$ to Eq. (\ref{s25}), we get
\begin{equation}{\dot \rho _{{\mathop{\rm int}} }} + 3H\left( {{\rho _{{\mathop{\rm int}} }} + \frac{{{p_{{\mathop{\rm int}} }}}}{{{c^2}}}} \right) = \sum\limits_i \rho_i \dot B_i\left( \phi  \right),\label{s25summ}
\end{equation}
where ${\rho _{\rm{int}}} = \sum\nolimits_i {{\rho _{i{\rm{int}}}}} $ and ${p _{\rm{int}}} = \sum\nolimits_i {{p _{i{\rm{int}}}}} $, respectively.

The sum of Eqs. (\ref{s24}) and (\ref{s25summ}) gives
\begin{equation}
{\dot \rho _{{\rm{eff}}}} + 3H\left( {{\rho _{{\rm{eff}}}} + \frac{{{p_{{\rm{eff}}}}}}{{{c^2}}}} \right) = 0,\label{s26}
\end{equation}
with the effective energy density of the scalar field ${\rho _{{\rm{eff}}}} = {\rho _\phi } + {\rho _{{\mathop{\rm int}} }}$, and the effective pressure of the scalar field ${p_{{\rm{eff}}}} = {p_\phi } + {p_{{\mathop{\rm int}} }}$, respectively. Thus, the effective energy of the scalar field is conserved. Eq. (\ref{s26}) implies that for a coupled scalar field the effective equation of state becomes
\begin{equation}
{w_{{\rm{eff}}}} \equiv \frac{{{p_{{\rm{eff}}}}}}{{{\rho _{{\rm{eff}}}}{c^2}}}.\label{s26pluse}
\end{equation}

Summing over all species matter $i$ to Eq. (\ref{s14}), we get
\begin{equation}
\dot \rho  + 3H\left( {\rho  + \frac{p}{{{c^2}}}} \right)=0,\label{s14summ}
\end{equation}
where ${\rho} = \sum\nolimits_i {{\rho _{i}}} $ and ${p } = \sum\nolimits_i {{p _{i}}} $, respectively. The conserved energy density implies that the corresponding equation of state can be introduced, \emph{i.e}., ${w} \equiv {p}/\left( {{\rho}{c^2}} \right)$. If we assume that all the matter species of non-interacting perfect fluids have the same parameter $w_i$, we have ${p} = {w_i}{\rho} {c^2}$. We prefer to use ${w_i}$ instead of ${w}$ to emphasize species of matter.

The sum of Eqs. (\ref{s14summ}) and (\ref{s26}), or the sum of Eqs. (\ref{s24})) and (\ref{s21summ}) results in a conservation equation of the total energy density by
\begin{equation}
{\dot \rho _{{\rm{total}}}} + 3H\left( {{\rho _{{\rm{total}}}} + \frac{{{p_{{\rm{total}}}}}}{{{c^2}}}} \right) = 0,\label{s27}
\end{equation}
where ${\rho _{{\rm{total}}}} \equiv \rho  + {\rho _\phi } + {\rho _{{\mathop{\rm int}} }} \equiv \rho  + {\rho _{{\rm{eff}}}} \equiv {\rho _{\rm{m}}} + {\rho _\phi }$, ${p_{{\rm{total}}}} \equiv p + {p_\phi } + {p_{{\mathop{\rm int}} }} \equiv p + {p_{{\rm{eff}}}} \equiv {p_{\rm{m}}} + {p_\phi }$. The total energy definition is different from the equation (4.82) in literature \cite {z12} because its total energy density is defined by $\rho+\rho_{\phi}$ and is considered being conserved shown in equation (4.83) in the literature. The total energy should be $\rho_{\rm{m}}+\rho_{\phi}$ rather than $\rho+\rho_{\phi}$.  Now, using ${A^{1 - 3{w_i}}}\left( \phi  \right)$ instead of  $B_i\left( \phi  \right)$, Eq. (\ref{s21summ}) becomes
\begin{equation}
{\dot \rho _{\rm{m}}} + 3H\left( {{\rho _{\rm{m}}} + \frac{{{p_{\rm{m}}}}}{{{c^2}}}} \right) = \sum\limits_i {{\rho _i}\frac{{d{A^{1 - 3{w_i}}}\left( \phi  \right)}}{{dt}}} ,\label{s21plus}
\end{equation}
where ${\rho _{\rm{m}}} = \sum\nolimits_i {{\rho _i}{A^{1 - 3{w_i}}}\left( \phi  \right)} $ is the physics density of matter, ${p_{\rm{m}}} = \sum\nolimits_i {{p_i}{A^{1 - 3{w_i}}}\left( \phi  \right)} $ is the physics pressure of matter fluid, respectively. And Eq. (\ref{s24}) becomes
\begin{equation}
{\dot \rho _\phi } + 3H\left( {{\rho _\phi } + \frac{{{p_\phi }}}{{{c^2}}}} \right) =  - \sum\limits_i {{\rho _i}\frac{{d{A^{1 - 3{w_i}}}\left( \phi  \right)}}{{dt}}} .\label{s24plus}
\end{equation}
Eqs. (\ref{s21plus}) and (\ref{s24plus}) describe the energy exchange between matter and the scalar field, while Eq. (\ref{s27}) describes the total energy conservation law. Since $d{A^{1 - 3{w_i}}}\left( \phi  \right){\rm{/}}dt = \left( {1 - 3{w_i}} \right){A^{ - 3{w_i}}}\dot A + \left( { - 3{{\dot w}_i}} \right){A^{1 - 3{w_i}}}\ln A$, one can from the equation get the physical image of the energy transfer between matter and the scalar field as follows: The first term on the right-hand of the equation is related to the work done upon the system of matter by the scalar field; The second term on the right-hand is related to the entropy variation of the system of matter due to the temperature-dependence of the equation of state for matter fluid.

At this time, one can see that, it is necessary and important to introduce the scalar-field-independent density of matter and the corresponding conservation law of Eq. (\ref{s14}).

\subsection{The average energy of an individual particle}\label{aveappb}

We now analyze further the meaning of the conservation law of Eq. (\ref{equ6}), \emph{i.e.}, Eq. (\ref{s14}). Its solution can be easily obtained as follows:
\begin{equation}\label{equ6re1}
{\rho _i} = \frac{{{\rho _{i0}}a_0^{3({w_i} + 1)}}}{{{a^{3({w_i} + 1)}}}},
\end{equation}
with subscript `0' marking the current time. Since the number of matter particles is not altered, the number density ${n_i}$ changed with the volume expansion is given as follows
\begin{equation}\label{equ6re2}
{n_i} = \frac{{{n_0}a_0^3}}{{{a^3}}}.
\end{equation}
If an average energy $\left\langle {{\varepsilon _{ci}}} \right\rangle$ per particle is introduced, one has
\begin{equation}\label{equ6re3}
{\rho _i} = {n_i}\left\langle {{\varepsilon _{ci}}} \right\rangle.
\end{equation}
The average energy of an individual particle corresponds to the thermal scalar-field-independent Compton energy of matter particle. The ratio of Eq. (\ref{equ6re1}) to Eq. (\ref{equ6re2}) gives
\begin{equation}\label{equ6re4}
\left\langle {{\varepsilon _{ci}}} \right\rangle  = {\left\langle {{\varepsilon _{ci}}} \right\rangle _0}\frac{{a_0^{3{w_i}}}}{{{a^{3{w_i}}}}},
\end{equation}
with ${\left\langle {{\varepsilon _{ci}}} \right\rangle _0} = {\rho _{i0}}/{n_{i0}}$. Eq. (\ref{equ6}) means that both the number density and the corresponding entropy are conserved. When the coupling to the scalar field is introduced, one has to describe not only the scalar-field-independent number density but also the scalar-field-dependent energy density of matter. Without the scalar field coupling, these two densities are essentially the same things. Therefore, invoking Eq. (\ref{equ6}) is by no means an expedient measure.

\subsection{The effective equation of state}\label{eesappb}

Based on Eq. (\ref{s26pluse}), we now discuss several important values of the effective equation of state for the scalar field as follows:
For ${w_i} = 0$ and $ \dot \phi  = 0$, one has
\begin{equation}\label{s26plusere1}
{w_{{\text{eff}}}} = \frac{{ - V({\phi _{\min }})}}{{V({\phi _{\min }}) + \rho {\hbar ^3}{c^5}\left[ {A({\phi _{\min }}) - 1} \right]}}.
\end{equation}
When matter density is large enough, \emph{i.e.}, $A({\phi _{\min }}) \approx 1$, one sees ${w_{{\text{eff}}}} \approx  - 1$, which mimic the cosmological constant before the current time in the pressureless case of matter source.

For ${w_i} = 1/3$, \emph{i.e.}, ${A^{1 - 3{w_i}}} = 1 $, and assuming that the scalar field has climbed along its self-interaction potential to the highest top where its kinetic energy is zero, from Eq. (\ref{s26pluse}) one has
\begin{equation}\label{s26plusere2}
{w_{{\text{eff}}}} = \frac{{ - V({\phi _{\max }})}}{{V({\phi _{\max }})}} =  - 1,
\end{equation}
which mimic the largest cosmological constant during inflation era. When the the kinetic energy density far larger than the potential energy density, \emph{i.e.,} ${\hbar ^2}{{\dot \phi }^2}/2 \gg V(\phi ) $, from Eq. (\ref{s26pluse}) one has
\begin{equation}\label{s26plusere3}
 {w_{{\text{eff}}}} = 1,
\end{equation}
which means that the scalar field can generate a great deceleration effect.

Thus, in general, $- 1 \leq {w_{{\text{eff}}}} \leq 1 $. When the scalar field oscillates around the minimum in high frequency in the case of $\left| {3H/2} \right| < {\omega _{\text{c}}}$, the time average gives ${w_{{\text{eff}}}} = 0$, which means that the scalar field behaves as pressureless matter fluid. In the case of $\left| {3H/2} \right| > {\omega _{\text{c}}}$, the over (negative) damped evolution of the scalar field occurs when the scalar field decouples with matter due to the ultrahigh temperature, and then the parameter ${w_{{\text{eff}}}}$ varies with time between -1 and 1.

\section{ The $\lambda $-independent effective mass and the changing rate of ${{\dot \phi }_{\min }}/{\phi _{\min }}$ }\label{appC}
\setcounter{section}{3}

When both the quartic self-interaction potential and the symmetry-breaking coupling function are chosen, it will be proven in this Appendix that the effective mass around the minimum of the effective potential is independent on the self-interaction potential of the scalar field. The $\lambda $-independent property is very important to circumvent Weinberg's no-go theorem, which has been demonstrated in section \ref{zeropoint}. Since the condition of the scalar field adiabatic following the the minimum of the effective potential is essential in obtaining the cosmological constant, the changing rate of ${{\dot \phi }_{\min }}/{\phi _{\min }}$ will also be discussed in this Appendix.

\subsection{The case of ${w_i} = 0$}\label{prlesappc}

For pressureless fluid of matter sources ${w_i} = 0$, the effective potential is the sum of
\begin{equation}
{V_{{\rm{eff}}}}(\phi ) = V(\phi ) + {V_{{\mathop{\rm int}} }},\label{s1}
\end{equation}
where
\begin{equation}
V(\phi ) = \frac{\lambda }{{\rm{4}}}{\phi ^4} \label{s2}
\end{equation}
and
\begin{equation}
{V_{{\mathop{\rm int}} }}{\rm{  = }}\rho {\hbar ^3}{c^5}\left[ {A(\phi ) - 1} \right] = \frac{{\rho {\hbar ^3}{c^5}}}{{{\rm{4}}{M_1}^{\rm{4}}{c^{\rm{8}}}}}{\left( {{\phi ^{\rm{2}}} - {M_2}^{\rm{2}}{c^{\rm{4}}}} \right)^2}.\label{s3}
\end{equation}
$A(\phi)$ with a symmetry-breaking shape is described by Eq. (\ref{equ16b}). The first derivative and second derivative of the effective potential with respect to the scalar field are:
\begin{equation}
{V_{{\rm{eff}},\phi}}(\phi )= \left( {\lambda  + \frac{{\rho {\hbar ^3}{c^5}}}{{{M_1}^{\rm{4}}{c^{\rm{8}}}}}} \right){\phi ^{\rm{3}}} - \left( {\frac{{\rho {\hbar ^3}{c^5}{M_2}^{\rm{2}}}}{{{M_1}^{\rm{4}}{c^{\rm{4}}}}}} \right)\phi,\label{s4}
\end{equation}

\begin{equation}
{V_{{\rm{eff}},\phi \phi }}(\phi )= {\rm{3}}\left( {\lambda  + \frac{{\rho {\hbar ^3}{c^5}}}{{{M_1}^{\rm{4}}{c^{\rm{8}}}}}} \right){\phi ^{\rm{2}}} - \left( {\frac{{\rho {\hbar ^3}{c^5}{M_2}^{\rm{2}}}}{{{M_1}^{\rm{4}}{c^{\rm{4}}}}}} \right).\label{s5}
\end{equation}

Let the first derivative of the effective potential equal to zero,\emph{ i.e.},
\begin{equation}
V_{,\phi}+\rho {\hbar ^3}{c^5} A_{,\phi}=0, \label{C5re}
\end{equation}
the extrema are obtained as:
\begin{subequations}\label{s6zong}
\begin{eqnarray}
{\phi _{_{\min }}^{\rm{2}}} & = & {\frac{{{{\rho {\hbar ^3}{M_2}^{\rm{2}}{c^4}}}}}{{\lambda{{M_1}^{\rm{4}}{c^{3}}}  + {{\rho {\hbar ^3}}}}}},\label{s6minus}\\
{\phi _{_{\max }}} & = & {0},\label{s6}
\end{eqnarray}
\end{subequations}
and then the effective mass about the minima is
\begin{equation}
m_{{\rm{eff}}}^{\rm{2}}\equiv \frac{{{V_{{\rm{eff}},\phi \phi }}({\phi _{\min }})}}{{{c^4}}}= {\frac{{2\rho {\hbar ^3}{M_2}^{\rm{2}}}}{{{M_1}^{\rm{4}}{c^{3}}}}}.\label{s8}
\end{equation}

The ${\mathbf{Z}_{\rm{2}}}$ symmetry is spontaneously broken as the scalar field choose one of the minima of the effective potential. The field values corresponding to the minima of the effective potential are nearly the same as that of the minima of the coupling function in the past of the Universe due to the coupled interaction and larger matter density. However, the difference is the minima of the effective potential are density-dependent while that of the coupling function are density-independent. This means that the minima of the effective potential will depart from the fixed minima of the coupling function in the future.

We choose $\lambda {\rm{ = 1/6}}$ so that the coefficient in Eq. (\ref{s2}) can be written naturally as $1/(4!)$. Besides the consideration of the theoretical naturalness, the choice of the value of the parameters ${M_2}$ is determined by fitting the cosmological constant. The range of ratio ${M_1}/{M_2} \geq 4 $ is obtained in the following requirements: The ratio should not only match the range of the current cosmological scale factor under the constraint of the approximately current density of the Universe, but also can correspond to a Compton wavelength of the scalar field as large as possible. The larger the ratio is, the shorter the wavelength is. As a concrete example, the ratio ${M_1}/{M_2}$ is selected in this paper to be a slightly larger integer of $2^{3}$ instead of the lower bound of $2^{2}$. Of course, the other choice is not forbidden as long as the inequality of ${M_1}/{M_2} \geq 4 $ is satisfied. Indeed, we can let ${M_1}/{M_2} = 2^{n}$ and regard $n$ as another adjustable parameter. We can find through calculation that $n=2$ to $n =4$ all works well in obtaining the cosmological constant, which means that the symmetry-breaking model is insensitive to the selection of the parameter. However, the range of the fifth interaction shown in Eq. (\ref{s8}) is sensitive to the selection of the parameter. Since the main purpose of this paper is to display the symmetry-breaking coupling function, the detailed determination of ${M_1}/{M_2}$ will be discussed elsewhere.

Due to the change of the matter density, the minimum position of the effective potential will change. The changing rate can be defined by ${{\dot \phi }_{\min }}/{\phi _{\min }}$. Differentiating Eq. (\ref{C5re}) with respect to time $t$, and using the conservation law of Eq. (\ref{s14summ}) with assuming pressureless matter source $p=0$, one has
\begin{subequations}\label{s8pzong}
\begin{eqnarray}
{\frac{{{{\dot \phi }_{{\text{min}}}}}}{{{\phi _{{\text{min}}}}}}} & = & { - H \cdot \frac{{3{V_{,\phi }}({\phi _{{\text{min}}}})}}{{{\phi _{{\text{min}}}}m_{{\text{eff}}}^2{c^4}}}} \label{s8plusr} \\
& = & { H \cdot \frac{{3c{\hbar ^3}\rho {A_{,\phi }}({\phi _{{\text{min}}}})}}{{{\phi _{{\text{min}}}}m_{{\text{eff}}}^2}}}.\label{s8plus}
\end{eqnarray}
\end{subequations}
Therefore, the changing rate of the minimum position of the effective potential can be expressed by the Hubble parameter.

Noticing Eq. (\ref{s6minus}), Eq. (\ref{s8pzong}) becomes
\begin{equation}
\frac{{{{\dot \phi }_{\min }}}}{{{\phi _{\min }}}} = - \frac{{3H\lambda M_1^4{c^3}}}{{2\left( {\lambda M_1^4{c^3} + \rho {\hbar ^3}} \right)}} .\label{s8plus2}
\end{equation}
The negative sign on the right-hand side of the equation describes the fact that the trend of the changing rate is opposite to that of the Hubble rate. For $\rho  \gg  {\lambda {M_1}^{\rm{4}}{c^3}}/{{\hbar ^3}} \sim  10^{-30}\,\rm{kg/m^3}$, the absolute value of the changing rate is much less than that of the Hubble rate. In general, one has
\begin{equation}
\left| {\frac{{{{\dot \phi }_{{\text{min}}}}}}{{{\phi _{{\text{min}}}}}}} \right| < \left| {\frac{{3H}}{2}} \right| .\label{s8plusre3}
\end{equation}
Here $\left| {3H/2} \right|$ denotes the damping rate in the oscillation equation of the scalar field. Thus, for pressureless matter source the scalar field can adiabatically follow the minimum [see also the adiabatic condition of Eq. (\ref{equ18replus})].

\subsection{The general case of ${w_i} \neq 0$ }\label{prappc}

We now consider the general case of ${w_i} \neq 0$. In this case the value of the scalar field at the minimum cannot be described by a simply analysis expression. One can only give the region $0 \leq \phi _{\min }^2 \leq M_2^2{c^4}$ which is the same as the case of ${w_i} = 0$. Since the self-interaction potential of Eq. (\ref{s2}) is a quartic form, the effective mass of the scalar field around the minimum can be derived just only from the interaction potential shown by Eq. (\ref{equ8plusa}), \emph{i.e.},
\begin{equation}
m_{{\text{eff}}}^2 = \frac{{ - 2{V_{{\text{int}},\phi \phi }}(\phi  = 0)}}{{{c^4}}}.\label{s8plusre4}
\end{equation}
For the sake of simplicity, we assume that all the species of non-interacting perfect fluids of matter sources have the same parameter $w_i$. Thus, Eq. (\ref{equ8plusa})) becomes as follows:
\begin{eqnarray}
V_{{\mathop{\rm int}} } =  {{\rho}{\hbar ^3}{c^5}\left[ {{A^{1 - 3{w_i}}}\left( \phi  \right) - 1} \right]} \label{equ8shape2}
\end{eqnarray}
with ${\rho} = \sum\nolimits_i {{\rho _{i}}} $. Combining Eqs. (\ref{s8plusre4}) and (\ref{equ8shape2}) with the symmetry-breaking coupling function of Eq. (\ref{equ16b}), one has
\begin{equation}
m_{{\text{eff}}}^2 = \frac{{2\rho {\hbar ^3}M_2^2\left( {1 - 3{w_i}} \right)}}{{M_1^4{c^3}}}{\left( {1 + \frac{{M_2^4}}{{4M_1^4}}} \right)^{ - 3{w_i}}}.\label{s8plusre5}
\end{equation}
When the fluid of matter source approaches to be relativistic,\emph{ i.e.}, ${w_i} \to 1/3$, the effective mass around the minimum approaches to zero. In the contraction process, since matter fluid evolves from norelativistic to relativistic, the scalar field must experience chirped under-negative-damping oscillations from higher frequency to lower one, and then shift to an over-negative-damping motion. In the expansion process the evolution of the Universe is in the reverse sequence. To describe whether the scalar field is able to follow adiabatically the minimum or not, by using the same method as in deriving Eq. (\ref{s8pzong}) and noticing the $\phi$-independence of ${w_i}$ denoted by Eq. (\ref{s18}), the changing rate of the minimum position of the effective potential is obtained as follows:
\begin{subequations}\label{s8plusre6zong}
\begin{eqnarray}
{\frac{{{{\dot \phi }_{\min }}}}{{{\phi _{\min }}}}} & = & { - 3H\left( {1 + {w_i}} \right)\frac{{{V_{,\phi }}\left( {{\phi _{\min }}} \right)}}{{{\phi _{\min }}m_{{\text{eff}}}^2{c^4}}}} \label{s8plusre6a} \\
  & = & { \frac{{3H\rho \left( {1 + {w_i}} \right){\hbar ^3}c\left( {1 - 3{w_i}} \right){A^{ - 3{w_i}}}\left( {{\phi _{\min }}} \right){A_{,\phi }}\left( {{\phi _{\min }}} \right)}}{{{\phi _{\min }}m_{{\text{eff}}}^2}}}.\label{s8plusre6}
\end{eqnarray}
\end{subequations}
Substituting that
\begin{subequations}\label{s8plusre7zong}
\begin{eqnarray}
{{V_{,\phi }}\left( {{\phi _{\min }}} \right)} & = & {\lambda \phi _{\min }^3}, \label{s8plusre7a}\\
{{A_{,\phi }}\left( {{\phi _{\min }}} \right)} & = & {\frac{{{\phi _{\min }}\left( {\phi _{\min }^2 - M_2^2{c^4}} \right)}}{{M_1^4{c^8}}}}, \label{s8plusre7}
\end{eqnarray}
\end{subequations}
into Eqs. (\ref{s8plusre6a}) and (\ref{s8plusre6}), respectively, one has
\begin{subequations}\label{s8plusre8zong}
\begin{eqnarray}
{\frac{{{{\dot \phi }_{\min }}}}{{{\phi _{\min }}}}} &=&  {\frac{{ - 3H\left( {1 + {w_i}} \right)\lambda \phi _{\min }^2}}{{m_{{\text{eff}}}^2{c^4}}}}, \label{s8plusre8a}\\
&=& { \frac{{3H\left( {1 + {w_i}} \right){{\left( {1 + \frac{{M_2^4}}{{4M_1^4}}} \right)}^{3{w_i}}}}}{{2{A^{3{w_i}}}\left( {{\phi _{\min }}} \right)}}\frac{{\left( {\phi _{\min }^2 - M_2^2{c^4}} \right)}}{{M_2^2{c^4}}}}.\label{s8plusre8}
\end{eqnarray}
\end{subequations}
Thus, with ${w_i} $ increasing from 0 to 1/3, the evolution of the scalar field must undergo a stage that $\left| {{{\dot \phi }_{{\text{min}}}}/{\phi _{{\text{min}}}}} \right|$ is larger than the damping rate $\left| {3H/2} \right|$. In this stage, the scalar field is not able to sit stably at the minimum. In the case of ${w_i} = 0$, from Eq. (\ref{s8plus2}) one can see that the stability of the minimum enhances with the density of matter increasing. In the case of ${w_i} \neq 0$, the stability of the minimum cannot enhance further with the density of matter increasing. In fact, the stability even decays with matter density increasing due to the decoupled effect between the scalar field and matter at the extremely temperature. The fragile stability of the adiabatic following can lead to a great result when the effective mass of the scalar field around the minimum approaches to zero. For example, in the contraction process, the magnitude of the scalar field will grow exponentially in the over-negative-damping case. This is in sharp contrast to the case when the Universe is at the maximum radius. In that time, Hubble rate is zero but the effective mass of the scalar field is not zero, though the mass is extremely small. Both the condition of adiabatic tracking the minimum and the oscillation condition can always hold in the case of ${w_i} = 0$. The two conditions cannot hold forever in the case of ${w_i} \neq 0$ (see also section \ref{startneda}). Periodically vibrational state of the scalar field only appears when the oscillation condition is satisfied but the adiabatic condition is not satisfied (see caption of Figure \ref{figure2}), which corresponds to the overshooting phenomenon in the under-(negative-)damped case. When both conditions are not satisfied, the field can neither adiabatically follow the minimum nor oscillate around the minimum, which corresponds to the overshooting phenomenon in the over-(negative-)damped case (see also section \ref{overshoot}).

In addition, it worth noting that the mass of the scalar field corresponding to the minimum radius of the Universe (the maximum absolute value of the scalar field) is not related to any minimum of the effective potential. The mass is essential the self-mass of the scalar field since only the self-interaction potential is considered. The evolution of the scalar field now belongs to over-(negative-)damped motion. In fact, it does not make sense to talk about adiabatic following some minimum in this case. One often uses so-called slow-roll conditions in this case. However, the requirement of the slow-roll conditions is not always reasonable (see also section \ref{concavevex1}).

\subsection{ The coupling coefficient $\beta $ }\label{betaappc}

The coupling coefficient $\beta $ is dependent on the ambient density, which corresponds to the linear part of the coupling function $A\left( \phi  \right)$. Using Eqs. (\ref{32plus0}) and (\ref{equ16b}), one has
\begin{eqnarray}
\beta \left( \phi  \right)= {M_{{\rm{Pl}}}}{c^2}\frac{{4{{{\phi}}}\left( {\phi^{\rm{2}} - {M_2}^{\rm{2}}{c^{\rm{4}}}} \right)}}{{{{{\rm{4}}{M_1}^{\rm{4}}{c^{\rm{8}}}}} + {{\left( {\phi ^{\rm{2}} - {M_2}^{\rm{2}}{c^{\rm{4}}}} \right)}^2}}}.\label{s9minus}
\end{eqnarray}
Therefore, the coupling coefficient $\beta $ is not a constant but depends on $\phi$. At $\phi_{\rm{min}}$, the coupling coefficient $\beta $ is
\begin{equation}
\beta \left( \phi_{\rm{min}}  \right)={M_{{\rm{Pl}}}}{c^2}\frac{{4{{{\phi _{\rm{min}}}}}\left( {\phi _{\min }^{\rm{2}} - {M_2}^{\rm{2}}{c^{\rm{4}}}} \right)}}{{{{\rm{4}}{M_1}^{\rm{4}}{c^{\rm{8}}}} + {{\left( {\phi _{\rm{min}}^{\rm{2}} - {M_2}^{\rm{2}}{c^{\rm{4}}}} \right)}^2}}}.\label{s9}
\end{equation}
Thus, $\beta\left( \phi_{\rm{min}}  \right)>0 $ if ${\phi _{\min }}<0$, and $\beta\left( \phi_{\rm{min}}  \right)<0 $ if ${\phi _{\min }}>0$.
Consider a space-variable source density embedded in a homogeneous ambient density ${\rho _{\rm{b}}}$, the coupling coefficient $\beta (\phi)$ can be approximated to $\beta (\phi_{\rm{b}})$ with ${\phi _{\rm{b}}}={\phi _{\min }}\left( {{\rho _{\rm{b}}}} \right)$. In fact, one often use $\beta =\beta (\phi_{\rm{b}})$ to denote the coupling coefficient.

The value of the coupling coefficient $\beta $ denotes the coupling strength between matter and the scalar field. For lower density of matter in the local environment, the coupling strength is much larger than gravitational. This means that the scalar fifth force might be detectable provided that future experiments are designed properly. If ${M_2} = 0$, one has ${\phi _{\min }} = 0$ and $\beta \left( \phi_{\rm{min}}  \right)= 0$ corresponding to a zero fifth force at $\phi_{\rm{min}}$.

\section{Fundamental physical constants}\label{appD}
\setcounter{section}{4}
Some physical constants we used in our calculation are:
The speed of light in vacuum  $c = 2.99792458 \times {10^8}\,{\rm{ m}} \cdot {{\rm{s}}^{ - 1}}$, the gravitational constant $G = 6.67430 \times {10^{ - 11}}\,{{\rm{m}}^3} \cdot {\rm{k}}{{\rm{g}}^{ - 1}} \cdot {{\rm{s}}^{ - 2}}$ , Planck's constant $h = 6.62607015 \times {10^{ - 34}}\,{\rm{ J}} \cdot {\rm{s}}$, the elementary charge $e = 1.602176624 \times {10^{ - 19}}\,{\rm{ C}}$ .

We do not use the units $c = \hbar  = 1$  but use the International System of Units in which $c$ and $\hbar$ appears explicitly. The purpose is not only for the calculation to compare with observation dada but also for the mark of the quantum effect in the cosmic scale. Quantum mechanics is often regarded as a theory for micro scale.


\begin{thebibliography}{100}
\bibitem{z65replus} If we extrapolate back to the early cosmological time when matter density is extremely high, a corollary is that the system temperature is extremely high and then matter is decoupled to the scalar field. The nearly fixed cosmological constant is no longer valid since the interaction potential trapping vanishes. The Universe evolution in the early period can be described by inflation models. For flat and open spaces, since the Hubble parameter cannot equal to zero [see Eq. (\ref{equ4})], one of the slow-roll conditions $| {\ddot \phi } | \ll 3| {H\dot \phi } |$ is reasonable as in \cite{z55, z26}. Thus, we need not to analyse these two cases. However, for closed space, since the Hubble parameter can pass through zero, the slow-roll condition above is not always valid and needs further investigation. The detailed discussion is shown in section \ref{contraction}.
\bibitem{z115} One should carefully distinguish the presentation of the Hubble constant in both the Jordan- and the Einsteinn-frame. Although the two frames can perhaps be used to relieve the Hubble tenstion, it is not an appropriate time to discuss the interesting issue. The discussion may lead to more confusion due to the model-dependent determination of some paramieters.
\end{thebibliography}
\end{document}